\documentclass[12pt]{article}
\begin{document}
\newcommand{\beq}{\begin{equation}}
\newcommand{\eeq}{\end{equation}}
\newcommand{\beqa}{\begin{eqnarray}}
\newcommand{\eeqa}{\end{eqnarray}}
\newcommand{\beqar}{\begin{eqnarray*}}
\newcommand{\eeqar}{\end{eqnarray*}}
\newcommand{\al}{\alpha}
\newcommand{\be}{\beta}
\newcommand{\del}{\delta}
\newcommand{\D}{\Delta}
\newcommand{\eps}{\epsilon}
\newcommand{\ga}{\gamma}
\newcommand{\Ga}{\Gamma}
\newcommand{\ka}{\kappa}
\newcommand{\nn}{\nonumber}
\newcommand{\inn}{\!\cdot\!}
\newcommand{\h}{\eta}
\newcommand{\ii}{\iota}
\newcommand{\kk}{\varphi}
\newcommand\F{{}_3F_2}
\newcommand{\la}{\lambda}
\newcommand{\La}{\Lambda}
\newcommand{\na}{\prt}
\newcommand{\Om}{\Omega}
\newcommand{\om}{\omega}
\newcommand{\p}{\phi}
\newcommand{\sig}{\sigma}
\renewcommand{\t}{\theta}
\newcommand{\z}{\zeta}
\newcommand{\ssc}{\scriptscriptstyle}
\newcommand{\eg}{{\it e.g.,}\ }
\newcommand{\ie}{{\it i.e.,}\ }
\newcommand{\labell}[1]{\label{#1}} 
\newcommand{\reef}[1]{(\ref{#1})}
\newcommand\prt{\partial}
\newcommand\veps{\varepsilon}
\newcommand{\pol}{\varepsilon}
\newcommand\vp{\varphi}
\newcommand\ls{\ell_s}
\newcommand\cF{{\cal F}}
\newcommand\cA{{\cal A}}
\newcommand\cS{{\cal S}}
\newcommand\cT{{\cal T}}
\newcommand\cV{{\cal V}}
\newcommand\cL{{\cal L}}
\newcommand\cM{{\cal M}}
\newcommand\cN{{\cal N}}
\newcommand\cG{{\cal G}}
\newcommand\cH{{\cal H}}
\newcommand\cI{{\cal I}}
\newcommand\cJ{{\cal J}}
\newcommand\cl{{\iota}}
\newcommand\cP{{\cal P}}
\newcommand\cQ{{\cal Q}}
\newcommand\cg{{\it g}}
\newcommand\cR{{\cal R}}
\newcommand\cB{{\cal B}}
\newcommand\cO{{\cal O}}
\newcommand\tcO{{\tilde {{\cal O}}}}
\newcommand\bg{\bar{g}}
\newcommand\bb{\bar{b}}
\newcommand\bH{\bar{H}}
\newcommand\bX{\bar{X}}
\newcommand\bK{\bar{K}}
\newcommand\bR{\bar{R}}
\newcommand\bZ{\bar{Z}}
\newcommand\bxi{\bar{\xi}}
\newcommand\bphi{\bar{\phi}}
\newcommand\bpsi{\bar{\psi}}
\newcommand\bprt{\bar{\prt}}
\newcommand\bet{\bar{\eta}}
\newcommand\btau{\bar{\tau}}
\newcommand\hF{\hat{F}}
\newcommand\hA{\hat{A}}
\newcommand\hT{\hat{T}}
\newcommand\htau{\hat{\tau}}
\newcommand\hD{\hat{D}}
\newcommand\hf{\hat{f}}
\newcommand\hg{\hat{g}}
\newcommand\hp{\hat{\phi}}
\newcommand\hi{\hat{i}}
\newcommand\ha{\hat{a}}
\newcommand\hb{\hat{b}}
\newcommand\hQ{\hat{Q}}
\newcommand\hP{\hat{\Phi}}
\newcommand\hS{\hat{S}}
\newcommand\hX{\hat{X}}
\newcommand\tL{\tilde{\cal L}}
\newcommand\hL{\hat{\cal L}}
\newcommand\tG{{\widetilde G}}
\newcommand\tg{{\widetilde g}}
\newcommand\tphi{{\widetilde \phi}}
\newcommand\tPhi{{\widetilde \Phi}}
\newcommand\te{{\tilde e}}
\newcommand\tk{{\tilde k}}
\newcommand\tf{{\tilde f}}
\newcommand\ta{{\tilde a}}
\newcommand\tb{{\tilde b}}
\newcommand\tR{{\tilde R}}
\newcommand\teta{{\tilde \eta}}
\newcommand\tF{{\widetilde F}}
\newcommand\tK{{\widetilde K}}
\newcommand\tE{{\widetilde E}}
\newcommand\tpsi{{\tilde \psi}}
\newcommand\tX{{\widetilde X}}
\newcommand\tD{{\widetilde D}}
\newcommand\tO{{\widetilde O}}
\newcommand\tS{{\tilde S}}
\newcommand\tB{{\widetilde B}}
\newcommand\tA{{\widetilde A}}
\newcommand\tT{{\widetilde T}}
\newcommand\tC{{\widetilde C}}
\newcommand\tV{{\widetilde V}}
\newcommand\thF{{\widetilde {\hat {F}}}}
\newcommand\Tr{{\rm Tr}}
\newcommand\tr{{\rm tr}}
\newcommand\STr{{\rm STr}}
\newcommand\hR{\hat{R}}
\newcommand\M[2]{M^{#1}{}_{#2}}

\newcommand\bS{\textbf{ S}}
\newcommand\bI{\textbf{ I}}
\newcommand\bJ{\textbf{ J}}

\begin{titlepage}
\begin{center}

\vskip 2 cm
{\LARGE \bf    Effective action of  type II superstring theories   \\  \vskip 0.25 cm  at order $\alpha'^3$: NS-NS couplings
 }\\
\vskip 1.25 cm
   Mohammad R. Garousi\footnote{garousi@um.ac.ir}

\vskip 1 cm
{{\it Department of Physics, Faculty of Science, Ferdowsi University of Mashhad\\}{\it P.O. Box 1436, Mashhad, Iran}\\}
\vskip .1 cm
 \end{center}

\begin{abstract}
 Recently, it has been shown  that  the minimum number of  gauge invariant  couplings  for   $B$-field, metric and dilaton at order $\alpha'^3$ is 872. These couplings,  in a particular scheme,  appear in 55 different structures.   In this paper, up to an overall factor,  we fix all parameters in type II supertirng theories by requiring the reduction of the couplings on a circle to be invariant under  T-duality transformations. We find that there are 445 non-zero couplings which appear  in 15 different structures. The couplings are fully consistent with the partial couplings that have been found in the literature by the four-point S-matrix element and by the non-linear Sigma model methods.
  
\end{abstract}
\end{titlepage}

\section{Introduction}

String theory is a candidate quantum theory for gravity   which includes  a finite number of massless fields and a  tower of infinite number of  massive fields reflecting the stringy nature of the gravity.
An efficient way to study different phenomena in this theory is to use an effective action  which includes     only  the massless fields \cite{Schwarz:1983qr,Schwarz:1983wa,Howe:1983sra,Carr:1986tk,Witten:1995ex}.
The effective action  has a double expansions. The genus-expansion which includes  the  classical tree-level  and a tower of  quantum  loop-level  corrections, and the   string-expansion which is an expansion in terms of  higher derivative couplings  at each loop level. 
The effective action   may be found  by applying   various  techniques   in string theory. One of them is T-duality  which shows up  when one compactifies the $D$-dimensional theory on $d$-dimensional tours $T^d$ \cite{Giveon:1994fu,Alvarez:1994dn}. 

The massless fields in the 26-dimensional   bosonic sting theory and in the 10-dimensional  heterotic string theory after truncating the Yang-Mills gauge fields,  are metric, dilaton and B-field. The corresponding effective actions at the tree-level and at the  two-derivative order, reveal   $O(d,d,R)$ symmetry after one   dimensionally reduces them on $T^d$ \cite{Veneziano:1991ek,Meissner:1991zj,Maharana:1992my}. The  number and the structure of couplings in the effective action at the higher orders of derivatives are not unique. They   are changed under field redefinitions, \ie they are scheme dependent \cite{Metsaev:1987zx}.  It is shown in \cite{Meissner:1996sa,Kaloper:1997ux} that the $O(d,d,R)$ symmetry also appears explicitly at the four-derivative order when one uses a specific scheme for the $D$-dimensional effective action. Using closed string field theory, it is proved in \cite{Sen:1991zi} that this symmetry in fact exists in the full tree-level effective actions including  all higher derivative couplings. Similar analysis has been done in \cite{Hohm:2014sxa} for the heterotic string theory. This global symmetry may then be used to constrain  the possible higher derivative couplings in the classical effective action of the string theory.

 At the most simple case when one dimensionally reduces the effective action on a circle, it has been shown in  \cite{Garousi:2017fbe,Razaghian:2017okr,Razaghian:2018svg,Garousi:2019wgz,Garousi:2019mca,Mashhadi:2020mzf} that invariance of the classical effective actions of string theory and its non-perturbative objects  under the $z_2$-subgroup of $O(1,1,R)$ can produces  many already known and unknown couplings in the  effective actions. In particular, it has been shown in \cite{Garousi:2019wgz,Garousi:2019mca} that this $Z_2$ symmetry  as well as various gauge symmetries corresponding to the massless fields  fix all four- and six-derivative couplings in the bosonic string theory up to an overall factor. The $Z_2$-transformations  are the Buscher rules \cite{Buscher:1987sk,Buscher:1987qj} plus their higher derivative corrections. The form of these corrections, however, depend on the scheme that one uses for the $D$-dimensional couplings \cite{Garousi:2019wgz}.
The $O(d,d,R)$ symmetry  for $d>1$ has been also used in \cite{Godazgar:2013bja,Hohm:2015doa,Eloy:2019hnl,Eloy:2020dko} to construct  the higher derivative couplings in the $D$-dimensional effective action. It has been shown in \cite{Baron:2017dvb,Eloy:2019hnl,Eloy:2020dko} that for $d>1$ the Green-Schwarz type mechanism is required to have $O(d,d,R)$ symmetry at four-derivative order.  Another T-duality based framework for constructing the higher derivative couplings  in string theory is the Double Field Theory \cite{Siegel:1993xq,Siegel:1993th,Siegel:1993bj,Hull:2009mi,Aldazabal:2013sca,Hohm:2010pp,Hohm:2014xsa,Marques:2015vua,Garousi:2018qes,Baron:2020xel} in which the dimension of spacetime  is doubled and the $O(D,D,R)$ symmetry is imposed on the $2D$-dimensional theory before using dimensional reduction.

The massless fields in the effective action of type II superstring theory appear in four sectors: NS-NS sector which has bosonic fields metric, dilaton and B-field, R-R sector which has bosonic  $n$-form with $n=0,1,2,3,4$, and fermionic  sectors NS-R and  R-NS  in which we are not interested.  Inspired by the bosonic and heterotic string theories \cite{Sen:1991zi,Hohm:2014sxa}, we speculate that the bosonic part of the classical  effective action of type II superstring theories is also invariant under   the $Z_2$ symmetry. 
At the two-derivative order, the NS-NS couplings are the same as the couplings in the bosonic string theory, hence they have the expected symmetry. It has been shown in \cite{Garousi:2019jbq} that the   R-R couplings at two-derivative order can  be written in the $Z_2$-invariant form after dimensionally reducing them on a circle. The first higher derivative corrections in this case are at eight-derivative order. In terms of order of R-R field strength, these couplings appear in 5 parts, \ie the couplings with $0,2,4,6$ and $8$ R-R field strengths. They may be found by imposing  the $Z_2$ symmetry. At the leading order of $\alpha'$, the order of R-R fields are not changed  under  the  $Z_2$-transformations\cite{Meessen:1998qm}, however, at the higher orders they may be changed.   In this paper, we are going to find the couplings in the first part, \ie the NS-NS couplings,  by imposing the  speculated $Z_2$ symmetry as well as various gauge symmetries, and leave the construction of the couplings involving the R-R fields to the future works.



The outline of the paper is as follows: In section 2, we write the  NS-NS gauge invariant couplings at order   $\alpha'^3$ in a minimal scheme that have been found in \cite{Garousi:2020mqn}.  In section 3, we impose the $Z_2$-symmetry on the gauge invariant couplings to find their corresponding parameters.  We have found both the effective action and the corresponding  T-duality transformations. However, since the expressions for the T-duality transformations are very lengthy we will   write only the effective action. We have found that there are  445 non-zero couplings in the effective action which appear in 15 different structures. A few of them  which have at lest four NS-NS fields are fully consistent with the S-matrix element of four NS-NS vertex operators and with non-linear sigma model.    In section 4, we briefly discuss our results.

  \section{Gauge invariance constraint}\label{sec.2}

 The classical effective action of type II superstring theories has the following   string-expansion or $\alpha'$-expansion in the string frame:
\beqa
\bS_{\rm eff}&=&\sum^\infty_{n=0}\alpha'^n\bS_n=\bS_0+\alpha'^3\bS_3+\alpha'^4 \bS_4+\cdots\ ; \quad \bS_n= -\frac{2}{\kappa^2}\int d^{10} x\sqrt{-G} \mathcal{L}_n\labell{seff}
\eeqa
The effective action must be invariant under the coordinate transformations, under the $B$-field  and R-R gauge transformations. $\cL_0$ is the Lagrangian of type II supergravity. Its bosonic part in democratic form is given by \cite{Fukuma:1999jt} 
 \beqa
\cL_0&=&e^{-2\Phi} \left(  R + 4\nabla_{a}\Phi \nabla^{a}\Phi-\frac{1}{12}H^2\right)+ \sum_{n=1}^{9}\frac{1}{n!}F^{(n)}\cdot F^{(n)}\labell{S0bf}
\eeqa
where $H$ is field strength of the $B$-field and $F^{(n)}$ is the R-R field strength. The first part has NS-NS couplings and the second part has metric and two R-R field strengths.
The R-R gauge symmetry dictates that the $\alpha'^3$-corrections has the following five parts:
\beqa
\cL_3&=&\cL_3^0+\cL_3^2+\cL_3^4+\cL_3^6+\cL_3^8
\eeqa
where $\cL_3^0$  has only NS-NS couplings, $\cL_3^2$ has NS-NS and two R-R fields, $\cL_3^4$ has NS-NS and four R-R fields, and so on. 
In this paper we are interested only in the tree-level couplings in the NS-NS sector. 

The NS-NS gauge symmetries dictate that there are 23996 couplings at order $\alpha'^3$ in 202 different structures, \ie $H^8$-structure, $R^4$-structure, $(\nabla\Phi)^8$-structure, and so on. However, many of couplings are related to each other by Bianchi identities, total derivative terms and field redefinitions. It has been shown in \cite{Garousi:2020mqn} that the minimum number of independent gauge invariant couplings is  872. These couplings in one particular scheme which appear in 55 different structures, have been found in \cite{Garousi:2020mqn}. They are  
\beqa
\cL_3^0&=&c_1 H_{\alpha}{}^{\delta \epsilon} H^{\alpha \beta \
\gamma} H_{\beta}{}^{\varepsilon \mu} \
H_{\gamma}{}^{\zeta \eta} H_{\delta \varepsilon}{}^{\theta} H_{\
\epsilon \zeta}{}^{\iota} H_{\mu \iota}{}^{\kappa} 
H_{\eta \theta \kappa} +\cdots+c_{520} H^{\alpha \beta \gamma} R_{\beta 
\mu \delta}{}^{\zeta} R_{\gamma \zeta 
\epsilon \varepsilon} \nabla^{\mu}\nabla^{\varepsilon}H_{\alpha}{}^{\delta \epsilon}\nn\\&& + c_{521}H^{\beta \gamma \delta} R_{\alpha \zeta \beta \epsilon} R_{\gamma \varepsilon \delta \mu} \nabla^{\alpha}\Phi \nabla^{\zeta}H^{\epsilon \varepsilon \mu}  +\cdots+ c_{872} (H^2)^2 \nabla_{\theta}H_{\mu \zeta \eta} \nabla^{\theta}H^{\mu \zeta \eta}\labell{T54}
\eeqa
where $c_1, \cdots, c_{872}$ are parameters that the gauge symmetry can not fix them. Except the coupling with coefficient $c_{520}$, all other couplings have no term with three derivatives.  We refer the interested reader to \cite{Garousi:2020mqn} for the explicit form of all couplings. 
A few  of these couplings have been found in \cite{Garousi:2020mqn} by comparing the above couplings with the $\alpha'$-expansion of the S-matrix element of four NS-NS vertex operators \cite{Gross:1986iv,Gross:1986mw}. In the next section, we show that imposition of the $Z_2$-symmetry   on the above couplings  can   fix all 872 parameters in the type II superstring theory in terms of an overall factor.

 \section{T-duality invariance constraint}\label{sec.3}

 The T-duality  constraint on the $D$-dimensional effective action $\bS_{\rm eff}$,  in the most simple form, is first to dimensionally reduce  theory on a circle with $U(1)$ isometry to produce its corresponding   $(D-1)$-dimensional effective action $S_{\rm eff}(\psi)$ where $\psi$ represents all massless fields in the $(D-1)$-dimensional base space. Then one has to transform it under the T-duality transformations to produce  $S_{\rm eff}(\psi')$ where $\psi'$ represents the T-duality transformations of the massless fields in the base space. The T-duality invariance constraint is then
 \beqa
 S_{\rm eff}(\psi)-S_{\rm eff}(\psi')&=&\int d^{D-1}x \sqrt{-\bg}\nabla_a(e^{-2\bphi}J^a(\psi))\labell{TST}
 \eeqa
where $\bg_{ab}$, $\bphi$ are the metric and dilaton in the    base space, and $J^a$ is an arbitrary covariant vector made of the $(D-1)$-dimensional fields and their derivatives. It has the following $\alpha'$-expansion:
\beqa
J^a&=&\sum^\infty_{n=0}\alpha'^nJ^a_n\labell{Ja}
\eeqa
 where $J^a_n$ is an arbitrary covariant vector at order $\alpha'^n$   made of the $(D-1)$-dimensional fields. The T-duality transformation $\psi'$ also has $\alpha'$-expansion, \ie
\beqa
\psi'&=&\sum^\infty_{n=0}\alpha'^n\psi'_n\labell{psi}
\eeqa
where $\psi'_0$ is the Buscher rules, $\psi'_1$ contains corrections to the Buscher rules at order $\alpha'$ and so on. The T-duality transformation forms a  $Z_2$-group, \ie
\beqa
\psi\rightarrow \psi'\rightarrow \psi
\eeqa
This $Z_2$-symmetry alone can not fix completely the corrections to the Buscher rules. However, this symmetry as well as the T-duality constraint \reef{TST} may fix the corrections completely. This indicates that the form of T-duality transformations depended on the scheme that one uses for the effective action \cite{Garousi:2019wgz}.

The  $\alpha'$-expansion of  $S_{\rm eff}(\psi)$ is
\beqa
S_{\rm eff}(\psi)&=&\sum^\infty_{n=0}\alpha'^n S_n(\psi)
\eeqa
where $S_n(\psi)$ is reduction of $\bS_n$ in \reef{seff}. However,  $S_{\rm eff}(\psi')$ in \reef{TST} has two $\alpha'$-expansions. One is the same as the above expansion which is  inherited from the expansion \reef{seff}, and the other one is corresponding to the $\alpha'$-expansion of the  T-duality transformations \reef{psi}. Using the following  $\alpha'$-expansion:
\beqa
S_n(\psi')=S_n(\psi_0'+\alpha'\psi_1'+\cdots)&=&\sum^\infty_{m=0}\alpha'^m\delta^{(m)} S_n(\psi_0')
\eeqa 
where $\delta^{(0)} S_n=S_n$, one can write 
the T-duality constraint \reef{TST} as
\beqa
\sum^\infty_{n=0}\alpha'^nS_n(\psi)-\sum^\infty_{n=0,m=0}\alpha'^{n+m}\delta^{(m)} S_n(\psi_0')&=&\sum^\infty_{n=0}\alpha'^n\int d^{D-1}x\sqrt{-\bg}\nabla_a(e^{-2\bphi}J_n^a(\psi))\labell{Tf}
\eeqa 
Using the above T-duality constraint at  orders  $\alpha'^0, \alpha',\alpha'^2,\cdots$, one may find the parameters in the $D$-dimensional effective actions $\bS_0,\bS_1,\bS_2,\cdots$, as well as the corresponding corrections to the Buscher rules and the boundary terms. We are, however, interested only on fixing the parameters in the gauge invariant couplings in the $D$-dimensional effective action. This constraint in the bosonic string theory  fixes the effective actions at orders $\alpha'$ and $\alpha'^2$ up to an overall pre-factor  \cite{Garousi:2019wgz,Garousi:2019mca}. The calculations in  \cite{Garousi:2019wgz,Garousi:2019mca} reveal that the constraint relations between the parameters of the $D$-dimensional effective action are independent of the geometry of the base space, \ie they are the same whether or not the base space is curved. Hence to simplify the calculations it is convenient to assume the base space is flat, \ie $\bg_{ab}=\eta_{ab}$.

  To have a background with $U(1)$ isometry,  it is convenient to use the following background for  the metric, $B$-field and dilaton \cite{Maharana:1992my}:
  \beqa
G_{\mu\nu}=\left(\matrix{\eta_{ab}+e^{\varphi}g_{a }g_{b }& e^{\varphi}g_{a }&\cr e^{\varphi}g_{b }&e^{\varphi}&}\right),\, B_{\mu\nu}= \left(\matrix{\bb_{ab}+\frac{1}{2}b_{a }g_{b }- \frac{1}{2}b_{b }g_{a }&b_{a }\cr - b_{b }&0&}\right),\,  \Phi=\bar{\phi}+\varphi/4\labell{reduc}\eeqa
where $ \bb_{ab}$  is  the   B-field in the base space, and $g_{a},\, b_{b}$ are two vectors  in this space. Inverse of the above $D$-dimensional metric is 
\beqa
G^{\mu\nu}=\left(\matrix{\eta^{ab} &  -g^{a }&\cr -g^{b }&e^{-\varphi}+g_{c}g^{c}&}\right)\labell{inver}
\eeqa
where $\eta^{ab}$ is  inverse of the base  metric which raises the index of the   vectors.

The T-duality transformations at the leading order of $\alpha'$ on the $(D-1)$-dimensional fields   are given by the  Buscher rules \cite{Buscher:1987sk,Buscher:1987qj}. In the above parametrisation, they  become the following $Z_2$-transformations:
\beqa
\varphi'= -\varphi
\,\,\,,\,\,g'_{a }= b_{a }\,\,\,,\,\, b'_{a }= g_{a } \,\,\,,\,\,\eta_{ab}'=\eta_{ab} \,\,\,,\,\,\bb_{ab}'=\bb_{ab} \,\,\,,\,\,  \bar{\phi}'= \bar{\phi}\labell{T2}
\eeqa
The reduction of field strength of B-field  in the parametrizations \reef{reduc} becomes
\beqa
H_{abc}&=& \bH_{abc}+g_{[a}W_{bc]}\nn\\
H_{ab y}&=&W_{ab}\labell{rH}
\eeqa
where   $W$ is field strength of the $U(1)$ gauge field $b_{a}$, \ie $W=db$, and the three-form $\bH$ which is torsion in the base space, is defined as
\beqa
 \bH_{abc}&\equiv &\hat{H}_{abc}-\frac{1}{2}g_{[a} W_{bc]}-\frac{1}{2}b_{[a} V_{bc]}\labell{bHH}
 \eeqa
 where $\hat{H}$ is field strength of the two-form $\bb_{ab}$ and  $V$ is field strength of the $U(1)$ gauge field $g_{\mu}$, \ie $\hat{H}=d\bb$, $V=dg$.  The three-form $\bH$ is invariant under the T-duality and under various  gauge transformations. Since $\bH$ is not exterior derivative of a two-form,  it satisfies  anomalous Bianchi identity, whereas the $W,V$ satisfy ordinary Bianchi identity, \ie
 \beqa
 \prt_{[a} \bH_{bcd]}&=&-V_{[ab}W_{cd]}\labell{anB}\\
 \prt_{[a} W_{bc]}&=&0\nn\\
  \prt_{[a} V_{bc]}&=&0\nn
 \eeqa
  Our notation for making  antisymmetry  is such that \eg $g_{[a}W_{bc]}=g_aW_{bc}-g_{b}W_{ac}-g_cW_{ba}$.
 
  At the higher orders of $\alpha'$, the  $Z_2$-transformations \reef{T2} receive higher derivative corrections. Since the higher derivative corrections to the leading order supergravity start at order $\alpha'^3$ in type II supersting theory, the higher derivative corrections to the above Buscher rules  also start at order $\alpha'^3$ in this theory, \ie
\beqa
&&\varphi'= -\varphi+\alpha'^3\Delta\vp^{(3)}(\psi)+\cdots
\,,\,g'_{a }= b_{a }+\alpha'^3e^{\vp/2}\Delta g^{(3)}_a(\psi)+\cdots
\,,\,\nn\\&&b'_{a }= g_{a }+\alpha'^3e^{-\vp/2}\Delta b^{(3)}_a(\psi)+\cdots
\,,\,\bg_{ab}'=\eta_{ab}+\alpha'^3\Delta \bg^{(3)}_{ab}(\psi)+\cdots
\,,\,\nn\\&&\bH_{abc}'=\bH_{abc}+\alpha'^3\Delta\bH^{(3)}_{abc}(\psi)+\cdots
\,,\,\bar{\phi}'= \bar{\phi}+\alpha'^3\Delta\bphi^{(3)}(\psi)+\cdots
\labell{T22}
\eeqa
The deformations $\Delta g^{(3)}_a(\psi)$  and $\Delta\bH^{(3)}_{abc}(\psi)$ are odd under the parity and all other  deformations are even  under the parity\footnote{Note that in the base space, the field that plays the role of torsion is $\bH_{abc}$ (see \reef{S0psi}). So it is convenient to find T-duality corrections to this field rather than the correction to $\bb_{ab}$. One can find, however,  the correction to $\bb_{ab}$ from the correction to  $\bH_{abc}$ using the equation \reef{bHH}.  }. They  must satisfy the $Z_2$-transformation.  This produces the following constraint between the corrections at order $\alpha'^3$:
\beqa
-\Delta\vp^{(3)}(\psi)+\Delta\vp^{(3)}(\psi_0') &=&0\nn\\
\Delta b_a^{(3)}(\psi)+\Delta g_a^{(3)}(\psi'_0) &=&0\nn\\
\Delta g_a^{(3)}(\psi)+\Delta b_a^{(3)}(\psi'_0)&=&0\nn\\
\Delta \bg_{ab}^{(3)}(\psi)+\Delta \bg_{ab}^{(3)}(\psi'_0)&=&0\nn\\
\Delta \bH_{abc}^{(3)}(\psi)+\Delta \bH_{abc}^{(3)}(\psi'_0) &=&0\nn\\
\Delta\bphi^{(3)}(\psi)+\Delta\bphi^{(3)}(\psi_0') &=&0\labell{Z22}
\eeqa
The deformations $\Delta b_a^{(3)},\,\Delta g_a^{(3)} $ and $\Delta \bH_{abc}^{(3)}$ must also satisfy  the Bianchi identity \reef{anB}, \ie 
\beqa
d(\bH+\alpha'^3\Delta\bH^{(3)}+\cdots)&=&-d(b+\alpha'^3e^{\vp/2}\Delta g^{(3)}+\cdots)\nn\\
&&\wedge d( g+\alpha'^3e^{-\vp/2}\Delta b^{(3)}  +\cdots)\labell{aa}
\eeqa
This relation at order $\alpha'^0$ gives the Bianchi identity \reef{anB}. At order $\alpha'^3$ it gives the following relation between the deformations at order $\alpha'^3$:
\beqa
\Delta\bH^{(3)}_{abc}&=&\tilde H^{(3)}_{abc}-e^{-\vp/2}W_{[ab}\Delta b^{(3)}_{c]}-e^{\vp/2}V_{[ab}\Delta g^{(3)}_{c]} \labell{DH}
\eeqa
where $\tilde H^{(3)}$ is a gauge invariant closed 3-form, \ie $d \tilde H^{(3)}=0$, at order $\alpha'^3$ which is odd under parity. 

The 3-form $\tilde{H}^{(3)}$ and the deformation   $\Delta g_a^{(3)} $  contain all  odd-parity contractions and the deformations $\Delta \vp^{(3)}$, $\Delta\bphi^{(3)}$, $\Delta b_a^{(3)}$,  $\Delta \bg_{ab}^{(3)}$ contain all  even-parity contractions of $\prt\vp$, $\prt\bphi$,  $e^{\vp/2}V, e^{-\vp/2}W$, $\bH$ and their  derivatives at order $\alpha'^3$ with  unknown coefficients. Using the package ''xAct'' \cite{Nutma:2013zea}, one finds $\tilde{H}^{(3)}$ has 171551 terms,  $\Delta \vp^{(3)}$ has 3371 terms, $\Delta\bphi^{(3)}$ has 3371 terms,  $\Delta b_a^{(3)}$ has 9054 terms, $\Delta g_a^{(3)} $ has 9054 terms, and  $\Delta \bg_{ab}^{(3)}$ has 17581 terms. The terms in $\tilde{H}^{(3)}$ must satisfy the $Z_2$-constraint $\tilde{H}^{(3)}(\psi)+\tilde{H}^{(3)}(\psi'_0)=0$ as well as the closed-form condition  $d \tilde H^{(3)}=0$. The terms in  $\Delta \vp^{(3)}$, $\Delta\bphi^{(3)}$, $\Delta b_a^{(3)}$, $\Delta g_a^{(3)} $, $\Delta \bg_{ab}^{(3)}$ must satisfy the $Z_2$-constraints \reef{Z22}. To impose these constraint to find some relations between the parameters of the deformations, one has to also use the Bianchi identities \reef{anB}. These constraints produce many relations between the parameters, however as expected, they can not fix them all, because the corrections to the Buscher rules depend on the scheme that one uses for the effective action \cite{Garousi:2019wgz}. Imposing these relations into the deformations, one finds the deformations that are consistent with the $Z_2$-symmetry and satisfy the Bianchi identity \reef{aa}. They should then be used in the T-duality constraint \reef{Tf} at order $\alpha'^3$.

\subsection{T-duality  at order $\alpha'^0$}

 The T-duality constraint \reef{Tf} at order $\alpha'^0$  when the base space is flat, is 
\beqa
S_0(\psi)-S_0(\psi'_0)&=& \int d^{D-1}x\, \prt_a(e^{-2\bphi}J_0^a(\psi))\labell{TS0}
\eeqa
The left-hand side is odd under the Buscher rules, hence, the vector $J_0^a$ on the right-hand side must be also odd under the Buscher rules.  

Reduction of the NS-NS couplings at order $\alpha'^0$    are the following:
\beqa
e^{-2\Phi}\sqrt{-G}&=&e^{-2\bphi}\nonumber\\
R&=&-\prt^a\prt_a\vp-\frac{1}{2}\prt_a\vp \prt^a\vp -\frac{1}{4}e^{\vp}V^2 \labell{R}\\
\nabla_{\mu}\Phi\nabla^{\mu}\Phi&=&\prt_a\bphi\prt^a \bphi+\frac{1}{2}\prt_a\bphi\prt^a\vp+\frac{1}{16}\prt_a\vp\prt^a\vp\nn\\
H^2&=&\bH_{abc}\bH^{abc}+3e^{-\vp}W^2 \nn
\eeqa
Then the reduction of $\bS_0$ becomes
\beqa
S_0(\psi)&=&\int d^{D-1}x\,e^{-2\bphi}\Big[-\prt^a\prt_a\vp-\frac{1}{4}\prt_a\vp \prt^a\vp+4\prt_a\bphi\prt^a \bphi+2\prt_a\bphi\prt^a\vp\nn\\
&&\qquad\qquad\qquad\qquad\qquad-\frac{1}{4}e^{\vp}V^2 -\frac{1}{4}e^{-\vp}W^2 -\frac{1}{12}\bH_{abc}\bH^{abc}\Big]\labell{S0psi}
\eeqa
It is even under the parity. It satisfies the T-duality constraint \reef{TS0} for the total derivative term with vector $J^a_0=-2\prt^a\vp$ which is odd under the Buscher rules and even under the parity, as expected. If spacetime has no boundary the total derivative term becomes zero using the Stokes's theorem. On the other hand, if the spaetime has boundary this total derivative term dictates the Hawking-Gibbons boundary term \cite{Garousi:2019xlf}. In this paper we assume the spacetime is closed, hence,  the total derivative terms can be ignored.

\subsection{T-duality  at order $\alpha'^3$}
The T-duality constraint \reef{Tf} at order $\alpha'^3$  when the base space is flat, is 
\beqa
S_3(\psi)-S_3(\psi'_0)-\delta^{(3)}S_0(\psi'_0)&=& \int d^{D-1}x\, \prt_a(e^{-2\bphi}J_3^a(\psi))\labell{TS}
\eeqa
where $S_3(\psi)$ is reduction of the gauge invariant couplings  \reef{T54} on a circle, $S_3(\psi_0')$ is its transformation under the Buscher rules \reef{T2} and    $\delta^{(3)}S_0(\psi'_0)$ is transformation of the perturbation of \reef{S0psi} under the Buscher rules  \reef{T2}, \ie
\beqa
 \delta^{(3)} S_0(\psi_0')
 &\!\!\!\!\!=\!\!\!\!\!& -\frac{2}{\kappa^2}\int d^{D-1}x e^{-2\bphi} \,  \Big[ -\Big(-2\prt^a\prt^b\bphi+\frac{1}{4}\prt^a\vp\prt^b\vp+\frac{1}{4}\bH^{acd}\bH^b{}_{cd} +\frac{1}{2}e^\vp V^{ac}V^b{}_{c}\nn\\
 &&+\frac{1}{2}e^{-\vp} W^{ac}W^b{}_{c}\Big)\Delta\bar{g}_{ab}^{(3)}-\Big(2\prt_c\prt^c\bphi-2\prt_c\bphi\prt^c\bphi -\frac{1}{8}\prt_c\vp\prt^c\vp-\frac{1}{24}\bH^2-\frac{1}{8}e^\vp V^2\nn\\
 &&-\frac{1}{8}e^{-\vp}W^2\Big)(\eta^{ab}\Delta\bar{g}_{ab}^{(3)}-4\Delta\bphi^{(3)})+\Big(\frac{1}{2}\prt_a\prt^a\vp-\prt_a\bphi\prt^a\vp-\frac{1}{4}e^\vp V^2+\frac{1}{4}e^{-\vp}W^2\Big)\Delta\vp^{(3)}\nn\\
 &&-e^{-\vp/2}\Big(- \prt_b W^{ab}+2\prt_b\bphi W^{ab}+\prt_b\vp W^{ab}\Big)\Delta g_a^{(3)} \nn\\
 &&-e^{\vp/2}\Big(- \prt_b V^{ab}+2\prt_b\bphi V^{ab}-\prt_b\vp V^{ab}\Big)\Delta b_a^{(3)}+\frac{1}{6}\bH^{abc}\Delta\bH_{abc}^{(3)}\Big] \labell{d3S}
 \eeqa
where we have used the relations \reef{Z22} and removed some total derivative terms in which we are not interested in this paper. In finding the above result for $\Delta\bar g_{ab}$ we assumed the metric of the base space is not flat. After perturbing the metric,  we set it $\bar g_{ab}=\eta_{ab}$. Note that   the extra factors of $e^{\vp/2}$ and $e^{-\vp/2}$ in \reef{T22} make it possible to have a factor of   $e^{\vp/2}$ in front of each $V$ and  a factor of   $e^{-\vp/2}$ in front of each $W$. Note  that $\delta^{(3)}S_0(\psi'_0)$ is odd under the Buscher rules and it is  even under parity. Hence, the left-hand side of \reef{TS} must be odd under the Buscher rules and even under the parity, \ie  the vector $J_3^a(\psi)$  must be even under parity and satisfies the following relation:
\beqa
J_3^a(\psi)+J_3^a(\psi_0')&=&0
\eeqa 
 The vector $J_3^a(\psi)$  contains all  even-parity contractions of $\prt\vp$, $\prt\bphi$,  $e^{\vp/2}V, e^{-\vp/2}W$, $\bH$ and their  derivatives at order $\alpha'^3$ with  unknown coefficients. Using the package ''xAct'', one finds it has 71678 terms. They should satisfy the above $Z_2$-constraint.

 The dimensional reduction of each gauge invariant term in \reef{T54} is a very lengthy calculation. However, the final result for the reduction of each  term when it is written in terms of the physical torsion $\bH_{abc}$, must be an invariant expression under the  $U(1)\times U(1)$ gauge transformations where the first $U(1)$ corresponds to $g_a$ gauge transformation and the second one corresponds to the $b_a$ gauge transformation. This fact has been used in \cite{Garousi:2019mca} to  simplify greatly the calculations of the reduction of the couplings at order $\alpha'^2$. We use the same simplification   for calculating reduction of the couplings at order $\alpha'^3$.
 
  The couplings in \reef{T54} have only Riemann curvature, $H$, $\nabla H$, $\nabla\nabla H$, $\nabla\Phi$ and $\nabla\nabla\Phi$. So we need to reduce these terms and then contract them with the metric \reef{inver}. In the  reduction of these terms, there are many terms which contains gauge field  $g_a$ without its field strength. These terms must be cancelled at the end of the day for the scalar couplings. Hence, to simplify the calculation one drops those terms in the reduction of  $R_{\mu\nu\alpha\beta}$, $H_{\mu\nu\alpha}$, $\nabla_{\mu} H_{\nu\alpha\beta}$, $\nabla_{\rho}\nabla_{\mu} H_{\nu\alpha\beta}$, $\nabla_{\mu}\Phi$, $\nabla_{\mu}\nabla_{\nu}\Phi$  and $G^{\mu\nu}$ which have the gauge field $g_a$.  Using this simplification, the reduction of  $\nabla_{\rho}\nabla_{\mu} H_{\nu\alpha\beta}$ becomes
 \beqa
\nabla_a\nabla_bH_{cde}&=& - \frac{1}{4} e^{\varphi} \bH_{def} V_{ac} V_{b}{}^{f} + 
\frac{1}{4} e^{\varphi} \bH_{cef} V_{ad} V_{b}{}^{f} -  
\frac{1}{4} e^{\varphi} \bH_{cdf} V_{ae} V_{b}{}^{f} -  
\frac{1}{4} e^{\varphi} \bH_{def} V_{ab} V_{c}{}^{f} \nn\\&&+ 
\frac{1}{4} e^{\varphi} \bH_{cef} V_{ab} V_{d}{}^{f} -  
\frac{1}{4} e^{\varphi} \bH_{cdf} V_{ab} V_{e}{}^{f} + 
\frac{1}{2} W_{de} \partial_{a}V_{bc} -  \frac{1}{2} W_{ce} 
\partial_{a}V_{bd} + \frac{1}{2} W_{cd} \partial_{a}V_{be}\nn\\&& + 
\frac{1}{2} V_{be} \partial_{a}W_{cd} -  \frac{1}{2} V_{bd} 
\partial_{a}W_{ce} + \frac{1}{2} V_{bc} \partial_{a}W_{de} + 
\partial_{a}\partial_{b}\bH_{cde} + \frac{1}{2} V_{ae} 
\partial_{b}W_{cd} -  \frac{1}{2} V_{ad} \partial_{b}W_{ce}\nn\\&& + 
\frac{1}{2} V_{ac} \partial_{b}W_{de} -  \frac{1}{4} V_{ae} 
W_{cd} \partial_{b}\varphi + \frac{1}{4} V_{ad} W_{ce} 
\partial_{b}\varphi -  \frac{1}{4} V_{ac} W_{de} 
\partial_{b}\varphi -  \frac{1}{4} V_{ab} W_{de} 
\partial_{c}\varphi\nn\\&& + \frac{1}{4} V_{ab} W_{ce} 
\partial_{d}\varphi -  \frac{1}{4} V_{ab} W_{cd} 
\partial_{e}\varphi\nn\\
 \nabla_a\nabla_b H_{cdy}&=&- \frac{1}{4} e^{\varphi} V_{a}{}^{e} V_{be} W_{cd} + 
\frac{1}{4} e^{\varphi} V_{a}{}^{e} V_{bd} W_{ce} -  \frac{1}{4} 
e^{\varphi} V_{ab} V_{d}{}^{e} W_{ce} -  \frac{1}{4} e^{\varphi} 
V_{a}{}^{e} V_{bc} W_{de}\nn\\&& + \frac{1}{4} e^{\varphi} V_{ab} 
V_{c}{}^{e} W_{de} -  \frac{1}{2} e^{\varphi} V_{b}{}^{e} 
\partial_{a}\bH_{cde} -  \frac{1}{2} e^{\varphi} \bH_{cd}{}^{e} 
\partial_{a}V_{be} -  \frac{1}{4} e^{\varphi} \bH_{cde} 
V_{b}{}^{e} \partial_{a}\varphi + \partial_{a}\partial_{b}W_{cd}\nn\\&& 
-  \frac{1}{2} W_{cd} \partial_{a}\partial_{b}\varphi -  
\frac{1}{2} e^{\varphi} V_{a}{}^{e} \partial_{b}\bH_{cde} -  
\frac{1}{2} \partial_{a}\varphi \partial_{b}W_{cd} -  
\frac{1}{2} \partial_{a}W_{cd} \partial_{b}\varphi + 
\frac{1}{4} W_{cd} \partial_{a}\varphi \partial_{b}\varphi\nn\\&& + 
\frac{1}{4} e^{\varphi} \bH_{cde} V_{ab} \partial^{e}\varphi\nn\\
\nabla_a\nabla_yH_{bcd}&=&\frac{1}{4} e^{\varphi} V_{a}{}^{e} V_{de} W_{bc} -  
\frac{1}{4} e^{\varphi} V_{a}{}^{e} V_{ce} W_{bd} -  \frac{1}{4} 
e^{\varphi} V_{ad} V_{c}{}^{e} W_{be} + \frac{1}{4} e^{\varphi} 
V_{ac} V_{d}{}^{e} W_{be}\nn\\&& + \frac{1}{4} e^{\varphi} V_{a}{}^{e} 
V_{be} W_{cd} + \frac{1}{4} e^{\varphi} V_{ad} V_{b}{}^{e} W_{ce} 
-  \frac{1}{4} e^{\varphi} V_{ab} V_{d}{}^{e} W_{ce} -  
\frac{1}{4} e^{\varphi} V_{ac} V_{b}{}^{e} W_{de}\nn\\&& + \frac{1}{4} 
e^{\varphi} V_{ab} V_{c}{}^{e} W_{de} -  \frac{1}{2} e^{\varphi} 
V_{d}{}^{e} \partial_{a}\bH_{bce} + \frac{1}{2} e^{\varphi} 
V_{c}{}^{e} \partial_{a}\bH_{bde} -  \frac{1}{2} e^{\varphi} 
V_{b}{}^{e} \partial_{a}\bH_{cde}\nn\\&& -  \frac{1}{2} e^{\varphi} 
\bH_{cd}{}^{e} \partial_{a}V_{be} + \frac{1}{2} e^{\varphi} 
\bH_{bd}{}^{e} \partial_{a}V_{ce} -  \frac{1}{2} e^{\varphi} 
\bH_{bc}{}^{e} \partial_{a}V_{de} -  \frac{1}{4} e^{\varphi} 
\bH_{cde} V_{b}{}^{e} \partial_{a}\varphi \nn\\&&+ \frac{1}{4} 
e^{\varphi} \bH_{bde} V_{c}{}^{e} \partial_{a}\varphi -  
\frac{1}{4} e^{\varphi} \bH_{bce} V_{d}{}^{e} \partial_{a}\varphi 
-  \frac{1}{2} W_{cd} \partial_{a}\partial_{b}\varphi + 
\frac{1}{2} W_{bd} \partial_{a}\partial_{c}\varphi\nn\\&& -  
\frac{1}{2} W_{bc} \partial_{a}\partial_{d}\varphi -  
\frac{1}{2} \partial_{a}W_{cd} \partial_{b}\varphi + 
\frac{1}{4} W_{cd} \partial_{a}\varphi \partial_{b}\varphi + 
\frac{1}{2} \partial_{a}W_{bd} \partial_{c}\varphi \nn\\&&-  
\frac{1}{4} W_{bd} \partial_{a}\varphi \partial_{c}\varphi -  
\frac{1}{2} \partial_{a}W_{bc} \partial_{d}\varphi + 
\frac{1}{4} W_{bc} \partial_{a}\varphi \partial_{d}\varphi -  
\frac{1}{2} e^{\varphi} V_{a}{}^{e} \partial_{e}\bH_{bcd}\nn\\&& + 
\frac{1}{4} e^{\varphi} \bH_{cde} V_{ab} \partial^{e}\varphi -  
\frac{1}{4} e^{\varphi} \bH_{bde} V_{ac} \partial^{e}\varphi + 
\frac{1}{4} e^{\varphi} \bH_{bce} V_{ad} \partial^{e}\varphi\nn
 \eeqa 
  \beqa
  \nabla_y\nabla_aH_{bcd}&=&- \frac{1}{4} e^{\varphi} V_{ad} V_{c}{}^{e} W_{be} + 
\frac{1}{4} e^{\varphi} V_{ac} V_{d}{}^{e} W_{be} + \frac{1}{4} 
e^{\varphi} V_{ad} V_{b}{}^{e} W_{ce} -  \frac{1}{4} e^{\varphi} 
V_{ab} V_{d}{}^{e} W_{ce}\nn\\&& -  \frac{1}{4} e^{\varphi} V_{ac} 
V_{b}{}^{e} W_{de} + \frac{1}{4} e^{\varphi} V_{ab} V_{c}{}^{e} 
W_{de} -  \frac{1}{2} e^{\varphi} V_{d}{}^{e} 
\partial_{a}\bH_{bce} + \frac{1}{2} e^{\varphi} V_{c}{}^{e} 
\partial_{a}\bH_{bde}\nn\\&& -  \frac{1}{2} e^{\varphi} V_{b}{}^{e} 
\partial_{a}\bH_{cde} + \frac{1}{4} e^{\varphi} \bH_{cde} 
V_{b}{}^{e} \partial_{a}\varphi -  \frac{1}{4} e^{\varphi} 
\bH_{bde} V_{c}{}^{e} \partial_{a}\varphi + \frac{1}{4} 
e^{\varphi} \bH_{bce} V_{d}{}^{e} \partial_{a}\varphi\nn\\&& + 
\frac{1}{4} e^{\varphi} \bH_{cde} V_{a}{}^{e} \partial_{b}\varphi 
-  \frac{1}{2} \partial_{a}W_{cd} \partial_{b}\varphi + 
\frac{1}{2} W_{cd} \partial_{a}\varphi \partial_{b}\varphi -  
\frac{1}{4} e^{\varphi} \bH_{bde} V_{a}{}^{e} \partial_{c}\varphi
+ \frac{1}{2} \partial_{a}W_{bd} \partial_{c}\varphi \nn\\&& -  
\frac{1}{2} W_{bd} \partial_{a}\varphi \partial_{c}\varphi + 
\frac{1}{4} e^{\varphi} \bH_{bce} V_{a}{}^{e} \partial_{d}\varphi 
-  \frac{1}{2} \partial_{a}W_{bc} \partial_{d}\varphi + 
\frac{1}{2} W_{bc} \partial_{a}\varphi \partial_{d}\varphi -  
\frac{1}{2} e^{\varphi} V_{a}{}^{e} \partial_{e}\bH_{bcd}\nn\\
\nabla_a\nabla_yH_{bcy}&=&\frac{1}{4} e^{2 \varphi} \bH_{cde} V_{a}{}^{d} V_{b}{}^{e} -  
\frac{1}{4} e^{2 \varphi} \bH_{bde} V_{a}{}^{d} V_{c}{}^{e} + 
\frac{1}{2} e^{2 \varphi} \bH_{bce} V_{a}{}^{d} V_{d}{}^{e} + 
\frac{1}{2} e^{\varphi} W_{c}{}^{d} \partial_{a}V_{bd}\nn\\&& -  
\frac{1}{2} e^{\varphi} W_{b}{}^{d} \partial_{a}V_{cd} -  
\frac{1}{2} e^{\varphi} V_{c}{}^{d} \partial_{a}W_{bd} + 
\frac{1}{2} e^{\varphi} V_{b}{}^{d} \partial_{a}W_{cd} + 
\frac{1}{2} e^{\varphi} \bH_{bc}{}^{d} 
\partial_{a}\partial_{d}\varphi\nn\\&& + \frac{1}{4} e^{\varphi} 
V_{a}{}^{d} W_{cd} \partial_{b}\varphi -  \frac{1}{4} 
e^{\varphi} V_{a}{}^{d} W_{bd} \partial_{c}\varphi -  
\frac{1}{2} e^{\varphi} V_{ad} \partial^{d}W_{bc} + \frac{1}{2} 
e^{\varphi} V_{ad} W_{bc} \partial^{d}\varphi\nn\\&& + \frac{1}{2} 
e^{\varphi} \partial_{a}\bH_{bcd} \partial^{d}\varphi\nn\\
\nabla_y\nabla_aH_{bcy}&=&\frac{1}{4} e^{2 \varphi} \bH_{cde} V_{a}{}^{d} V_{b}{}^{e} -  
\frac{1}{4} e^{2 \varphi} \bH_{bde} V_{a}{}^{d} V_{c}{}^{e} + 
\frac{1}{4} e^{2 \varphi} \bH_{bce} V_{a}{}^{d} V_{d}{}^{e} -  
\frac{1}{2} e^{\varphi} V_{c}{}^{d} \partial_{a}W_{bd}\nn\\&& + 
\frac{1}{2} e^{\varphi} V_{b}{}^{d} \partial_{a}W_{cd} + 
\frac{1}{2} e^{\varphi} V_{c}{}^{d} W_{bd} \partial_{a}\varphi - 
 \frac{1}{2} e^{\varphi} V_{b}{}^{d} W_{cd} \partial_{a}\varphi 
-  \frac{1}{2} e^{\varphi} V_{ad} \partial^{d}W_{bc}\nn\\&& + 
\frac{1}{2} e^{\varphi} V_{ad} W_{bc} \partial^{d}\varphi -  
\frac{1}{4} e^{\varphi} V_{ac} W_{bd} \partial^{d}\varphi + 
\frac{1}{4} e^{\varphi} V_{ab} W_{cd} \partial^{d}\varphi + 
\frac{1}{2} e^{\varphi} \partial_{a}\bH_{bcd} 
\partial^{d}\varphi\nn\\&& -  \frac{1}{4} e^{\varphi} \bH_{bcd} 
\partial_{a}\varphi \partial^{d}\varphi\nn\\
\nabla_y\nabla_yH_{abc}&=&\frac{1}{2} e^{2 \varphi} \bH_{cde} V_{a}{}^{d} V_{b}{}^{e} -  
\frac{1}{2} e^{2 \varphi} \bH_{bde} V_{a}{}^{d} V_{c}{}^{e} + 
\frac{1}{2} e^{2 \varphi} \bH_{ade} V_{b}{}^{d} V_{c}{}^{e} + 
\frac{1}{4} e^{2 \varphi} \bH_{bce} V_{a}{}^{d} V_{d}{}^{e}\nn\\&& -  
\frac{1}{4} e^{2 \varphi} \bH_{ace} V_{b}{}^{d} V_{d}{}^{e} + 
\frac{1}{4} e^{2 \varphi} \bH_{abe} V_{c}{}^{d} V_{d}{}^{e} + 
\frac{1}{2} e^{\varphi} V_{c}{}^{d} W_{bd} \partial_{a}\varphi - 
 \frac{1}{2} e^{\varphi} V_{b}{}^{d} W_{cd} \partial_{a}\varphi\nn\\&& 
-  \frac{1}{2} e^{\varphi} V_{c}{}^{d} W_{ad} 
\partial_{b}\varphi + \frac{1}{2} e^{\varphi} V_{a}{}^{d} W_{cd} 
\partial_{b}\varphi + \frac{1}{2} e^{\varphi} V_{b}{}^{d} W_{ad} 
\partial_{c}\varphi -  \frac{1}{2} e^{\varphi} V_{a}{}^{d} 
W_{bd} \partial_{c}\varphi\nn\\&& -  \frac{1}{4} e^{\varphi} \bH_{bcd} 
\partial_{a}\varphi \partial^{d}\varphi + \frac{1}{4} 
e^{\varphi} \bH_{acd} \partial_{b}\varphi \partial^{d}\varphi -  
\frac{1}{4} e^{\varphi} \bH_{abd} \partial_{c}\varphi 
\partial^{d}\varphi + \frac{1}{2} e^{\varphi} 
\partial_{d}\bH_{abc} \partial^{d}\varphi\nn\\
\nabla_y\nabla_y H_{aby}&=&\frac{1}{4} e^{2 \varphi} V_{b}{}^{c} V_{c}{}^{d} W_{ad} -  
\frac{1}{4} e^{2 \varphi} V_{a}{}^{c} V_{c}{}^{d} W_{bd} + 
\frac{1}{2} e^{2 \varphi} V_{a}{}^{c} V_{b}{}^{d} W_{cd} -  
\frac{1}{2} e^{2 \varphi} \bH_{bcd} V_{a}{}^{d} 
\partial^{c}\varphi\nn\\&& + \frac{1}{2} e^{2 \varphi} \bH_{acd} 
V_{b}{}^{d} \partial^{c}\varphi -  \frac{1}{2} e^{2 \varphi} 
\bH_{abd} V_{c}{}^{d} \partial^{c}\varphi -  \frac{1}{4} 
e^{\varphi} W_{bc} \partial_{a}\varphi \partial^{c}\varphi + 
\frac{1}{4} e^{\varphi} W_{ac} \partial_{b}\varphi \partial^{c}
\varphi\nn\\&& + \frac{1}{2} e^{\varphi} \partial_{c}W_{ab} 
\partial^{c}\varphi -  \frac{1}{2} e^{\varphi} W_{ab} 
\partial_{c}\varphi \partial^{c}\varphi
  \eeqa
The indices are raised by $\eta^{ab}$. The reduction of  $R_{\mu\nu\alpha\beta}$, $H_{\mu\nu\alpha}$, $\nabla_{\mu} H_{\nu\alpha\beta}$, $\nabla_{\mu}\Phi$, and $\nabla_{\mu}\nabla_{\nu}\Phi$  are calculated in \cite{Garousi:2019mca}. 
 The reduction of inverse of the $D$-dimensional metric becomes
 \beqa
G^{\mu\nu}=\left(\matrix{\eta^{ab} & 0&\cr 0&e^{-\varphi}&}\right)\labell{inver1}
\eeqa
Using above reductions, one can calculate the reduction of different gauge invariant  terms in \reef{T54} to find $S_3(\psi)$ and its corresponding $S_3(\psi_0')$.

Then using the constraint \reef{TS}, one finds some   equations involving the 872 parameters in \reef{T54}, the arbitrary parameters in $J_3^a$ and  in  $\Delta\vp^{(3)},\, \Delta g_a^{(3)},\, \Delta b_a^{(3)}, \,\Delta \bg_{ab}^{(3)}$, $\Delta\bphi^{(3)}$, $\tilde{H}^{(3)}$.  To solve this  constraint, one has to also impose the Bianchi identities \reef{anB}. To impose the last two Bianchi identities in \reef{anB}, we write $W$ and $V$ in the derivative terms which appear in \reef{TS}, \ie in $\prt W$, $\prt\prt W$ and so on, in terms of potential, \ie $W_{ab}=\prt_{[a}b_{b]}$,  $V_{ab}=\prt_{[a}g_{b]}$. To impose the first identity in \reef{anB}, we make all even-parity contraction of $\prt_{[a}\bH_{bcd]}+V_{[ab}W_{cd]}$ and its derivatives with  $\prt\vp$, $\prt\bphi$,  $e^{\vp/2}V, e^{-\vp/2}W$, $\bH$ and their  derivatives to produce terms at order $\alpha'^3$. We then add them with arbitrary coefficients to the constraint \reef{TS}. 

One then finds 143146 linear algebraic equations involving all the parameters.  Solving the resulting equations, one  finds many relations between the parameters. Since there are too many terms in the vector $J_3^a$ and in the T-duality corrections  $\Delta\vp^{(3)},\, \Delta g_a^{(3)},\, \Delta b_a^{(3)}, \,\Delta \bg_{ab}^{(3)}$, $\Delta\bphi^{(3)}$ and $\tilde{H}^{(3)}$, we are not interested in their corresponding parameters. Interestingly, when one solves the linear algebraic equations, one would find  there are 871 relations between ${\it only}$  the parameters in the $D$-dimensional Lagrangian \reef{T54}. It means the T-duality constraint can fix the parameters in \reef{T54} up to an overall factor. In the next subsection we write these relations, and compare some of the non-zero couplings with the partial couplings that have been found by other methods.

\subsection{Results}

Our calculation indicates that 427 parameters in \reef{T54} are zero. All other parameters can be written in terms of one of them. The 445 non-zero couplings appear in  only 15 structures.  We find that when there is no B-field, there are only two non-zero couplings. Writing the effective action as 
\beqa
\bS_3&=&-\frac{2c}{\kappa^2}\int d^{10}x\,\sqrt{-G}e^{-2\Phi}\cL_3\labell{S3}
\eeqa
where $c$ is the overall parameter, the couplings  involving only metric and dilaton are the following:
\beqa
{\cal L}_3(G,\Phi)&=& 2 R_{\alpha}{}^{\epsilon}{}_{\gamma}{}^{\varepsilon} R^{\alpha \beta \gamma \delta} R_{\beta}{}^{\mu}{}_{\epsilon}{}^{\zeta} R_{\delta \zeta \varepsilon \mu} + R_{\alpha \beta}{}^{\epsilon \varepsilon} R^{\alpha \beta \gamma \delta} R_{\gamma}{}^{\mu}{}_{\epsilon}{}^{\zeta} R_{\delta \zeta \varepsilon \mu}\labell{RRf}
\eeqa  
It is exactly the couplings that have been found by the S-matrix and sigma-model calculations \cite{Gross:1986mw,Myers:1987qx,Grisaru:1986vi,Freeman:1986zh,Policastro:2008hg,Paban:1998ea,Garousi:2013zca} provided that   one chooses the overall parameters to be $c=-\z(3)/2^6$. The above couplings have been also found in \cite{Razaghian:2018svg} by the T-duality constraint when metric is diagonal, B-field is zero and the base space metric is curved. In that case, if one assumes the base space is flat, then the T-duality could not fix the gravity and dilaton couplings. However, using the  assumption that the dilaton appears only as the overall factor $e^{-2\Phi}$, one can fix the couplings up to two parameters  if the  base space is curved  \cite{Razaghian:2018svg}. 

All other 443 non-zero couplings  involve B-field. We have found 8  couplings with structure  $(\nabla H)^4$. They are the following:
\beqa
{\cal L}_3^{(\prt H)^4}&=&\frac{1}{8} \nabla_{\delta}H_{\gamma}{}^{\mu \zeta} \nabla^{
\delta}H^{\alpha \beta \gamma} \nabla_{\varepsilon}H_{\epsilon 
\mu \zeta} \nabla^{\varepsilon}H_{\alpha \beta}{}^{\epsilon} - 
 \frac{1}{8} \nabla^{\delta}H^{\alpha \beta \gamma} 
\nabla^{\varepsilon}H_{\alpha \beta}{}^{\epsilon} 
\nabla_{\zeta}H_{\epsilon \varepsilon \mu} 
\nabla^{\zeta}H_{\gamma \delta}{}^{\mu}\labell{PH4}\\&& -  \frac{1}{4} \
\nabla^{\delta}H^{\alpha \beta \gamma} \
\nabla_{\epsilon}H_{\delta \mu \zeta} \
\nabla^{\varepsilon}H_{\alpha \beta}{}^{\epsilon} \
\nabla^{\zeta}H_{\gamma \varepsilon}{}^{\mu} -  \frac{1}{8} \
\nabla^{\delta}H^{\alpha \beta \gamma} \nabla^{\varepsilon}H_{\
\alpha \beta}{}^{\epsilon} \nabla_{\zeta}H_{\delta \epsilon \
\mu} \nabla^{\zeta}H_{\gamma \varepsilon}{}^{\mu} \nn\\&&-  \
\frac{1}{16} \nabla_{\delta}H_{\alpha \beta}{}^{\epsilon} \
\nabla^{\delta}H^{\alpha \beta \gamma} \
\nabla_{\zeta}H_{\epsilon \varepsilon \mu} \
\nabla^{\zeta}H_{\gamma}{}^{\varepsilon \mu} -  \frac{3}{16} \
\nabla_{\gamma}H_{\varepsilon \mu \zeta} \
\nabla^{\delta}H^{\alpha \beta \gamma} \
\nabla^{\epsilon}H_{\alpha \beta \delta} \
\nabla^{\zeta}H_{\epsilon}{}^{\varepsilon \mu}\nn\\&& + \frac{1}{144} \
\nabla_{\delta}H_{\alpha \beta \gamma} \
\nabla^{\delta}H^{\alpha \beta \gamma} \
\nabla_{\zeta}H_{\epsilon \varepsilon \mu} \
\nabla^{\zeta}H^{\epsilon \varepsilon \mu} + \frac{1}{4} \
\nabla^{\delta}H^{\alpha \beta \gamma} \nabla^{\varepsilon}H_{\
\alpha \beta}{}^{\epsilon} \nabla^{\zeta}H_{\gamma \
\varepsilon}{}^{\mu} \nabla_{\mu}H_{\delta \epsilon \zeta}\nn
\eeqa
They are exactly the couplings that are produced by the S-matrix element of four NS-NS vertex operators \cite{Gross:1986mw} using the scheme \reef{T54} for the field theory couplings \cite{Garousi:2020mqn}.

We have found 22 couplings with structure  $R^2(\nabla H)^2$, \ie
\beqa
{\cal L}_3^{R^2(\prt H)^2}&=&\frac{10}{3} R^{\alpha \beta \gamma \delta} \
R_{\gamma}{}^{\epsilon \varepsilon \mu} \
\nabla_{\beta}H_{\delta \mu \zeta} \
\nabla_{\varepsilon}H_{\alpha \epsilon}{}^{\zeta} -  \
\frac{8}{3} R^{\alpha \beta \gamma \delta} R^{\epsilon \
\varepsilon}{}_{\gamma}{}^{\mu} \nabla_{\epsilon}H_{\alpha \
\delta}{}^{\zeta} \nabla_{\varepsilon}H_{\beta \mu \zeta}\nn\\&& + \
\frac{1}{3} R^{\alpha \beta \gamma \delta} R^{\epsilon \
\varepsilon \mu \zeta} \nabla_{\gamma}H_{\alpha \beta \
\epsilon} \nabla_{\varepsilon}H_{\delta \mu \zeta} -  \
\frac{7}{3} R_{\alpha}{}^{\epsilon}{}_{\gamma}{}^{\varepsilon} \
R^{\alpha \beta \gamma \delta} \
\nabla_{\epsilon}H_{\beta}{}^{\mu \zeta} \
\nabla_{\varepsilon}H_{\delta \mu \zeta}\nn\\&& + \frac{5}{3} \
R^{\alpha \beta \gamma \delta} \
R_{\gamma}{}^{\epsilon}{}_{\alpha}{}^{\varepsilon} \
\nabla_{\epsilon}H_{\beta}{}^{\mu \zeta} \
\nabla_{\varepsilon}H_{\delta \mu \zeta} + \frac{4}{3} \
R^{\alpha \beta \gamma \delta} R^{\epsilon \
\varepsilon}{}_{\gamma}{}^{\mu} \nabla_{\beta}H_{\alpha \
\delta}{}^{\zeta} \nabla_{\varepsilon}H_{\epsilon \mu \zeta}\nn\\&& + \
\frac{5}{3} R^{\alpha \beta \gamma \delta} \
R_{\gamma}{}^{\epsilon}{}_{\alpha}{}^{\varepsilon} \
\nabla_{\delta}H_{\beta}{}^{\mu \zeta} \nabla_{\varepsilon}H_{\
\epsilon \mu \zeta} -  \frac{7}{3} R^{\alpha \beta \gamma \
\delta} R_{\gamma}{}^{\epsilon \varepsilon \mu} \
\nabla_{\beta}H_{\varepsilon \mu \zeta} \
\nabla^{\zeta}H_{\alpha \delta \epsilon} \nn\\&&+ 4 R^{\alpha \beta \
\gamma \delta} R^{\epsilon \varepsilon}{}_{\gamma}{}^{\mu} \
\nabla_{\varepsilon}H_{\beta \mu \zeta} \
\nabla^{\zeta}H_{\alpha \delta \epsilon} + \frac{16}{3} \
R_{\alpha}{}^{\epsilon}{}_{\gamma}{}^{\varepsilon} R^{\alpha \
\beta \gamma \delta} \nabla_{\varepsilon}H_{\epsilon \mu \
\zeta} \nabla^{\zeta}H_{\beta \delta}{}^{\mu}\nn\\&& + \frac{4}{3} \
R_{\alpha}{}^{\epsilon}{}_{\gamma}{}^{\varepsilon} R^{\alpha \
\beta \gamma \delta} \nabla_{\zeta}H_{\epsilon \varepsilon \
\mu} \nabla^{\zeta}H_{\beta \delta}{}^{\mu} + \frac{2}{3} R_{\
\alpha}{}^{\epsilon}{}_{\gamma}{}^{\varepsilon} R^{\alpha \beta \
\gamma \delta} \nabla_{\varepsilon}H_{\delta \mu \zeta} \
\nabla^{\zeta}H_{\beta \epsilon}{}^{\mu}\nn\\&& -  \frac{22}{3} \
R^{\alpha \beta \gamma \delta} \
R_{\gamma}{}^{\epsilon}{}_{\alpha}{}^{\varepsilon} \
\nabla_{\varepsilon}H_{\delta \mu \zeta} \
\nabla^{\zeta}H_{\beta \epsilon}{}^{\mu} + \
R_{\alpha}{}^{\epsilon}{}_{\gamma}{}^{\varepsilon} R^{\alpha \
\beta \gamma \delta} \nabla_{\zeta}H_{\delta \varepsilon \mu} \
\nabla^{\zeta}H_{\beta \epsilon}{}^{\mu}\nn\\&& -  \frac{4}{3} \
R^{\alpha \beta \gamma \delta} \
R_{\gamma}{}^{\epsilon}{}_{\alpha}{}^{\varepsilon} \
\nabla_{\zeta}H_{\delta \varepsilon \mu} \
\nabla^{\zeta}H_{\beta \epsilon}{}^{\mu} - 4 R^{\alpha \beta \
\gamma \delta} R_{\gamma}{}^{\epsilon}{}_{\alpha \delta} \
\nabla_{\epsilon}H_{\varepsilon \mu \zeta} \
\nabla^{\zeta}H_{\beta}{}^{\varepsilon \mu}\nn\\&& + \frac{2}{3} \
R^{\alpha \beta \gamma \delta} \
R_{\gamma}{}^{\epsilon}{}_{\alpha \delta} \
\nabla_{\zeta}H_{\epsilon \varepsilon \mu} \
\nabla^{\zeta}H_{\beta}{}^{\varepsilon \mu} + \frac{5}{36} R_{\
\alpha \gamma \beta \delta} R^{\alpha \beta \gamma \delta} \
\nabla_{\zeta}H_{\epsilon \varepsilon \mu} \
\nabla^{\zeta}H^{\epsilon \varepsilon \mu}\nn\\&& + \frac{2}{3} \
R^{\alpha \beta \gamma \delta} R^{\epsilon \varepsilon \mu \
\zeta} \nabla_{\varepsilon}H_{\beta \delta \zeta} \
\nabla_{\mu}H_{\alpha \gamma \epsilon} - 2 R^{\alpha \beta \
\gamma \delta} R^{\epsilon \varepsilon \mu \zeta} \
\nabla_{\zeta}H_{\beta \delta \varepsilon} \
\nabla_{\mu}H_{\alpha \gamma \epsilon}\nn\\&& -  \frac{20}{3} \
R^{\alpha \beta \gamma \delta} R_{\gamma}{}^{\epsilon \
\varepsilon \mu} \nabla^{\zeta}H_{\alpha \delta \epsilon} \
\nabla_{\mu}H_{\beta \varepsilon \zeta} + \frac{7}{3} 
R^{\alpha \beta \gamma \delta} R_{\gamma}{}^{\epsilon 
\varepsilon \mu} \nabla_{\epsilon}H_{\alpha \beta}{}^{\zeta} 
\nabla_{\mu}H_{\delta \varepsilon \zeta}\labell{R2PH2}
\eeqa
They are consistent with the couplings that are produced by the S-matrix element of four NS-NS vertex operators using the scheme \reef{T54} for the field theory couplings \cite{Garousi:2020mqn}. However, the four-point S-matrix calculation can not fix all 22 parameters in \reef{T54} which involve $R^2(\nabla H)^2$ couplings \cite{Garousi:2020mqn}. They can be fixed by studying five-point S-matrix elements. The T-duality constraint, however, fixes all 22 parameters in these couplings. It would be interesting to study the S-matrix element of five NS-NS vertex operators and check that its low energy limit reproduces the above couplings. This five-point S-matrix element has been calculated in \cite{Liu:2019ses} from which the couplings with structure $H^2R^3$ have been extracted.

Our calculation produce no couplings involving two dilatons and two B-fields, and no couplings involving one dilaton, one graviton and two B-fields in the string frame. They are consistent with four-point S-matrix element which produces no such couplings in the string frame \cite{Garousi:2020mqn}. 
 
All other couplings that the T-duality constraint produces involve more than four fields which can not be compared with the four-point S-matrix elements. We find there are no couplings involving more than one dilaton. The couplings involving $\nabla\nabla \Phi$ appears in three structures. We have found 10 couplings with structure $H^6\nabla\nabla\Phi$, \ie
\beqa
{\cal L}_3^{H^6\prt\prt\Phi}&=&\Big[- \frac{265}{64} H_{\alpha}{}^{\gamma \delta} \
H_{\beta}{}^{\epsilon \varepsilon} H_{\gamma}{}^{\mu \zeta} H_{\
\delta}{}^{\eta \theta} H_{\epsilon \mu \zeta} H_{\varepsilon \
\eta \theta} \nn\\&&-  \
\frac{399}{32} H_{\alpha}{}^{\gamma \delta} \
H_{\beta}{}^{\epsilon \varepsilon} H_{\gamma \epsilon}{}^{\mu} \
H_{\delta \mu}{}^{\zeta} H_{\varepsilon}{}^{\eta \theta} \
H_{\zeta \eta \theta}  + \
\frac{5299}{384} H_{\alpha}{}^{\gamma \delta} \
H_{\beta}{}^{\epsilon \varepsilon} H_{\gamma \delta}{}^{\mu} \
H_{\epsilon \mu}{}^{\zeta} H_{\varepsilon}{}^{\eta \theta} \
H_{\zeta \eta \theta} \nn\\&& -  \
\frac{159}{32} H_{\alpha}{}^{\gamma \delta} H_{\beta \
\gamma}{}^{\epsilon} H_{\delta}{}^{\varepsilon \mu} \
H_{\epsilon}{}^{\zeta \eta} H_{\varepsilon \mu}{}^{\theta} \
H_{\zeta \eta \theta} + \
\frac{1169}{3072} H_{\alpha}{}^{\gamma \delta} \
H_{\beta}{}^{\epsilon \varepsilon} H_{\gamma \delta}{}^{\mu} \
H_{\epsilon \varepsilon \mu} H_{\zeta \eta \theta} H^{\zeta \
\eta \theta} \nn\\&& + \
\frac{1169}{1152} H_{\alpha}{}^{\gamma \delta} H_{\beta \
\gamma}{}^{\epsilon} H_{\delta}{}^{\varepsilon \mu} H_{\epsilon \
\varepsilon \mu} H_{\zeta \eta \theta} H^{\zeta \eta \theta} \
-  \frac{399}{64} \
H_{\alpha}{}^{\gamma \delta} H_{\beta}{}^{\epsilon \varepsilon} \
H_{\gamma \delta}{}^{\mu} H_{\epsilon}{}^{\zeta \eta} \
H_{\varepsilon \zeta}{}^{\theta} H_{\mu \eta \theta} \
\nn\\&& + \frac{399}{32} \
H_{\alpha}{}^{\gamma \delta} H_{\beta}{}^{\epsilon \varepsilon} \
H_{\gamma \epsilon}{}^{\mu} H_{\delta \varepsilon}{}^{\zeta} \
H_{\zeta \eta \theta} H_{\mu}{}^{\eta \theta} \
 + \frac{1097}{768} \
H_{\alpha}{}^{\gamma \delta} H_{\beta}{}^{\epsilon \varepsilon} \
H_{\gamma \delta}{}^{\mu} H_{\epsilon \varepsilon}{}^{\zeta} \
H_{\zeta \eta \theta} H_{\mu}{}^{\eta \theta} \
\nn\\&& + \frac{1193}{1024} \
H_{\alpha}{}^{\gamma \delta} H_{\beta \gamma \delta} \
H_{\epsilon \varepsilon}{}^{\zeta} H^{\epsilon \varepsilon \mu} \
H_{\zeta \eta \theta} H_{\mu}{}^{\eta \theta}\Big]
\nabla^{\beta}\nabla^{\alpha}\Phi
\eeqa
They can be reproduced by studying seven-point S-matrix element which is very difficult to calculate it. We have found 27 couplings with structure $H^4R^2\nabla\nabla\Phi$, \ie
\beqa
{\cal L}_3^{H^4R\prt\prt\Phi}&=&\Big[- \frac{1169}{192} H_{\gamma \delta}{}^{\varepsilon} H^{\gamma \
\delta \epsilon} H_{\epsilon}{}^{\mu \zeta} H_{\mu \
\zeta}{}^{\eta} R_{\alpha \varepsilon \beta \eta} + \
\frac{399}{16} H_{\gamma}{}^{\varepsilon \mu} H^{\gamma \delta \
\epsilon} H_{\delta \varepsilon}{}^{\zeta} H_{\epsilon \mu}{}^{\
\eta} R_{\alpha \zeta \beta \eta}\nn\\&& -  \frac{1061}{96} \
H_{\gamma \delta}{}^{\varepsilon} H^{\gamma \delta \epsilon} \
H_{\epsilon}{}^{\mu \zeta} H_{\varepsilon \mu}{}^{\eta} \
R_{\alpha \zeta \beta \eta} + \frac{1169}{2304} H_{\gamma \
\delta \epsilon} H^{\gamma \delta \epsilon} H_{\varepsilon \
\mu}{}^{\eta} H^{\varepsilon \mu \zeta} R_{\alpha \zeta \beta \
\eta}\nn\\&& + \frac{1169}{192} H_{\alpha}{}^{\gamma \delta} \
H_{\gamma}{}^{\epsilon \varepsilon} H_{\mu \zeta \eta} H^{\mu \
\zeta \eta} R_{\beta \epsilon \delta \varepsilon} + \
\frac{443}{24} H_{\alpha}{}^{\gamma \delta} \
H_{\gamma}{}^{\epsilon \varepsilon} H_{\epsilon}{}^{\mu \zeta} \
H_{\mu \zeta}{}^{\eta} R_{\beta \varepsilon \delta \eta}\nn\\&& + \
\frac{1197}{8} H_{\alpha}{}^{\gamma \delta} \
H_{\gamma}{}^{\epsilon \varepsilon} H_{\delta}{}^{\mu \zeta} \
H_{\epsilon \mu}{}^{\eta} R_{\beta \varepsilon \zeta \eta} + \
\frac{1463}{24} H_{\alpha}{}^{\gamma \delta} \
H_{\gamma}{}^{\epsilon \varepsilon} H_{\epsilon}{}^{\mu \zeta} \
H_{\varepsilon \mu}{}^{\eta} R_{\beta \zeta \delta \eta}\nn\\&& -  \
\frac{795}{16} H_{\alpha}{}^{\gamma \delta} \
H_{\gamma}{}^{\epsilon \varepsilon} H_{\epsilon \varepsilon}{}^{\
\mu} H_{\mu}{}^{\zeta \eta} R_{\beta \zeta \delta \eta} -  \
\frac{1169}{32} H_{\alpha}{}^{\gamma \delta} H_{\gamma \
\delta}{}^{\epsilon} H_{\varepsilon \mu}{}^{\eta} \
H^{\varepsilon \mu \zeta} R_{\beta \zeta \epsilon \eta}\nn\\&&  + \
\frac{266}{3} H_{\alpha}{}^{\gamma \delta} \
H_{\gamma}{}^{\epsilon \varepsilon} H_{\delta \epsilon}{}^{\mu} \
H_{\varepsilon}{}^{\zeta \eta} R_{\beta \zeta \mu \eta}-  \
\frac{247}{6} H_{\alpha}{}^{\gamma \delta} H_{\gamma \
\delta}{}^{\epsilon} H_{\epsilon}{}^{\varepsilon \mu} \
H_{\varepsilon}{}^{\zeta \eta} R_{\beta \zeta \mu \eta}\nn\\&&  -  \
\frac{371}{16} H_{\alpha}{}^{\gamma \delta} \
H_{\gamma}{}^{\epsilon \varepsilon} H_{\epsilon}{}^{\mu \zeta} \
H_{\mu \zeta}{}^{\eta} R_{\beta \eta \delta \varepsilon} + \
\frac{583}{32} H_{\alpha}{}^{\gamma \delta} \
H_{\gamma}{}^{\epsilon \varepsilon} H_{\delta}{}^{\mu \zeta} \
H_{\epsilon \varepsilon}{}^{\eta} R_{\beta \eta \mu \zeta}\nn\\&&  -  \
\frac{1169}{576} H_{\alpha}{}^{\gamma \delta} \
H_{\beta}{}^{\epsilon \varepsilon} H_{\mu \zeta \eta} H^{\mu \
\zeta \eta} R_{\gamma \epsilon \delta \varepsilon} -  \
\frac{159}{8} H_{\alpha}{}^{\gamma \delta} H_{\beta \
\gamma}{}^{\epsilon} H_{\varepsilon \mu}{}^{\eta} \
H^{\varepsilon \mu \zeta} R_{\delta \zeta \epsilon \eta} \nn\\&& + \
\frac{67}{2} H_{\alpha}{}^{\gamma \delta} \
H_{\beta}{}^{\epsilon \varepsilon} H_{\gamma}{}^{\mu \zeta} H_{\
\epsilon \mu}{}^{\eta} R_{\delta \zeta \varepsilon \eta} -  \
\frac{399}{8} H_{\alpha}{}^{\gamma \delta} \
H_{\beta}{}^{\epsilon \varepsilon} H_{\gamma \epsilon}{}^{\mu} \
H_{\mu}{}^{\zeta \eta} R_{\delta \zeta \varepsilon \eta}\nn\\&&  -  \
\frac{413}{32} H_{\alpha}{}^{\gamma \delta} \
H_{\beta}{}^{\epsilon \varepsilon} H_{\gamma}{}^{\mu \zeta} H_{\
\mu \zeta}{}^{\eta} R_{\delta \eta \epsilon \varepsilon} -  \
\frac{629}{24} H_{\alpha}{}^{\gamma \delta} \
H_{\beta}{}^{\epsilon \varepsilon} H_{\gamma}{}^{\mu \zeta} H_{\
\epsilon \mu}{}^{\eta} R_{\delta \eta \varepsilon \zeta}\nn\\&&  -  \
\frac{67}{4} H_{\alpha}{}^{\gamma \delta} \
H_{\beta}{}^{\epsilon \varepsilon} H_{\gamma}{}^{\mu \zeta} H_{\
\delta \mu}{}^{\eta} R_{\epsilon \zeta \varepsilon \eta} + \
\frac{1817}{48} H_{\alpha}{}^{\gamma \delta} \
H_{\beta}{}^{\epsilon \varepsilon} H_{\gamma \delta}{}^{\mu} \
H_{\mu}{}^{\zeta \eta} R_{\epsilon \zeta \varepsilon \eta}\nn\\&&  -  \
\frac{629}{24} H_{\alpha}{}^{\gamma \delta} H_{\beta \
\gamma}{}^{\epsilon} H_{\delta}{}^{\varepsilon \mu} \
H_{\varepsilon}{}^{\zeta \eta} R_{\epsilon \zeta \mu \eta} -  \
\frac{399}{4} H_{\alpha}{}^{\gamma \delta} \
H_{\beta}{}^{\epsilon \varepsilon} H_{\gamma \epsilon}{}^{\mu} \
H_{\delta}{}^{\zeta \eta} R_{\varepsilon \zeta \mu \eta}\nn\\&&  + \
\frac{247}{6} H_{\alpha}{}^{\gamma \delta} \
H_{\beta}{}^{\epsilon \varepsilon} H_{\gamma \delta}{}^{\mu} \
H_{\epsilon}{}^{\zeta \eta} R_{\varepsilon \zeta \mu \eta} + \
\frac{265}{8} H_{\alpha}{}^{\gamma \delta} H_{\beta \
\gamma}{}^{\epsilon} H_{\delta}{}^{\varepsilon \mu} \
H_{\epsilon}{}^{\zeta \eta} R_{\varepsilon \zeta \mu \eta}\nn\\&&  -  \
\frac{511}{48} H_{\alpha}{}^{\gamma \delta} H_{\beta \gamma \
\delta} H_{\epsilon}{}^{\zeta \eta} H^{\epsilon \varepsilon \
\mu} R_{\varepsilon \zeta \mu \eta}\Big]
\nabla^{\beta}\nabla^{\alpha}\Phi
\eeqa
They can be reproduced by studying six-point S-matrix element. We have found 15 couplings with structure $H^2R^2\nabla\nabla\Phi$, \ie
\beqa
{\cal L}_3^{H^2R^2\prt\prt\Phi}&=&\Big[- \frac{17}{6} H^{\gamma \delta \epsilon} H^{\varepsilon \mu \
\zeta} R_{\alpha \gamma \varepsilon \mu} R_{\beta \zeta \
\delta \epsilon} -  \frac{10}{3} H_{\gamma}{}^{\varepsilon \
\mu} H^{\gamma \delta \epsilon} R_{\alpha}{}^{\zeta}{}_{\delta \
\varepsilon} R_{\beta \zeta \epsilon \mu}\nn\\&& + \frac{17}{6} \
H_{\gamma}{}^{\varepsilon \mu} H^{\gamma \delta \epsilon} \
R_{\alpha}{}^{\zeta}{}_{\delta \epsilon} R_{\beta \zeta \
\varepsilon \mu} -  \frac{17}{3} H_{\gamma}{}^{\varepsilon \
\mu} H^{\gamma \delta \epsilon} R_{\alpha \delta \
\varepsilon}{}^{\zeta} R_{\beta \mu \epsilon \zeta}\nn\\&& -  \
\frac{7}{6} H_{\gamma \delta}{}^{\varepsilon} H^{\gamma \delta \
\epsilon} R_{\alpha}{}^{\mu}{}_{\epsilon}{}^{\zeta} R_{\beta \
\mu \varepsilon \zeta} -  \frac{29}{3} H_{\alpha}{}^{\gamma \
\delta} H^{\epsilon \varepsilon \mu} \
R_{\beta}{}^{\zeta}{}_{\epsilon \varepsilon} R_{\gamma \mu \
\delta \zeta}\nn\\&& -  \frac{17}{3} H_{\alpha}{}^{\gamma \delta} H^{\
\epsilon \varepsilon \mu} R_{\beta \epsilon \gamma}{}^{\zeta} \
R_{\delta \zeta \varepsilon \mu} + 4 H_{\alpha}{}^{\gamma \
\delta} H^{\epsilon \varepsilon \mu} \
R_{\beta}{}^{\zeta}{}_{\gamma \epsilon} R_{\delta \zeta \
\varepsilon \mu} \nn\\&&+ 8 H_{\alpha}{}^{\gamma \delta} \
H_{\gamma}{}^{\epsilon \varepsilon} \
R_{\beta}{}^{\mu}{}_{\epsilon}{}^{\zeta} R_{\delta \zeta \
\varepsilon \mu} -  \frac{7}{3} H_{\alpha}{}^{\gamma \delta} \
H_{\beta}{}^{\epsilon \varepsilon} \
R_{\gamma}{}^{\mu}{}_{\epsilon}{}^{\zeta} R_{\delta \zeta \
\varepsilon \mu}\nn\\&& -  \frac{10}{3} H_{\alpha}{}^{\gamma \delta} \
H_{\beta}{}^{\epsilon \varepsilon} \
R_{\gamma}{}^{\mu}{}_{\epsilon}{}^{\zeta} R_{\delta \mu \
\varepsilon \zeta} + \frac{17}{6} H_{\gamma}{}^{\varepsilon \
\mu} H^{\gamma \delta \epsilon} R_{\alpha \delta \
\beta}{}^{\zeta} R_{\epsilon \zeta \varepsilon \mu}\nn\\&& -  \
\frac{7}{3} H_{\alpha}{}^{\gamma \delta} \
H_{\gamma}{}^{\epsilon \varepsilon} \
R_{\beta}{}^{\mu}{}_{\delta}{}^{\zeta} R_{\epsilon \mu \
\varepsilon \zeta} + \frac{17}{3} H_{\alpha}{}^{\gamma \delta} \
H_{\beta}{}^{\epsilon \varepsilon} \
R_{\gamma}{}^{\mu}{}_{\delta}{}^{\zeta} R_{\epsilon \mu \
\varepsilon \zeta}\nn\\&& -  \frac{7}{3} H_{\alpha}{}^{\gamma \delta} \
H_{\beta \gamma}{}^{\epsilon} R_{\delta}{}^{\varepsilon \mu \
\zeta} R_{\epsilon \mu \varepsilon \zeta}\Big]
\nabla^{\beta}\nabla^{\alpha}\Phi
\eeqa
They can be reproduced by studying the five-point S-matrix element.

 The couplings involving $\nabla \Phi$ appear in three structures. We have found 30 couplings with structure $H^5\nabla H\nabla\Phi$, \ie
 \beqa
{\cal L}_3^{H^5\prt H\prt\Phi}&\!\!\!\!\!=\!\!\!\!\!&\Big[ - \frac{399}{32} H_{\alpha}{}^{\beta \gamma} \
H_{\beta}{}^{\delta \epsilon} H_{\gamma \delta}{}^{\varepsilon} \
H_{\epsilon}{}^{\mu \zeta} H_{\mu}{}^{\eta \theta} \
\nabla_{\zeta}H_{\varepsilon \eta \theta} -  \frac{211}{384} \
H_{\alpha}{}^{\beta \gamma} H_{\beta \gamma}{}^{\delta} \
H_{\delta}{}^{\epsilon \varepsilon} H_{\epsilon}{}^{\mu \zeta} \
H_{\mu}{}^{\eta \theta} \nabla_{\zeta}H_{\varepsilon \eta \
\theta}\nn\\&& + \frac{2087}{192} H_{\beta \gamma}{}^{\epsilon} \
H^{\beta \gamma \delta} H_{\delta}{}^{\varepsilon \mu} \
H_{\epsilon}{}^{\zeta \eta} H_{\varepsilon \mu}{}^{\theta} \
\nabla_{\eta}H_{\alpha \zeta \theta} -  \frac{399}{16} \
H_{\alpha}{}^{\beta \gamma} H_{\beta}{}^{\delta \epsilon} \
H_{\delta}{}^{\varepsilon \mu} H_{\epsilon}{}^{\zeta \eta} \
H_{\varepsilon \mu}{}^{\theta} \nabla_{\eta}H_{\gamma \zeta \
\theta}\nn\\&& + \frac{399}{32} H_{\beta \gamma}{}^{\epsilon} \
H^{\beta \gamma \delta} H_{\delta}{}^{\varepsilon \mu} \
H_{\varepsilon}{}^{\zeta \eta} H_{\mu \zeta}{}^{\theta} \
\nabla_{\theta}H_{\alpha \epsilon \eta} + \frac{1241}{3072} \
H_{\beta \gamma}{}^{\epsilon} H^{\beta \gamma \delta} \
H_{\delta}{}^{\varepsilon \mu} H_{\epsilon \varepsilon \mu} H^{\
\zeta \eta \theta} \nabla_{\theta}H_{\alpha \zeta \eta}\nn\\&& + \
\frac{399}{16} H_{\beta}{}^{\epsilon \varepsilon} H^{\beta \
\gamma \delta} H_{\gamma \epsilon}{}^{\mu} H_{\delta \
\varepsilon}{}^{\zeta} H_{\mu}{}^{\eta \theta} \
\nabla_{\theta}H_{\alpha \zeta \eta} + \frac{529}{96} \
H_{\beta \gamma}{}^{\epsilon} H^{\beta \gamma \delta} \
H_{\delta}{}^{\varepsilon \mu} H_{\epsilon \
\varepsilon}{}^{\zeta} H_{\mu}{}^{\eta \theta} \
\nabla_{\theta}H_{\alpha \zeta \eta}\nn\\&& + \frac{1169}{2304} \
H_{\beta \gamma \delta} H^{\beta \gamma \delta} H_{\epsilon \
\varepsilon}{}^{\zeta} H^{\epsilon \varepsilon \mu} H_{\mu}{}^{\
\eta \theta} \nabla_{\theta}H_{\alpha \zeta \eta} + \
\frac{399}{32} H_{\beta \gamma}{}^{\epsilon} H^{\beta \gamma \
\delta} H_{\delta}{}^{\varepsilon \mu} H_{\epsilon}{}^{\zeta \
\eta} H_{\varepsilon \zeta}{}^{\theta} \
\nabla_{\theta}H_{\alpha \mu \eta} \nn\\&&-  \frac{399}{32} \
H_{\alpha}{}^{\beta \gamma} H_{\delta}{}^{\mu \zeta} H^{\delta \
\epsilon \varepsilon} H_{\epsilon \mu}{}^{\eta} H_{\varepsilon \
\zeta}{}^{\theta} \nabla_{\theta}H_{\beta \gamma \eta} -  \
\frac{529}{192} H_{\alpha}{}^{\beta \gamma} H_{\delta \
\epsilon}{}^{\mu} H^{\delta \epsilon \varepsilon} \
H_{\varepsilon}{}^{\zeta \eta} H_{\mu \zeta}{}^{\theta} \
\nabla_{\theta}H_{\beta \gamma \eta}\nn\\&& + \frac{1169}{4608} \
H_{\alpha}{}^{\beta \gamma} H_{\delta \epsilon \varepsilon} H^{\
\delta \epsilon \varepsilon} H_{\mu \zeta}{}^{\theta} H^{\mu \
\zeta \eta} \nabla_{\theta}H_{\beta \gamma \eta} + \
\frac{1133}{384} H_{\alpha}{}^{\beta \gamma} H_{\delta \
\epsilon}{}^{\mu} H^{\delta \epsilon \varepsilon} \
H_{\varepsilon}{}^{\zeta \eta} H_{\zeta \eta}{}^{\theta} \
\nabla_{\theta}H_{\beta \gamma \mu}\nn\\&& + \frac{265}{32} \
H_{\alpha}{}^{\beta \gamma} H_{\beta}{}^{\delta \epsilon} \
H_{\delta \epsilon}{}^{\varepsilon} H_{\mu \zeta}{}^{\theta} \
H^{\mu \zeta \eta} \nabla_{\theta}H_{\gamma \varepsilon \eta} \
+ \frac{53}{16} H_{\alpha}{}^{\beta \gamma} \
H_{\beta}{}^{\delta \epsilon} H_{\delta}{}^{\varepsilon \mu} \
H_{\epsilon}{}^{\zeta \eta} H_{\varepsilon \mu}{}^{\theta} \
\nabla_{\theta}H_{\gamma \zeta \eta}\nn\\&& -  \frac{399}{32} \
H_{\alpha}{}^{\beta \gamma} H_{\beta}{}^{\delta \epsilon} \
H_{\delta \epsilon}{}^{\varepsilon} H_{\varepsilon}{}^{\mu \
\zeta} H_{\mu}{}^{\eta \theta} \nabla_{\theta}H_{\gamma \zeta \
\eta} + \frac{1169}{3072} H_{\alpha}{}^{\beta \gamma} H_{\beta \
\gamma}{}^{\delta} H_{\epsilon \varepsilon \mu} H^{\epsilon \
\varepsilon \mu} H^{\zeta \eta \theta} \
\nabla_{\theta}H_{\delta \zeta \eta}\nn\\&& -  \frac{1973}{96} \
H_{\alpha}{}^{\beta \gamma} H_{\beta \gamma}{}^{\delta} \
H_{\epsilon \varepsilon}{}^{\zeta} H^{\epsilon \varepsilon \mu} \
H_{\mu}{}^{\eta \theta} \nabla_{\theta}H_{\delta \zeta \eta} \
-  \frac{399}{64} H_{\alpha}{}^{\beta \gamma} H_{\beta \
\gamma}{}^{\delta} H_{\epsilon}{}^{\zeta \eta} H^{\epsilon \
\varepsilon \mu} H_{\varepsilon \zeta}{}^{\theta} \
\nabla_{\theta}H_{\delta \mu \eta}\nn\\&& + \frac{5773}{768} \
H_{\alpha}{}^{\beta \gamma} H_{\beta \gamma}{}^{\delta} \
H_{\delta}{}^{\epsilon \varepsilon} H_{\mu \zeta}{}^{\theta} \
H^{\mu \zeta \eta} \nabla_{\theta}H_{\epsilon \varepsilon \
\eta} + \frac{53}{16} H_{\alpha}{}^{\beta \gamma} \
H_{\beta}{}^{\delta \epsilon} H_{\gamma}{}^{\varepsilon \mu} \
H_{\delta}{}^{\zeta \eta} H_{\zeta \eta}{}^{\theta} \
\nabla_{\theta}H_{\epsilon \varepsilon \mu}\nn\\&& + \frac{399}{32} \
H_{\alpha}{}^{\beta \gamma} H_{\beta \gamma}{}^{\delta} \
H_{\delta}{}^{\epsilon \varepsilon} H_{\epsilon}{}^{\mu \zeta} \
H_{\mu}{}^{\eta \theta} \nabla_{\theta}H_{\varepsilon \zeta \
\eta} + \frac{399}{8} H_{\alpha}{}^{\beta \gamma} \
H_{\beta}{}^{\delta \epsilon} H_{\gamma}{}^{\varepsilon \mu} \
H_{\delta \varepsilon}{}^{\zeta} H_{\epsilon}{}^{\eta \theta} \
\nabla_{\theta}H_{\mu \zeta \eta}\nn\\&& -  \frac{399}{32} \
H_{\alpha}{}^{\beta \gamma} H_{\beta}{}^{\delta \epsilon} \
H_{\gamma}{}^{\varepsilon \mu} H_{\delta \epsilon}{}^{\zeta} \
H_{\varepsilon}{}^{\eta \theta} \nabla_{\theta}H_{\mu \zeta \
\eta} + \frac{399}{32} H_{\alpha}{}^{\beta \gamma} \
H_{\beta}{}^{\delta \epsilon} H_{\gamma \delta}{}^{\varepsilon} \
H_{\epsilon}{}^{\mu \zeta} H_{\varepsilon}{}^{\eta \theta} \
\nabla_{\theta}H_{\mu \zeta \eta}\nn\\&& + \frac{211}{384} \
H_{\alpha}{}^{\beta \gamma} H_{\beta \gamma}{}^{\delta} \
H_{\delta}{}^{\epsilon \varepsilon} H_{\epsilon}{}^{\mu \zeta} \
H_{\varepsilon}{}^{\eta \theta} \nabla_{\theta}H_{\mu \zeta \
\eta} + \frac{421}{192} H_{\alpha}{}^{\beta \gamma} \
H_{\beta}{}^{\delta \epsilon} H_{\gamma}{}^{\varepsilon \mu} \
H_{\delta \epsilon \varepsilon} H^{\zeta \eta \theta} \nabla_{\
\theta}H_{\mu \zeta \eta}\nn\\&& + \frac{5701}{1536} \
H_{\alpha}{}^{\beta \gamma} H_{\beta \gamma}{}^{\delta} \
H_{\delta}{}^{\epsilon \varepsilon} H_{\epsilon \varepsilon}{}^{\
\mu} H^{\zeta \eta \theta} \nabla_{\theta}H_{\mu \zeta \eta} \nn\\&&
+ \frac{399}{16} H_{\alpha}{}^{\beta \gamma} \
H_{\beta}{}^{\delta \epsilon} H_{\gamma}{}^{\varepsilon \mu} \
H_{\delta \varepsilon}{}^{\zeta} H_{\zeta}{}^{\eta \theta} \
\nabla_{\mu}H_{\epsilon \eta \theta}\Big]\nabla^{\alpha}\Phi
 \eeqa
 They can be reproduced by studying the seven-point S-matrix elements.  We have found 79 couplings with structure $H^3\nabla H R\nabla\Phi$, \ie
 \beqa
{\cal L}_3^{H^3\prt HR\prt\Phi}&\!\!\!\!\!=\!\!\!\!\!&\Big[ - \frac{629}{96} H_{\alpha}{}^{\beta \gamma} \
H_{\delta}{}^{\mu \zeta} H^{\delta \epsilon \varepsilon} \
R_{\varepsilon \eta \mu \zeta} \nabla_{\epsilon}H_{\beta \
\gamma}{}^{\eta} + \frac{265}{8} H_{\alpha}{}^{\beta \gamma} \
H_{\beta}{}^{\delta \epsilon} H_{\delta}{}^{\varepsilon \mu} \
R_{\varepsilon \zeta \mu \eta} \nabla_{\epsilon}H_{\gamma}{}^{\
\zeta \eta}\nn\\&& -  \frac{511}{48} H_{\beta \gamma}{}^{\epsilon} \
H^{\beta \gamma \delta} H^{\varepsilon \mu \zeta} R_{\alpha \
\eta \mu \zeta} \nabla_{\epsilon}H_{\delta \
\varepsilon}{}^{\eta} + \frac{223}{16} H_{\beta \
\gamma}{}^{\epsilon} H^{\beta \gamma \delta} \
H_{\delta}{}^{\varepsilon \mu} R_{\epsilon \zeta \mu \eta} \
\nabla_{\varepsilon}H_{\alpha}{}^{\zeta \eta}\nn\\&& -  \frac{399}{8} \
H_{\alpha}{}^{\beta \gamma} H_{\delta}{}^{\mu \zeta} H^{\delta \
\epsilon \varepsilon} R_{\gamma \eta \mu \zeta} \
\nabla_{\varepsilon}H_{\beta \epsilon}{}^{\eta} + \
\frac{753}{16} H_{\alpha}{}^{\beta \gamma} H_{\delta \
\epsilon}{}^{\mu} H^{\delta \epsilon \varepsilon} R_{\gamma \
\zeta \mu \eta} \nabla_{\varepsilon}H_{\beta}{}^{\zeta \eta}\nn\\&& \
-  \frac{67}{2} H_{\alpha}{}^{\beta \gamma} \
H_{\beta}{}^{\delta \epsilon} H_{\delta}{}^{\varepsilon \mu} \
R_{\epsilon \zeta \mu \eta} \nabla_{\varepsilon}H_{\gamma}{}^{\
\zeta \eta} -  \frac{167}{8} H_{\alpha}{}^{\beta \gamma} \
H_{\beta}{}^{\delta \epsilon} H^{\varepsilon \mu \zeta} \
R_{\gamma \eta \mu \zeta} \nabla_{\varepsilon}H_{\delta \
\epsilon}{}^{\eta}\nn\\&& -  \frac{131}{8} H_{\alpha}{}^{\beta \
\gamma} H_{\beta}{}^{\delta \epsilon} H_{\gamma}{}^{\varepsilon \
\mu} R_{\epsilon \zeta \mu \eta} \
\nabla_{\varepsilon}H_{\delta}{}^{\zeta \eta} -  \
\frac{227}{48} H_{\beta \gamma}{}^{\epsilon} H^{\beta \gamma \
\delta} H_{\delta}{}^{\varepsilon \mu} R_{\alpha \zeta \mu \
\eta} \nabla_{\varepsilon}H_{\epsilon}{}^{\zeta \eta} \nn\\&&+ \
\frac{175}{24} H_{\alpha}{}^{\beta \gamma} H_{\beta}{}^{\delta \
\epsilon} H_{\delta}{}^{\varepsilon \mu} R_{\gamma \zeta \mu \
\eta} \nabla_{\varepsilon}H_{\epsilon}{}^{\zeta \eta} + \
\frac{1415}{48} H_{\beta}{}^{\epsilon \varepsilon} H^{\beta \
\gamma \delta} H^{\mu \zeta \eta} R_{\delta \eta \epsilon \
\varepsilon} \nabla_{\zeta}H_{\alpha \gamma \mu}\nn\\&& + \
\frac{931}{48} H_{\beta}{}^{\epsilon \varepsilon} H^{\beta \
\gamma \delta} H^{\mu \zeta \eta} R_{\alpha \eta \epsilon \
\varepsilon} \nabla_{\zeta}H_{\gamma \delta \mu} -  \
\frac{119}{48} H_{\beta}{}^{\epsilon \varepsilon} H^{\beta \
\gamma \delta} H^{\mu \zeta \eta} R_{\alpha \eta \delta \
\varepsilon} \nabla_{\zeta}H_{\gamma \epsilon \mu}\nn\\&& -  \
\frac{449}{48} H_{\alpha}{}^{\beta \gamma} H^{\delta \epsilon \
\varepsilon} H^{\mu \zeta \eta} R_{\beta \varepsilon \gamma \
\eta} \nabla_{\zeta}H_{\delta \epsilon \mu} -  \
\frac{875}{192} H_{\beta}{}^{\epsilon \varepsilon} H^{\beta \
\gamma \delta} H^{\mu \zeta \eta} R_{\gamma \epsilon \delta \
\varepsilon} \nabla_{\eta}H_{\alpha \mu \zeta}\nn\\&& + \
\frac{875}{96} H_{\beta}{}^{\epsilon \varepsilon} H^{\beta \
\gamma \delta} H^{\mu \zeta \eta} R_{\alpha \epsilon \delta \
\varepsilon} \nabla_{\eta}H_{\gamma \mu \zeta} -  \
\frac{1415}{96} H_{\alpha}{}^{\beta \gamma} H_{\delta}{}^{\mu \
\zeta} H^{\delta \epsilon \varepsilon} R_{\varepsilon \eta \mu \
\zeta} \nabla^{\eta}H_{\beta \gamma \epsilon} \nn\\&&-  \
\frac{1169}{48} H_{\alpha}{}^{\beta \gamma} H_{\delta \
\epsilon}{}^{\mu} H^{\delta \epsilon \varepsilon} \
R_{\varepsilon \zeta \mu \eta} \nabla^{\eta}H_{\beta \
\gamma}{}^{\zeta} + \frac{2923}{96} H_{\alpha}{}^{\beta \
\gamma} H_{\delta}{}^{\mu \zeta} H^{\delta \epsilon \
\varepsilon} R_{\gamma \eta \mu \zeta} \nabla^{\eta}H_{\beta \
\epsilon \varepsilon}\nn\\&& -  \frac{399}{16} H_{\alpha}{}^{\beta \
\gamma} H_{\delta}{}^{\mu \zeta} H^{\delta \epsilon \
\varepsilon} R_{\gamma \eta \varepsilon \zeta} \
\nabla^{\eta}H_{\beta \epsilon \mu} + \frac{435}{16} \
H_{\alpha}{}^{\beta \gamma} H_{\delta \epsilon}{}^{\mu} \
H^{\delta \epsilon \varepsilon} R_{\gamma \zeta \mu \eta} \
\nabla^{\eta}H_{\beta \varepsilon}{}^{\zeta}\nn\\&& -  \frac{117}{16} \
H_{\alpha}{}^{\beta \gamma} H_{\delta \epsilon}{}^{\mu} \
H^{\delta \epsilon \varepsilon} R_{\gamma \eta \mu \zeta} \
\nabla^{\eta}H_{\beta \varepsilon}{}^{\zeta} -  \
\frac{1169}{576} H_{\alpha}{}^{\beta \gamma} H_{\delta \
\epsilon \varepsilon} H^{\delta \epsilon \varepsilon} R_{\gamma \
\eta \mu \zeta} \nabla^{\eta}H_{\beta}{}^{\mu \zeta}\nn\\&& + \
\frac{265}{8} H_{\alpha}{}^{\beta \gamma} H_{\beta}{}^{\delta \
\epsilon} H^{\varepsilon \mu \zeta} R_{\epsilon \eta \mu \
\zeta} \nabla^{\eta}H_{\gamma \delta \varepsilon} + \
\frac{399}{8} H_{\alpha}{}^{\beta \gamma} H_{\beta}{}^{\delta \
\epsilon} H_{\delta}{}^{\varepsilon \mu} R_{\varepsilon \zeta \
\mu \eta} \nabla^{\eta}H_{\gamma \epsilon}{}^{\zeta} \nn\\&&-  \
\frac{667}{8} H_{\alpha}{}^{\beta \gamma} H_{\beta}{}^{\delta \
\epsilon} H_{\delta}{}^{\varepsilon \mu} R_{\epsilon \zeta \mu \
\eta} \nabla^{\eta}H_{\gamma \varepsilon}{}^{\zeta} + \
\frac{1465}{8} H_{\alpha}{}^{\beta \gamma} H_{\beta}{}^{\delta \
\epsilon} H_{\delta}{}^{\varepsilon \mu} R_{\epsilon \eta \mu \
\zeta} \nabla^{\eta}H_{\gamma \varepsilon}{}^{\zeta}\nn\\&& + \
\frac{491}{24} H_{\alpha}{}^{\beta \gamma} H_{\beta}{}^{\delta \
\epsilon} H^{\varepsilon \mu \zeta} R_{\delta \zeta \epsilon \
\eta} \nabla^{\eta}H_{\gamma \varepsilon \mu} -  \
\frac{159}{16} H_{\alpha}{}^{\beta \gamma} H_{\beta}{}^{\delta \
\epsilon} H_{\delta \epsilon}{}^{\varepsilon} R_{\varepsilon \
\eta \mu \zeta} \nabla^{\eta}H_{\gamma}{}^{\mu \zeta} \nn\\&&+ \
\frac{931}{96} H_{\beta}{}^{\epsilon \varepsilon} H^{\beta \
\gamma \delta} H_{\gamma}{}^{\mu \zeta} R_{\alpha \eta \mu \
\zeta} \nabla^{\eta}H_{\delta \epsilon \varepsilon} + \
\frac{229}{8} H_{\alpha}{}^{\beta \gamma} H_{\beta}{}^{\delta \
\epsilon} H^{\varepsilon \mu \zeta} R_{\gamma \eta \mu \zeta} \
\nabla^{\eta}H_{\delta \epsilon \varepsilon}\nn\\&& -  \frac{643}{12} \
H_{\alpha}{}^{\beta \gamma} H_{\beta}{}^{\delta \epsilon} \
H_{\gamma}{}^{\varepsilon \mu} R_{\varepsilon \zeta \mu \eta} \
\nabla^{\eta}H_{\delta \epsilon}{}^{\zeta} -  \frac{545}{24} \
H_{\alpha}{}^{\beta \gamma} H_{\beta \gamma}{}^{\delta} \
H^{\epsilon \varepsilon \mu} R_{\varepsilon \zeta \mu \eta} \
\nabla^{\eta}H_{\delta \epsilon}{}^{\zeta} \nn\\&&-  \frac{1463}{48} \
H_{\beta}{}^{\epsilon \varepsilon} H^{\beta \gamma \delta} \
H_{\gamma \epsilon}{}^{\mu} R_{\alpha \zeta \mu \eta} \
\nabla^{\eta}H_{\delta \varepsilon}{}^{\zeta} + \frac{133}{24} \
H_{\beta}{}^{\epsilon \varepsilon} H^{\beta \gamma \delta} \
H_{\gamma \epsilon}{}^{\mu} R_{\alpha \eta \mu \zeta} \
\nabla^{\eta}H_{\delta \varepsilon}{}^{\zeta}\nn\\&& -  \frac{399}{4} \
H_{\alpha}{}^{\beta \gamma} H_{\beta}{}^{\delta \epsilon} \
H_{\gamma}{}^{\varepsilon \mu} R_{\epsilon \zeta \mu \eta} \
\nabla^{\eta}H_{\delta \varepsilon}{}^{\zeta} -  \
\frac{3563}{192} H_{\beta \gamma}{}^{\epsilon} H^{\beta \gamma \
\delta} H^{\varepsilon \mu \zeta} R_{\alpha \eta \epsilon \
\zeta} \nabla^{\eta}H_{\delta \varepsilon \mu}\nn\\&& + \
\frac{491}{24} H_{\alpha}{}^{\beta \gamma} H_{\beta}{}^{\delta \
\epsilon} H^{\varepsilon \mu \zeta} R_{\gamma \zeta \epsilon \
\eta} \nabla^{\eta}H_{\delta \varepsilon \mu} + \frac{13}{16} \
H_{\beta \gamma}{}^{\epsilon} H^{\beta \gamma \delta} \
H_{\delta}{}^{\varepsilon \mu} R_{\alpha \zeta \mu \eta} \
\nabla^{\eta}H_{\epsilon \varepsilon}{}^{\zeta} \nn\\&&+ \
\frac{983}{96} H_{\beta \gamma}{}^{\epsilon} H^{\beta \gamma \
\delta} H_{\delta}{}^{\varepsilon \mu} R_{\alpha \eta \mu \
\zeta} \nabla^{\eta}H_{\epsilon \varepsilon}{}^{\zeta} -  \
\frac{1169}{2304} H_{\beta \gamma \delta} H^{\beta \gamma \
\delta} H^{\epsilon \varepsilon \mu} R_{\alpha \eta \mu \
\zeta} \nabla^{\eta}H_{\epsilon \varepsilon}{}^{\zeta}\nn\\&& -  \
\frac{511}{12} H_{\alpha}{}^{\beta \gamma} H_{\beta}{}^{\delta \
\epsilon} H_{\delta}{}^{\varepsilon \mu} R_{\gamma \zeta \mu \
\eta} \nabla^{\eta}H_{\epsilon \varepsilon}{}^{\zeta} -  \
\frac{175}{24} H_{\alpha}{}^{\beta \gamma} H_{\beta}{}^{\delta \
\epsilon} H_{\delta}{}^{\varepsilon \mu} R_{\gamma \eta \mu \
\zeta} \nabla^{\eta}H_{\epsilon \varepsilon}{}^{\zeta}\nn\\&& + \
\frac{609}{16} H_{\alpha}{}^{\beta \gamma} H_{\beta \
\gamma}{}^{\delta} H^{\epsilon \varepsilon \mu} R_{\delta \
\zeta \mu \eta} \nabla^{\eta}H_{\epsilon \
\varepsilon}{}^{\zeta} -  \frac{329}{24} H_{\alpha}{}^{\beta \
\gamma} H_{\beta \gamma}{}^{\delta} H^{\epsilon \varepsilon \
\mu} R_{\delta \eta \mu \zeta} \nabla^{\eta}H_{\epsilon \
\varepsilon}{}^{\zeta}\nn\\&& -  \frac{1197}{32} H_{\beta}{}^{\epsilon \
\varepsilon} H^{\beta \gamma \delta} H_{\gamma}{}^{\mu \zeta} \
R_{\alpha \eta \delta \zeta} \nabla^{\eta}H_{\epsilon \
\varepsilon \mu} + \frac{399}{16} H_{\alpha}{}^{\beta \gamma} \
H_{\delta}{}^{\mu \zeta} H^{\delta \epsilon \varepsilon} \
R_{\beta \zeta \gamma \eta} \nabla^{\eta}H_{\epsilon \
\varepsilon \mu}\nn\\&& -  \frac{399}{16} H_{\alpha}{}^{\beta \gamma} \
H_{\beta}{}^{\delta \epsilon} H_{\gamma \delta}{}^{\varepsilon} \
R_{\varepsilon \eta \mu \zeta} \
\nabla^{\eta}H_{\epsilon}{}^{\mu \zeta} + \frac{159}{16} \
H_{\alpha}{}^{\beta \gamma} H_{\beta \gamma}{}^{\delta} \
H_{\delta}{}^{\epsilon \varepsilon} R_{\varepsilon \eta \mu \
\zeta} \nabla^{\eta}H_{\epsilon}{}^{\mu \zeta}\nn\\&& + \
\frac{1169}{384} H_{\beta \gamma}{}^{\epsilon} H^{\beta \gamma \
\delta} H^{\varepsilon \mu \zeta} R_{\alpha \delta \epsilon \
\eta} \nabla^{\eta}H_{\varepsilon \mu \zeta} -  \
\frac{215}{288} H_{\alpha}{}^{\beta \gamma} \
H_{\beta}{}^{\delta \epsilon} H^{\varepsilon \mu \zeta} \
R_{\gamma \eta \delta \epsilon} \nabla^{\eta}H_{\varepsilon \
\mu \zeta}\nn\\&& + \frac{67}{16} H_{\beta \gamma}{}^{\epsilon} \
H^{\beta \gamma \delta} H_{\delta}{}^{\varepsilon \mu} \
R_{\alpha \zeta \epsilon \eta} \nabla^{\eta}H_{\varepsilon \
\mu}{}^{\zeta} + \frac{421}{192} H_{\beta \gamma}{}^{\epsilon} \
H^{\beta \gamma \delta} H_{\delta}{}^{\varepsilon \mu} \
R_{\alpha \eta \epsilon \zeta} \nabla^{\eta}H_{\varepsilon \
\mu}{}^{\zeta} \nn\\&&-  \frac{399}{16} H_{\alpha}{}^{\beta \gamma} \
H_{\beta}{}^{\delta \epsilon} H_{\delta}{}^{\varepsilon \mu} \
R_{\gamma \zeta \epsilon \eta} \nabla^{\eta}H_{\varepsilon \
\mu}{}^{\zeta} + \frac{215}{48} H_{\alpha}{}^{\beta \gamma} \
H_{\beta}{}^{\delta \epsilon} H_{\delta}{}^{\varepsilon \mu} \
R_{\gamma \eta \epsilon \zeta} \nabla^{\eta}H_{\varepsilon \
\mu}{}^{\zeta}\nn\\&& + \frac{265}{16} H_{\alpha}{}^{\beta \gamma} \
H_{\beta}{}^{\delta \epsilon} H_{\delta \
\epsilon}{}^{\varepsilon} R_{\gamma \eta \mu \zeta} \
\nabla^{\eta}H_{\varepsilon}{}^{\mu \zeta} + \frac{227}{48} \
H_{\beta}{}^{\epsilon \varepsilon} H^{\beta \gamma \delta} \
H^{\mu \zeta \eta} R_{\epsilon \zeta \varepsilon \eta} \
\nabla_{\mu}H_{\alpha \gamma \delta}\nn\\&& + \frac{265}{8} \
H_{\beta}{}^{\epsilon \varepsilon} H^{\beta \gamma \delta} \
H^{\mu \zeta \eta} R_{\delta \zeta \varepsilon \eta} \nabla_{\
\mu}H_{\alpha \gamma \epsilon} + 166 H_{\beta}{}^{\epsilon \
\varepsilon} H^{\beta \gamma \delta} H_{\gamma}{}^{\mu \zeta} \
R_{\delta \zeta \varepsilon \eta} \nabla_{\mu}H_{\alpha \
\epsilon}{}^{\eta}\nn\\&& -  \frac{1765}{24} H_{\beta}{}^{\epsilon \
\varepsilon} H^{\beta \gamma \delta} H_{\gamma}{}^{\mu \zeta} \
R_{\delta \eta \varepsilon \zeta} \nabla_{\mu}H_{\alpha \
\epsilon}{}^{\eta} -  \frac{173}{6} H_{\beta \
\gamma}{}^{\epsilon} H^{\beta \gamma \delta} H^{\varepsilon \
\mu \zeta} R_{\delta \zeta \epsilon \eta} \
\nabla_{\mu}H_{\alpha \varepsilon}{}^{\eta}\nn\\&& + \frac{511}{24} \
H_{\alpha}{}^{\beta \gamma} H^{\delta \epsilon \varepsilon} H^{\
\mu \zeta \eta} R_{\epsilon \zeta \varepsilon \eta} \
\nabla_{\mu}H_{\beta \gamma \delta} + \frac{215}{48} \
H_{\alpha}{}^{\beta \gamma} H^{\delta \epsilon \varepsilon} H^{\
\mu \zeta \eta} R_{\gamma \zeta \varepsilon \eta} \
\nabla_{\mu}H_{\beta \delta \epsilon}\nn\\&& + \frac{629}{24} \
H_{\alpha}{}^{\beta \gamma} H_{\delta}{}^{\mu \zeta} H^{\delta \
\epsilon \varepsilon} R_{\gamma \zeta \varepsilon \eta} \
\nabla_{\mu}H_{\beta \epsilon}{}^{\eta} + \frac{175}{24} \
H_{\alpha}{}^{\beta \gamma} H_{\delta}{}^{\mu \zeta} H^{\delta \
\epsilon \varepsilon} R_{\gamma \eta \varepsilon \zeta} \
\nabla_{\mu}H_{\beta \epsilon}{}^{\eta}\nn\\&& + \frac{931}{48} \
H_{\beta}{}^{\epsilon \varepsilon} H^{\beta \gamma \delta} \
H^{\mu \zeta \eta} R_{\alpha \zeta \varepsilon \eta} \nabla_{\
\mu}H_{\gamma \delta \epsilon} -  \frac{399}{8} H_{\alpha}{}^{\
\beta \gamma} H_{\beta}{}^{\delta \epsilon} H^{\varepsilon \mu \
\zeta} R_{\delta \zeta \epsilon \eta} \nabla_{\mu}H_{\gamma \
\varepsilon}{}^{\eta}\nn\\&& + \frac{421}{288} H_{\alpha}{}^{\beta \
\gamma} H^{\delta \epsilon \varepsilon} H^{\mu \zeta \eta} R_{\
\beta \zeta \gamma \eta} \nabla_{\mu}H_{\delta \epsilon \
\varepsilon} -  \frac{266}{3} H_{\beta}{}^{\epsilon \
\varepsilon} H^{\beta \gamma \delta} H_{\gamma}{}^{\mu \zeta} \
R_{\alpha \zeta \varepsilon \eta} \nabla_{\mu}H_{\delta \
\epsilon}{}^{\eta} \nn\\&&+ \frac{399}{8} H_{\beta}{}^{\epsilon \
\varepsilon} H^{\beta \gamma \delta} H_{\gamma}{}^{\mu \zeta} \
R_{\alpha \eta \varepsilon \zeta} \nabla_{\mu}H_{\delta \
\epsilon}{}^{\eta} + \frac{173}{8} H_{\beta \
\gamma}{}^{\epsilon} H^{\beta \gamma \delta} H^{\varepsilon \
\mu \zeta} R_{\alpha \zeta \epsilon \eta} \
\nabla_{\mu}H_{\delta \varepsilon}{}^{\eta}\nn\\&& + \frac{89}{3} H_{\
\beta \gamma}{}^{\epsilon} H^{\beta \gamma \delta} \
H^{\varepsilon \mu \zeta} R_{\alpha \eta \epsilon \zeta} \
\nabla_{\mu}H_{\delta \varepsilon}{}^{\eta} -  \frac{89}{12} \
H_{\alpha}{}^{\beta \gamma} H_{\beta}{}^{\delta \epsilon} \
H^{\varepsilon \mu \zeta} R_{\gamma \zeta \epsilon \eta} \
\nabla_{\mu}H_{\delta \varepsilon}{}^{\eta}\nn\\&& -  \frac{1157}{24} \
H_{\alpha}{}^{\beta \gamma} H_{\beta}{}^{\delta \epsilon} \
H^{\varepsilon \mu \zeta} R_{\gamma \eta \epsilon \zeta} \
\nabla_{\mu}H_{\delta \varepsilon}{}^{\eta} -  \frac{399}{16} \
H_{\beta}{}^{\epsilon \varepsilon} H^{\beta \gamma \delta} \
H_{\gamma}{}^{\mu \zeta} R_{\alpha \zeta \delta \eta} \
\nabla_{\mu}H_{\epsilon \varepsilon}{}^{\eta}\nn\\&& + \frac{847}{16} \
H_{\beta}{}^{\epsilon \varepsilon} H^{\beta \gamma \delta} \
H_{\gamma}{}^{\mu \zeta} R_{\alpha \eta \delta \zeta} \
\nabla_{\mu}H_{\epsilon \varepsilon}{}^{\eta} -  \
\frac{187}{12} H_{\alpha}{}^{\beta \gamma} H_{\delta}{}^{\mu \
\zeta} H^{\delta \epsilon \varepsilon} R_{\beta \zeta \gamma \
\eta} \nabla_{\mu}H_{\epsilon \varepsilon}{}^{\eta}\nn\\&& + \
\frac{265}{16} H_{\alpha}{}^{\beta \gamma} H_{\delta \
\epsilon}{}^{\mu} H^{\delta \epsilon \varepsilon} R_{\beta \
\zeta \gamma \eta} \nabla_{\mu}H_{\varepsilon}{}^{\zeta \eta}\Big]\nabla^{\alpha}\Phi
 \eeqa
 They can be reproduced by studying the six-point S-matrix element. We have found 23 couplings with structure $H\nabla H R^2\nabla\Phi$, \ie 
 \beqa
{\cal L}_3^{H\prt HR^2\prt\Phi}&\!\!\!=\!\!\!&\Big[ \frac{7}{6} H^{\beta \gamma \delta} R_{\delta}{}^{\varepsilon \
\mu \zeta} R_{\epsilon \mu \varepsilon \zeta} \
\nabla^{\epsilon}H_{\alpha \beta \gamma} + \frac{7}{3} \
H^{\beta \gamma \delta} \
R_{\gamma}{}^{\mu}{}_{\epsilon}{}^{\zeta} R_{\delta \zeta \
\varepsilon \mu} \nabla^{\varepsilon}H_{\alpha \
\beta}{}^{\epsilon}\nn\\&& + \frac{20}{3} H^{\beta \gamma \delta} R_{\
\gamma}{}^{\mu}{}_{\epsilon}{}^{\zeta} R_{\delta \mu \
\varepsilon \zeta} \nabla^{\varepsilon}H_{\alpha \
\beta}{}^{\epsilon} - 9 H^{\beta \gamma \delta} \
R_{\gamma}{}^{\mu}{}_{\delta}{}^{\zeta} R_{\epsilon \mu \
\varepsilon \zeta} \nabla^{\varepsilon}H_{\alpha \
\beta}{}^{\epsilon}\nn\\&& -  \frac{17}{6} H^{\beta \gamma \delta} \
R_{\alpha}{}^{\mu}{}_{\varepsilon}{}^{\zeta} R_{\delta \zeta \
\epsilon \mu} \nabla^{\varepsilon}H_{\beta \
\gamma}{}^{\epsilon} + \frac{5}{3} H^{\beta \gamma \delta} R_{\
\alpha}{}^{\mu}{}_{\epsilon}{}^{\zeta} R_{\delta \zeta \
\varepsilon \mu} \nabla^{\varepsilon}H_{\beta \
\gamma}{}^{\epsilon}\nn\\&& -  \frac{7}{6} H^{\beta \gamma \delta} \
R_{\alpha}{}^{\mu}{}_{\delta}{}^{\zeta} R_{\epsilon \zeta \
\varepsilon \mu} \nabla^{\varepsilon}H_{\beta \
\gamma}{}^{\epsilon} + \frac{7}{3} H_{\alpha}{}^{\beta \gamma} \
R_{\gamma}{}^{\mu}{}_{\delta}{}^{\zeta} R_{\epsilon \zeta \
\varepsilon \mu} \nabla^{\varepsilon}H_{\beta}{}^{\delta \
\epsilon}\nn\\&& -  \frac{14}{3} H_{\alpha}{}^{\beta \gamma} \
R_{\gamma}{}^{\mu}{}_{\delta}{}^{\zeta} R_{\epsilon \mu \
\varepsilon \zeta} \nabla^{\varepsilon}H_{\beta}{}^{\delta \
\epsilon} + \frac{34}{3} H^{\beta \gamma \delta} R_{\alpha \
\epsilon \beta \varepsilon} R_{\gamma \mu \delta \zeta} \
\nabla^{\zeta}H^{\epsilon \varepsilon \mu} \nn\\&& + \frac{17}{6} \
H^{\beta \gamma \delta} R_{\alpha \epsilon \beta \gamma} \
R_{\delta \zeta \varepsilon \mu} \nabla^{\zeta}H^{\epsilon \
\varepsilon \mu}-  \frac{7}{6} H^{\beta \gamma \delta} \
R_{\beta \mu \gamma}{}^{\zeta} R_{\delta \zeta \epsilon \
\varepsilon} \nabla^{\mu}H_{\alpha}{}^{\epsilon \varepsilon}  \nn\\&&+ \
9 H^{\beta \gamma \delta} R_{\beta \epsilon \gamma}{}^{\zeta} \
R_{\delta \zeta \varepsilon \mu} \
\nabla^{\mu}H_{\alpha}{}^{\epsilon \varepsilon} - 9 H^{\beta \
\gamma \delta} R_{\beta \epsilon \gamma}{}^{\zeta} R_{\delta \
\mu \varepsilon \zeta} \nabla^{\mu}H_{\alpha}{}^{\epsilon \
\varepsilon} \nn\\&& + \frac{17}{6} H^{\beta \gamma \delta} \
R_{\alpha}{}^{\zeta}{}_{\epsilon \varepsilon} R_{\gamma \mu \
\delta \zeta} \nabla^{\mu}H_{\beta}{}^{\epsilon \varepsilon} + \
\frac{17}{3} H^{\beta \gamma \delta} R_{\alpha \epsilon \
\gamma}{}^{\zeta} R_{\delta \zeta \varepsilon \mu} \
\nabla^{\mu}H_{\beta}{}^{\epsilon \varepsilon} \nn\\&& + \frac{20}{3} \
H^{\beta \gamma \delta} R_{\alpha}{}^{\zeta}{}_{\gamma \
\epsilon} R_{\delta \zeta \varepsilon \mu} \
\nabla^{\mu}H_{\beta}{}^{\epsilon \varepsilon} + \frac{17}{3} \
H^{\beta \gamma \delta} R_{\alpha \epsilon \gamma}{}^{\zeta} \
R_{\delta \mu \varepsilon \zeta} \
\nabla^{\mu}H_{\beta}{}^{\epsilon \varepsilon} \nn\\&& -  \frac{20}{3} \
H^{\beta \gamma \delta} R_{\alpha}{}^{\zeta}{}_{\gamma \
\epsilon} R_{\delta \mu \varepsilon \zeta} \
\nabla^{\mu}H_{\beta}{}^{\epsilon \varepsilon} + \frac{17}{3} \
H^{\beta \gamma \delta} R_{\alpha}{}^{\zeta}{}_{\gamma \delta} \
R_{\epsilon \mu \varepsilon \zeta} \
\nabla^{\mu}H_{\beta}{}^{\epsilon \varepsilon}  \nn\\&&+ \frac{7}{6} \
H_{\alpha}{}^{\beta \gamma} R_{\beta \mu \delta}{}^{\zeta} R_{\
\gamma \zeta \epsilon \varepsilon} \nabla^{\mu}H^{\delta \
\epsilon \varepsilon} -  \frac{9}{2} H_{\alpha}{}^{\beta \
\gamma} R_{\beta}{}^{\zeta}{}_{\delta \epsilon} R_{\gamma \
\zeta \varepsilon \mu} \nabla^{\mu}H^{\delta \epsilon \
\varepsilon} \nn\\&& -  \frac{41}{3} H_{\alpha}{}^{\beta \gamma} \
R_{\beta \delta \gamma}{}^{\zeta} R_{\epsilon \mu \varepsilon \
\zeta} \nabla^{\mu}H^{\delta \epsilon \varepsilon}\Big]\nabla^{\alpha}\Phi
 \eeqa
 They can be reproduced by studying the five-point S-matrix element.
 
 The other couplings have no dilaton. They appear in six structures.  We have found 5 couplings with structure $H^8$, \ie
 \beqa
{\cal L}_3^{H^8}&=& 
\frac{1}{48} H_{\alpha}{}^{\delta \epsilon} H^{\alpha \beta \
\gamma} H_{\beta}{}^{\varepsilon \mu} H_{\gamma}{}^{\zeta \
\eta} H_{\delta \varepsilon}{}^{\theta} H_{\epsilon \
\zeta}{}^{\iota} H_{\theta \iota \kappa} H_{\mu \
\eta}{}^{\kappa}\nn\\&&  - 
\frac{9}{128} H_{\alpha}{}^{\delta \epsilon} H^{\alpha \beta \
\gamma} H_{\beta}{}^{\varepsilon \mu} H_{\gamma}{}^{\zeta \
\eta} H_{\delta \varepsilon}{}^{\theta} H_{\epsilon \
\zeta}{}^{\iota} H_{\eta \theta \kappa} H_{\mu \
\iota}{}^{\kappa} \nn\\&&+ \frac{17}{384} H_{\alpha \beta}{}^{\delta} \
H^{\alpha \beta \gamma} H_{\gamma}{}^{\epsilon \varepsilon} H_{\
\delta}{}^{\mu \zeta} H_{\epsilon \varepsilon}{}^{\eta} \
H_{\zeta \theta}{}^{\kappa} H_{\eta \iota \kappa} \
H_{\mu}{}^{\theta \iota}\nn\\&& + \frac{1}{64} H_{\alpha \
\beta}{}^{\delta} H^{\alpha \beta \gamma} \
H_{\gamma}{}^{\epsilon \varepsilon} H_{\delta \epsilon}{}^{\mu} \
H_{\varepsilon}{}^{\zeta \eta} H_{\zeta \theta}{}^{\kappa} \
H_{\eta \iota \kappa} H_{\mu}{}^{\theta \iota}\nn\\&&+ \frac{1}{16} H_{\alpha \beta}{}^{\delta} H^{\alpha \beta \
\gamma} H_{\gamma}{}^{\epsilon \varepsilon} H_{\delta}{}^{\mu \
\zeta} H_{\epsilon \mu}{}^{\eta} H_{\varepsilon}{}^{\theta \
\iota} H_{\zeta \theta}{}^{\kappa} H_{\eta \iota \kappa}\labell{H8}
 \eeqa
 They can be reproduced by studying the eight-point S-matrix element which is extremely difficult to calculate it. We have found 78 couplings with structure $H^4(\nabla H)^2$, \ie 
 \beqa
{\cal L}_3^{H^4(\prt H)^2}&\!\!\!\!\!=\!\!\!\!\!& \frac{5}{16} H_{\alpha}{}^{\delta \epsilon} H^{\alpha \beta 
\gamma} H_{\beta}{}^{\varepsilon \mu} H_{\delta 
\varepsilon}{}^{\zeta} \nabla_{\epsilon}H_{\gamma}{}^{\eta 
\theta} \nabla_{\zeta}H_{\mu \eta \theta} -  \frac{3}{8} 
H_{\alpha}{}^{\delta \epsilon} H^{\alpha \beta \gamma} 
H_{\beta}{}^{\varepsilon \mu} H_{\delta}{}^{\zeta \eta} 
\nabla_{\varepsilon}H_{\gamma \zeta}{}^{\theta} 
\nabla_{\eta}H_{\epsilon \mu \theta}\nn\\&\!\!\!\!\!\!\!\!\!& -  \frac{25}{8} 
H_{\alpha}{}^{\delta \epsilon} H^{\alpha \beta \gamma} 
H_{\beta}{}^{\varepsilon \mu} H_{\delta}{}^{\zeta \eta} 
\nabla_{\epsilon}H_{\gamma \varepsilon}{}^{\theta} 
\nabla_{\eta}H_{\mu \zeta \theta} + \frac{1}{8} 
H_{\alpha}{}^{\delta \epsilon} H^{\alpha \beta \gamma} 
H_{\varepsilon}{}^{\eta \theta} H^{\varepsilon \mu \zeta} 
\nabla_{\eta}H_{\beta \delta \mu} \nabla_{\theta}H_{\gamma 
\epsilon \zeta}\nn\\&\!\!\!\!\!\!\!\!\!& -  \frac{11}{8} H_{\alpha}{}^{\delta \epsilon} 
H^{\alpha \beta \gamma} H_{\varepsilon}{}^{\eta \theta} 
H^{\varepsilon \mu \zeta} \nabla_{\zeta}H_{\beta \delta \mu} 
\nabla_{\theta}H_{\gamma \epsilon \eta} + \frac{5}{32} 
H_{\alpha}{}^{\delta \epsilon} H^{\alpha \beta \gamma} 
H_{\varepsilon}{}^{\eta \theta} H^{\varepsilon \mu \zeta} 
\nabla_{\eta}H_{\beta \gamma \mu} \nabla_{\theta}H_{\delta 
\epsilon \zeta}\nn\\&\!\!\!\!\!\!\!\!\!& + \frac{797}{192} H_{\alpha}{}^{\delta 
\epsilon} H^{\alpha \beta \gamma} H_{\varepsilon}{}^{\eta 
\theta} H^{\varepsilon \mu \zeta} \nabla_{\zeta}H_{\beta 
\gamma \mu} \nabla_{\theta}H_{\delta \epsilon \eta} + 
\frac{83}{4} H_{\alpha \beta}{}^{\delta} H^{\alpha \beta 
\gamma} H^{\epsilon \varepsilon \mu} H^{\zeta \eta \theta} 
\nabla_{\varepsilon}H_{\gamma \epsilon \zeta} 
\nabla_{\theta}H_{\delta \mu \eta}\nn\\&\!\!\!\!\!\!\!\!\!& -  \frac{161}{16} 
H_{\alpha \beta}{}^{\delta} H^{\alpha \beta \gamma} 
H^{\epsilon \varepsilon \mu} H^{\zeta \eta \theta} 
\nabla_{\zeta}H_{\gamma \epsilon \varepsilon} 
\nabla_{\theta}H_{\delta \mu \eta} + \frac{281}{96} 
H_{\alpha}{}^{\delta \epsilon} H^{\alpha \beta \gamma} 
H_{\beta}{}^{\varepsilon \mu} H^{\zeta \eta \theta} 
\nabla_{\delta}H_{\gamma \zeta \eta} 
\nabla_{\theta}H_{\epsilon \varepsilon \mu}\nn\\&\!\!\!\!\!\!\!\!\!& -  \frac{73}{6} 
H_{\alpha}{}^{\delta \epsilon} H^{\alpha \beta \gamma} 
H_{\beta}{}^{\varepsilon \mu} H^{\zeta \eta \theta} 
\nabla_{\eta}H_{\gamma \delta \zeta} 
\nabla_{\theta}H_{\epsilon \varepsilon \mu} + \frac{665}{96} 
H_{\alpha}{}^{\delta \epsilon} H^{\alpha \beta \gamma} 
H_{\varepsilon}{}^{\eta \theta} H^{\varepsilon \mu \zeta} 
\nabla_{\delta}H_{\beta \gamma \mu} 
\nabla_{\theta}H_{\epsilon \zeta \eta}\nn\\&\!\!\!\!\!\!\!\!\!& + \frac{5}{4} 
H_{\alpha}{}^{\delta \epsilon} H^{\alpha \beta \gamma} 
H_{\beta}{}^{\varepsilon \mu} H^{\zeta \eta \theta} 
\nabla_{\varepsilon}H_{\gamma \delta \zeta} \nabla_{\theta}H_{
\epsilon \mu \eta} -  \frac{83}{8} H_{\alpha 
\beta}{}^{\delta} H^{\alpha \beta \gamma} H^{\epsilon 
\varepsilon \mu} H^{\zeta \eta \theta} 
\nabla_{\delta}H_{\gamma \epsilon \zeta} 
\nabla_{\theta}H_{\varepsilon \mu \eta}\nn\\&\!\!\!\!\!\!\!\!\!& + \frac{3}{4} 
H_{\alpha}{}^{\delta \epsilon} H^{\alpha \beta \gamma} 
H_{\beta}{}^{\varepsilon \mu} H^{\zeta \eta \theta} 
\nabla_{\epsilon}H_{\gamma \delta \zeta} 
\nabla_{\theta}H_{\varepsilon \mu \eta} + \frac{731}{96} 
H_{\alpha}{}^{\delta \epsilon} H^{\alpha \beta \gamma} 
H_{\beta}{}^{\varepsilon \mu} H^{\zeta \eta \theta} 
\nabla_{\zeta}H_{\gamma \delta \epsilon} 
\nabla_{\theta}H_{\varepsilon \mu \eta}\nn\\&\!\!\!\!\!\!\!\!\!& -  \frac{1169}{384} 
H_{\alpha \beta}{}^{\delta} H^{\alpha \beta \gamma} 
H^{\epsilon \varepsilon \mu} H^{\zeta \eta \theta} 
\nabla_{\delta}H_{\gamma \epsilon \varepsilon} 
\nabla_{\theta}H_{\mu \zeta \eta} -  \frac{291}{64} H_{\alpha 
\beta}{}^{\delta} H^{\alpha \beta \gamma} 
H_{\epsilon}{}^{\zeta \eta} H^{\epsilon \varepsilon \mu} 
\nabla_{\delta}H_{\gamma \varepsilon}{}^{\theta} 
\nabla_{\theta}H_{\mu \zeta \eta} \nn\\&\!\!\!\!\!\!\!\!\!&+ \frac{67}{16} 
H_{\alpha}{}^{\delta \epsilon} H^{\alpha \beta \gamma} 
H_{\beta}{}^{\varepsilon \mu} H^{\zeta \eta \theta} 
\nabla_{\epsilon}H_{\gamma \delta \varepsilon} 
\nabla_{\theta}H_{\mu \zeta \eta} -  \frac{13}{32} 
H_{\alpha}{}^{\delta \epsilon} H^{\alpha \beta \gamma} 
H_{\beta}{}^{\varepsilon \mu} H^{\zeta \eta \theta} 
\nabla_{\varepsilon}H_{\gamma \delta \epsilon} 
\nabla_{\theta}H_{\mu \zeta \eta}\nn\\&\!\!\!\!\!\!\!\!\!& + \frac{1}{2} 
H_{\alpha}{}^{\delta \epsilon} H^{\alpha \beta \gamma} 
H_{\beta}{}^{\varepsilon \mu} H_{\delta}{}^{\zeta \eta} 
\nabla_{\varepsilon}H_{\gamma \epsilon}{}^{\theta} 
\nabla_{\theta}H_{\mu \zeta \eta} -  \frac{139}{12} H_{\alpha 
\beta}{}^{\delta} H^{\alpha \beta \gamma} 
H_{\gamma}{}^{\epsilon \varepsilon} H_{\epsilon}{}^{\mu \zeta} 
\nabla_{\varepsilon}H_{\delta}{}^{\eta \theta} 
\nabla_{\theta}H_{\mu \zeta \eta}\nn\\&\!\!\!\!\!\!\!\!\!& + \frac{13}{6} 
H_{\alpha}{}^{\delta \epsilon} H^{\alpha \beta \gamma} 
H_{\beta \delta}{}^{\varepsilon} H^{\mu \zeta \eta} 
\nabla_{\eta}H_{\varepsilon \zeta \theta} 
\nabla^{\theta}H_{\gamma \epsilon \mu} -  \frac{23}{48} 
H_{\alpha \beta}{}^{\delta} H^{\alpha \beta \gamma} 
H_{\epsilon}{}^{\zeta \eta} H^{\epsilon \varepsilon \mu} 
\nabla_{\eta}H_{\delta \mu \theta} \nabla^{\theta}H_{\gamma 
\varepsilon \zeta}\nn\\&\!\!\!\!\!\!\!\!\!& -  \frac{1}{2} H_{\alpha}{}^{\delta 
\epsilon} H^{\alpha \beta \gamma} H_{\beta}{}^{\varepsilon 
\mu} H_{\delta}{}^{\zeta \eta} \nabla_{\eta}H_{\epsilon \mu 
\theta} \nabla^{\theta}H_{\gamma \varepsilon \zeta} + 
\frac{13}{48} H_{\alpha \beta}{}^{\delta} H^{\alpha \beta 
\gamma} H_{\epsilon}{}^{\zeta \eta} H^{\epsilon \varepsilon 
\mu} \nabla_{\theta}H_{\delta \mu \eta} 
\nabla^{\theta}H_{\gamma \varepsilon \zeta} \nn\\&\!\!\!\!\!\!\!\!\!&-  \frac{1}{4} H_{
\alpha}{}^{\delta \epsilon} H^{\alpha \beta \gamma} 
H_{\beta}{}^{\varepsilon \mu} H_{\delta}{}^{\zeta \eta} 
\nabla_{\theta}H_{\epsilon \mu \eta} \nabla^{\theta}H_{\gamma 
\varepsilon \zeta} -  \frac{515}{96} H_{\alpha}{}^{\delta 
\epsilon} H^{\alpha \beta \gamma} H_{\beta}{}^{\varepsilon 
\mu} H_{\delta}{}^{\zeta \eta} \nabla_{\eta}H_{\epsilon \zeta 
\theta} \nabla^{\theta}H_{\gamma \varepsilon \mu} \nn\\&\!\!\!\!\!\!\!\!\!&+ 
\frac{665}{192} H_{\alpha}{}^{\delta \epsilon} H^{\alpha \beta 
\gamma} H_{\beta}{}^{\varepsilon \mu} H_{\delta}{}^{\zeta 
\eta} \nabla_{\theta}H_{\epsilon \zeta \eta} 
\nabla^{\theta}H_{\gamma \varepsilon \mu} -  \frac{229}{24} 
H_{\alpha}{}^{\delta \epsilon} H^{\alpha \beta \gamma} 
H_{\beta}{}^{\varepsilon \mu} H_{\delta}{}^{\zeta \eta} 
\nabla_{\theta}H_{\epsilon \varepsilon \mu} \nabla^{\theta}H_{
\gamma \zeta \eta} \nn\\&\!\!\!\!\!\!\!\!\!&-  \frac{1}{8} H_{\alpha}{}^{\delta 
\epsilon} H^{\alpha \beta \gamma} H_{\beta 
\delta}{}^{\varepsilon} H^{\mu \zeta \eta} 
\nabla_{\varepsilon}H_{\epsilon \eta \theta} 
\nabla^{\theta}H_{\gamma \mu \zeta} + \frac{1283}{384} 
H_{\alpha \beta}{}^{\delta} H^{\alpha \beta \gamma} 
H_{\epsilon \varepsilon}{}^{\zeta} H^{\epsilon \varepsilon \mu} 
\nabla_{\eta}H_{\delta \zeta \theta} \nabla^{\theta}H_{\gamma 
\mu}{}^{\eta}\nn\\&\!\!\!\!\!\!\!\!\!& -  \frac{1283}{384} H_{\alpha \beta}{}^{\delta} 
H^{\alpha \beta \gamma} H_{\epsilon \varepsilon}{}^{\zeta} 
H^{\epsilon \varepsilon \mu} \nabla_{\theta}H_{\delta \zeta 
\eta} \nabla^{\theta}H_{\gamma \mu}{}^{\eta} -  
\frac{289}{24} H_{\alpha}{}^{\delta \epsilon} H^{\alpha \beta 
\gamma} H_{\beta}{}^{\varepsilon \mu} H_{\gamma}{}^{\zeta 
\eta} \nabla_{\eta}H_{\mu \zeta \theta} 
\nabla^{\theta}H_{\delta \epsilon \varepsilon}\nn\\&\!\!\!\!\!\!\!\!\!& + 
\frac{283}{48} H_{\alpha}{}^{\delta \epsilon} H^{\alpha \beta 
\gamma} H_{\beta}{}^{\varepsilon \mu} H_{\gamma}{}^{\zeta 
\eta} \nabla_{\theta}H_{\mu \zeta \eta} 
\nabla^{\theta}H_{\delta \epsilon \varepsilon} -  
\frac{263}{12} H_{\alpha \beta}{}^{\delta} H^{\alpha \beta 
\gamma} H_{\gamma}{}^{\epsilon \varepsilon} H^{\mu \zeta \eta} 
\nabla_{\eta}H_{\varepsilon \zeta \theta} \
\nabla^{\theta}H_{\delta \epsilon \mu}\nn\\&\!\!\!\!\!\!\!\!\!& + \frac{2207}{576} 
H_{\alpha \beta \gamma} H^{\alpha \beta \gamma} H^{\delta 
\epsilon \varepsilon} H^{\mu \zeta \eta} 
\nabla_{\eta}H_{\varepsilon \zeta \theta} 
\nabla^{\theta}H_{\delta \epsilon \mu} + \frac{95}{12} 
H_{\alpha \beta}{}^{\delta} H^{\alpha \beta \gamma} 
H_{\gamma}{}^{\epsilon \varepsilon} H^{\mu \zeta \eta} 
\nabla_{\theta}H_{\varepsilon \zeta \eta} 
\nabla^{\theta}H_{\delta \epsilon \mu}\nn\\&\!\!\!\!\!\!\!\!\!& -  \frac{1177}{1152} 
H_{\alpha \beta \gamma} H^{\alpha \beta \gamma} H^{\delta 
\epsilon \varepsilon} H^{\mu \zeta \eta} 
\nabla_{\theta}H_{\varepsilon \zeta \eta} 
\nabla^{\theta}H_{\delta \epsilon \mu} + \frac{1301}{96} 
H_{\alpha \beta}{}^{\delta} H^{\alpha \beta \gamma} 
H_{\gamma}{}^{\epsilon \varepsilon} H_{\epsilon}{}^{\mu \zeta} 
\nabla_{\eta}H_{\mu \zeta \theta} \nabla^{\theta}H_{\delta 
\varepsilon}{}^{\eta}\nn\\&\!\!\!\!\!\!\!\!\!& -  \frac{99}{8} H_{\alpha 
\beta}{}^{\delta} H^{\alpha \beta \gamma} 
H_{\gamma}{}^{\epsilon \varepsilon} H_{\epsilon}{}^{\mu \zeta} 
\nabla_{\theta}H_{\mu \zeta \eta} \nabla^{\theta}H_{\delta 
\varepsilon}{}^{\eta} + \frac{37}{48} H_{\alpha 
\beta}{}^{\delta} H^{\alpha \beta \gamma} 
H_{\gamma}{}^{\epsilon \varepsilon} H_{\epsilon \varepsilon}{}^{
\mu} \nabla_{\eta}H_{\mu \zeta \theta} 
\nabla^{\theta}H_{\delta}{}^{\zeta \eta} \nn\\&\!\!\!\!\!\!\!\!\!&-  \frac{1}{32} 
H_{\alpha \beta}{}^{\delta} H^{\alpha \beta \gamma} 
H_{\gamma}{}^{\epsilon \varepsilon} H_{\epsilon \varepsilon}{}^{
\mu} \nabla_{\theta}H_{\mu \zeta \eta} 
\nabla^{\theta}H_{\delta}{}^{\zeta \eta}-  \frac{463}{192} 
H_{\alpha \beta}{}^{\delta} H^{\alpha \beta \gamma} 
H_{\gamma}{}^{\epsilon \varepsilon} H^{\mu \zeta \eta} 
\nabla_{\eta}H_{\epsilon \varepsilon \theta} 
\nabla^{\theta}H_{\delta \mu \zeta} \nn\\&\!\!\!\!\!\!\!\!\!& + \frac{281}{384} 
H_{\alpha \beta}{}^{\delta} H^{\alpha \beta \gamma} 
H_{\gamma}{}^{\epsilon \varepsilon} H^{\mu \zeta \eta} 
\nabla_{\theta}H_{\epsilon \varepsilon \eta} 
\nabla^{\theta}H_{\delta \mu \zeta} -  \frac{5}{2} H_{\alpha 
\beta}{}^{\delta} H^{\alpha \beta \gamma} 
H_{\gamma}{}^{\epsilon \varepsilon} H_{\epsilon}{}^{\mu \zeta} 
\nabla_{\zeta}H_{\varepsilon \eta \theta} 
\nabla^{\theta}H_{\delta \mu}{}^{\eta}  \nn\\&\!\!\!\!\!\!\!\!\!&+ \frac{5}{2} 
H_{\alpha \beta}{}^{\delta} H^{\alpha \beta \gamma} 
H_{\gamma}{}^{\epsilon \varepsilon} H_{\epsilon}{}^{\mu \zeta} 
\nabla_{\eta}H_{\varepsilon \zeta \theta} 
\nabla^{\theta}H_{\delta \mu}{}^{\eta} -  \frac{5}{2} 
H_{\alpha \beta}{}^{\delta} H^{\alpha \beta \gamma} 
H_{\gamma}{}^{\epsilon \varepsilon} H_{\epsilon}{}^{\mu \zeta} 
\nabla_{\theta}H_{\varepsilon \zeta \eta} 
\nabla^{\theta}H_{\delta \mu}{}^{\eta} \nn\\&\!\!\!\!\!\!\!\!\!& + \frac{13}{48} 
H_{\alpha}{}^{\delta \epsilon} H^{\alpha \beta \gamma} 
H_{\beta \delta}{}^{\varepsilon} H_{\gamma}{}^{\mu \zeta} 
\nabla_{\eta}H_{\mu \zeta \theta} \nabla^{\theta}H_{\epsilon 
\varepsilon}{}^{\eta} -  \frac{557}{192} H_{\alpha 
\beta}{}^{\delta} H^{\alpha \beta \gamma} 
H_{\gamma}{}^{\epsilon \varepsilon} H_{\delta}{}^{\mu \zeta} 
\nabla_{\eta}H_{\mu \zeta \theta} \nabla^{\theta}H_{\epsilon 
\varepsilon}{}^{\eta} \nn\\&\!\!\!\!\!\!\!\!\!& + \frac{1571}{2304} H_{\alpha \beta 
\gamma} H^{\alpha \beta \gamma} H_{\delta}{}^{\mu \zeta} 
H^{\delta \epsilon \varepsilon} \nabla_{\eta}H_{\mu \zeta 
\theta} \nabla^{\theta}H_{\epsilon \varepsilon}{}^{\eta} + 
\frac{1}{12} H_{\alpha}{}^{\delta \epsilon} H^{\alpha \beta 
\gamma} H_{\beta \delta}{}^{\varepsilon} H_{\gamma}{}^{\mu 
\zeta} \nabla_{\theta}H_{\mu \zeta \eta} 
\nabla^{\theta}H_{\epsilon \varepsilon}{}^{\eta} \nn\\&\!\!\!\!\!\!\!\!\!& + 
\frac{307}{384} H_{\alpha \beta}{}^{\delta} H^{\alpha \beta 
\gamma} H_{\gamma}{}^{\epsilon \varepsilon} H_{\delta}{}^{\mu 
\zeta} \nabla_{\theta}H_{\mu \zeta \eta} 
\nabla^{\theta}H_{\epsilon \varepsilon}{}^{\eta} -  
\frac{1531}{2304} H_{\alpha \beta \gamma} H^{\alpha \beta 
\gamma} H_{\delta}{}^{\mu \zeta} H^{\delta \epsilon 
\varepsilon} \nabla_{\theta}H_{\mu \zeta \eta} 
\nabla^{\theta}H_{\epsilon \varepsilon}{}^{\eta} \nn\\&\!\!\!\!\!\!\!\!\!& + 
\frac{121}{32} H_{\alpha \beta}{}^{\delta} H^{\alpha \beta 
\gamma} H_{\gamma}{}^{\epsilon \varepsilon} H^{\mu \zeta \eta} 
\nabla_{\delta}H_{\zeta \eta \theta} 
\nabla^{\theta}H_{\epsilon \varepsilon \mu} + \frac{16}{3} H_{
\alpha \beta}{}^{\delta} H^{\alpha \beta \gamma} 
H_{\gamma}{}^{\epsilon \varepsilon} H^{\mu \zeta \eta} 
\nabla_{\delta}H_{\varepsilon \eta \theta} 
\nabla^{\theta}H_{\epsilon \mu \zeta} \nn\\&\!\!\!\!\!\!\!\!\!& + \frac{31}{8} 
H_{\alpha}{}^{\delta \epsilon} H^{\alpha \beta \gamma} 
H_{\beta \delta}{}^{\varepsilon} H_{\gamma}{}^{\mu \zeta} 
\nabla_{\zeta}H_{\varepsilon \eta \theta} 
\nabla^{\theta}H_{\epsilon \mu}{}^{\eta} -  \frac{23}{8} 
H_{\alpha}{}^{\delta \epsilon} H^{\alpha \beta \gamma} 
H_{\beta \delta}{}^{\varepsilon} H_{\gamma}{}^{\mu \zeta} 
\nabla_{\eta}H_{\varepsilon \zeta \theta} 
\nabla^{\theta}H_{\epsilon \mu}{}^{\eta} \nn\\&\!\!\!\!\!\!\!\!\!& + \frac{125}{48} 
H_{\alpha \beta}{}^{\delta} H^{\alpha \beta \gamma} 
H_{\gamma}{}^{\epsilon \varepsilon} H_{\delta}{}^{\mu \zeta} 
\nabla_{\eta}H_{\varepsilon \zeta \theta} 
\nabla^{\theta}H_{\epsilon \mu}{}^{\eta} -  \frac{133}{128} 
H_{\alpha \beta \gamma} H^{\alpha \beta \gamma} H_{\delta}{}^{
\mu \zeta} H^{\delta \epsilon \varepsilon} 
\nabla_{\eta}H_{\varepsilon \zeta \theta} 
\nabla^{\theta}H_{\epsilon \mu}{}^{\eta} \nn\\&\!\!\!\!\!\!\!\!\!& + \frac{25}{8} 
H_{\alpha}{}^{\delta \epsilon} H^{\alpha \beta \gamma} 
H_{\beta \delta}{}^{\varepsilon} H_{\gamma}{}^{\mu \zeta} 
\nabla_{\theta}H_{\varepsilon \zeta \eta} 
\nabla^{\theta}H_{\epsilon \mu}{}^{\eta} -  \frac{125}{48} H_{
\alpha \beta}{}^{\delta} H^{\alpha \beta \gamma} 
H_{\gamma}{}^{\epsilon \varepsilon} H_{\delta}{}^{\mu \zeta} 
\nabla_{\theta}H_{\varepsilon \zeta \eta} 
\nabla^{\theta}H_{\epsilon \mu}{}^{\eta} \nn\\&\!\!\!\!\!\!\!\!\!& + \frac{133}{128} H_{
\alpha \beta \gamma} H^{\alpha \beta \gamma} 
H_{\delta}{}^{\mu \zeta} H^{\delta \epsilon \varepsilon} 
\nabla_{\theta}H_{\varepsilon \zeta \eta} 
\nabla^{\theta}H_{\epsilon \mu}{}^{\eta} + \frac{1}{4} 
H_{\alpha}{}^{\delta \epsilon} H^{\alpha \beta \gamma} 
H_{\beta \delta}{}^{\varepsilon} H_{\gamma \epsilon}{}^{\mu} 
\nabla_{\eta}H_{\mu \zeta \theta} 
\nabla^{\theta}H_{\varepsilon}{}^{\zeta \eta} \nn\\&\!\!\!\!\!\!\!\!\!& -  
\frac{179}{48} H_{\alpha \beta}{}^{\delta} H^{\alpha \beta 
\gamma} H_{\gamma}{}^{\epsilon \varepsilon} H_{\delta 
\epsilon}{}^{\mu} \nabla_{\eta}H_{\mu \zeta \theta} 
\nabla^{\theta}H_{\varepsilon}{}^{\zeta \eta} -  
\frac{193}{576} H_{\alpha \beta \gamma} H^{\alpha \beta 
\gamma} H_{\delta \epsilon}{}^{\mu} H^{\delta \epsilon 
\varepsilon} \nabla_{\eta}H_{\mu \zeta \theta} 
\nabla^{\theta}H_{\varepsilon}{}^{\zeta \eta} \nn\\&\!\!\!\!\!\!\!\!\!& -  \frac{3}{32} 
H_{\alpha}{}^{\delta \epsilon} H^{\alpha \beta \gamma} 
H_{\beta \delta}{}^{\varepsilon} H_{\gamma \epsilon}{}^{\mu} 
\nabla_{\theta}H_{\mu \zeta \eta} 
\nabla^{\theta}H_{\varepsilon}{}^{\zeta \eta} + \frac{11}{6} 
H_{\alpha \beta}{}^{\delta} H^{\alpha \beta \gamma} 
H_{\gamma}{}^{\epsilon \varepsilon} H_{\delta \epsilon}{}^{\mu} 
\nabla_{\theta}H_{\mu \zeta \eta} 
\nabla^{\theta}H_{\varepsilon}{}^{\zeta \eta} \nn\\&\!\!\!\!\!\!\!\!\!& + 
\frac{1181}{2304} H_{\alpha \beta \gamma} H^{\alpha \beta 
\gamma} H_{\delta \epsilon}{}^{\mu} H^{\delta \epsilon 
\varepsilon} \nabla_{\theta}H_{\mu \zeta \eta} 
\nabla^{\theta}H_{\varepsilon}{}^{\zeta \eta} -  
\frac{485}{24} H_{\alpha \beta}{}^{\delta} H^{\alpha \beta 
\gamma} H_{\gamma}{}^{\epsilon \varepsilon} H_{\epsilon}{}^{\mu 
\zeta} \nabla_{\delta}H_{\zeta \eta \theta} 
\nabla^{\theta}H_{\varepsilon \mu}{}^{\eta} \nn\\&\!\!\!\!\!\!\!\!\!& + \frac{1}{192} 
H_{\alpha \beta}{}^{\delta} H^{\alpha \beta \gamma} 
H_{\gamma}{}^{\epsilon \varepsilon} H_{\delta \epsilon 
\varepsilon} \nabla_{\theta}H_{\mu \zeta \eta} 
\nabla^{\theta}H^{\mu \zeta \eta} -  \frac{1169}{82944} 
H_{\alpha \beta \gamma} H^{\alpha \beta \gamma} H_{\delta 
\epsilon \varepsilon} H^{\delta \epsilon \varepsilon} 
\nabla_{\theta}H_{\mu \zeta \eta} \nabla^{\theta}H^{\mu 
\zeta \eta} \nn\\&\!\!\!\!\!\!\!\!\!& + \frac{25}{16} H_{\alpha}{}^{\delta \epsilon} H^{
\alpha \beta \gamma} H_{\varepsilon}{}^{\eta \theta} 
H^{\varepsilon \mu \zeta} \nabla_{\theta}H_{\epsilon \zeta 
\eta} \nabla_{\mu}H_{\beta \gamma \delta} + \frac{1169}{384} 
H_{\alpha \beta}{}^{\delta} H^{\alpha \beta \gamma} 
H^{\epsilon \varepsilon \mu} H^{\zeta \eta \theta} 
\nabla_{\theta}H_{\delta \zeta \eta} \nabla_{\mu}H_{\gamma 
\epsilon \varepsilon} \nn\\&\!\!\!\!\!\!\!\!\!& -  \frac{4}{3} H_{\alpha}{}^{\delta 
\epsilon} H^{\alpha \beta \gamma} H_{\beta 
\delta}{}^{\varepsilon} H^{\mu \zeta \eta} 
\nabla_{\eta}H_{\varepsilon \zeta \theta} 
\nabla_{\mu}H_{\gamma \epsilon}{}^{\theta} + \frac{913}{48} 
H_{\alpha}{}^{\delta \epsilon} H^{\alpha \beta \gamma} 
H_{\beta}{}^{\varepsilon \mu} H_{\delta}{}^{\zeta \eta} 
\nabla^{\theta}H_{\gamma \zeta \eta} \nabla_{\mu}H_{\epsilon 
\varepsilon \theta} \nn\\&\!\!\!\!\!\!\!\!\!& -  \frac{1}{4} H_{\alpha}{}^{\delta 
\epsilon} H^{\alpha \beta \gamma} H_{\beta}{}^{\varepsilon 
\mu} H_{\delta}{}^{\zeta \eta} \nabla^{\theta}H_{\gamma 
\varepsilon \zeta} \nabla_{\mu}H_{\epsilon \eta \theta} -  
\frac{379}{192} H_{\alpha \beta}{}^{\delta} H^{\alpha \beta 
\gamma} H_{\gamma}{}^{\epsilon \varepsilon} H_{\delta}{}^{\mu 
\zeta} \nabla_{\zeta}H_{\varepsilon \eta \theta} 
\nabla_{\mu}H_{\epsilon}{}^{\eta \theta}
 \eeqa
 They can be reproduced by studying the six-point S-matrix element. We have found 25 couplings with structure $H^6 R$, \ie 
 \beqa
{\cal L}_3^{H^6R}&=&  \frac{271}{96} H_{\alpha \beta}{}^{\delta} H^{\alpha \beta \
\gamma} H_{\gamma}{}^{\epsilon \varepsilon} H_{\epsilon}{}^{\mu \
\zeta} H_{\eta \theta}{}^{\iota} H_{\mu}{}^{\eta \theta} \
R_{\delta \zeta \varepsilon \iota} + \frac{83}{4} H_{\alpha \
\beta}{}^{\delta} H^{\alpha \beta \gamma} \
H_{\gamma}{}^{\epsilon \varepsilon} H_{\epsilon}{}^{\mu \zeta} \
H_{\varepsilon}{}^{\eta \theta} H_{\mu \eta}{}^{\iota} \
R_{\delta \zeta \theta \iota}\nn\\&& + \frac{211}{12} H_{\alpha \
\beta}{}^{\delta} H^{\alpha \beta \gamma} \
H_{\gamma}{}^{\epsilon \varepsilon} H_{\epsilon}{}^{\mu \zeta} \
H_{\zeta \eta}{}^{\iota} H_{\mu}{}^{\eta \theta} R_{\delta \
\theta \varepsilon \iota} + \frac{29}{48} H_{\alpha \beta}{}^{\
\delta} H^{\alpha \beta \gamma} H_{\gamma}{}^{\epsilon \
\varepsilon} H_{\epsilon}{}^{\mu \zeta} H_{\varepsilon \mu}{}^{\
\eta} H_{\zeta}{}^{\theta \iota} R_{\delta \theta \eta \iota} \nn\\&&
-  \frac{389}{96} H_{\alpha \beta}{}^{\delta} H^{\alpha \beta \
\gamma} H_{\gamma}{}^{\epsilon \varepsilon} H_{\epsilon}{}^{\mu \
\zeta} H_{\eta \theta}{}^{\iota} H_{\mu}{}^{\eta \theta} \
R_{\delta \iota \varepsilon \zeta} -  \frac{649}{384} \
H_{\alpha \beta}{}^{\delta} H^{\alpha \beta \gamma} \
H_{\gamma}{}^{\epsilon \varepsilon} H_{\epsilon}{}^{\mu \zeta} \
H_{\varepsilon}{}^{\eta \theta} H_{\mu \zeta}{}^{\iota} \
R_{\delta \iota \eta \theta} \nn\\&&+ \frac{1}{2} \
H_{\alpha}{}^{\delta \epsilon} H^{\alpha \beta \gamma} \
H_{\beta}{}^{\varepsilon \mu} H_{\gamma}{}^{\zeta \eta} \
H_{\delta \varepsilon}{}^{\theta} H_{\zeta \theta}{}^{\iota} \
R_{\epsilon \eta \mu \iota} + \frac{25}{8} \
H_{\alpha}{}^{\delta \epsilon} H^{\alpha \beta \gamma} \
H_{\beta \delta}{}^{\varepsilon} H_{\gamma}{}^{\mu \zeta} \
H_{\zeta \eta}{}^{\iota} H_{\mu}{}^{\eta \theta} R_{\epsilon \
\theta \varepsilon \iota}\nn\\&& -  \frac{17}{8} H_{\alpha}{}^{\delta \
\epsilon} H^{\alpha \beta \gamma} H_{\beta \
\delta}{}^{\varepsilon} H_{\gamma}{}^{\mu \zeta} \
H_{\epsilon}{}^{\eta \theta} H_{\mu \eta}{}^{\iota} \
R_{\varepsilon \zeta \theta \iota} -  \frac{1}{16} H_{\alpha \
\beta}{}^{\delta} H^{\alpha \beta \gamma} \
H_{\gamma}{}^{\epsilon \varepsilon} H_{\delta}{}^{\mu \zeta} \
H_{\epsilon \mu}{}^{\eta} H_{\eta}{}^{\theta \iota} \
R_{\varepsilon \theta \zeta \iota}\nn\\&& + \frac{79}{12} H_{\alpha \
\beta}{}^{\delta} H^{\alpha \beta \gamma} \
H_{\gamma}{}^{\epsilon \varepsilon} H_{\delta}{}^{\mu \zeta} \
H_{\epsilon}{}^{\eta \theta} H_{\mu \eta}{}^{\iota} \
R_{\varepsilon \theta \zeta \iota} -  \frac{1}{4} \
H_{\alpha}{}^{\delta \epsilon} H^{\alpha \beta \gamma} \
H_{\beta \delta}{}^{\varepsilon} H_{\gamma}{}^{\mu \zeta} \
H_{\epsilon \mu}{}^{\eta} H_{\zeta}{}^{\theta \iota} \
R_{\varepsilon \theta \eta \iota}\nn\\&& -  \frac{715}{192} H_{\alpha \
\beta}{}^{\delta} H^{\alpha \beta \gamma} \
H_{\gamma}{}^{\epsilon \varepsilon} H_{\delta \epsilon}{}^{\mu} \
H_{\zeta \eta}{}^{\iota} H^{\zeta \eta \theta} R_{\varepsilon \
\theta \mu \iota} -  \frac{41}{6} H_{\alpha \beta}{}^{\delta} \
H^{\alpha \beta \gamma} H_{\gamma}{}^{\epsilon \varepsilon} H_{\
\delta}{}^{\mu \zeta} H_{\epsilon}{}^{\eta \theta} H_{\mu \
\eta}{}^{\iota} R_{\varepsilon \iota \zeta \theta}\nn\\&& -  
\frac{799}{192} H_{\alpha \beta}{}^{\delta} H^{\alpha \beta \
\gamma} H_{\gamma}{}^{\epsilon \varepsilon} H_{\delta}{}^{\mu \
\zeta} H_{\epsilon}{}^{\eta \theta} H_{\eta \theta}{}^{\iota} \
R_{\varepsilon \iota \mu \zeta} -  \frac{5}{4} \
H_{\alpha}{}^{\delta \epsilon} H^{\alpha \beta \gamma} \
H_{\beta \delta}{}^{\varepsilon} H_{\gamma}{}^{\mu \zeta} \
H_{\epsilon \mu}{}^{\eta} H_{\varepsilon}{}^{\theta \iota} \
R_{\zeta \theta \eta \iota} \nn\\&&-  \frac{1}{4} H_{\alpha \
\beta}{}^{\delta} H^{\alpha \beta \gamma} \
H_{\gamma}{}^{\epsilon \varepsilon} H_{\delta}{}^{\mu \zeta} \
H_{\epsilon \mu}{}^{\eta} H_{\varepsilon}{}^{\theta \iota} \
R_{\zeta \theta \eta \iota} + \frac{5}{16} H_{\alpha \
\beta}{}^{\delta} H^{\alpha \beta \gamma} \
H_{\gamma}{}^{\epsilon \varepsilon} H_{\delta}{}^{\mu \zeta} \
H_{\epsilon \varepsilon}{}^{\eta} H_{\mu}{}^{\theta \iota} \
R_{\zeta \theta \eta \iota}\nn\\&& -  \frac{1}{16} \
H_{\alpha}{}^{\delta \epsilon} H^{\alpha \beta \gamma} \
H_{\beta \delta}{}^{\varepsilon} H_{\gamma \epsilon}{}^{\mu} \
H_{\varepsilon}{}^{\zeta \eta} H_{\mu}{}^{\theta \iota} \
R_{\zeta \theta \eta \iota} -  \frac{5}{24} H_{\alpha \
\beta}{}^{\delta} H^{\alpha \beta \gamma} \
H_{\gamma}{}^{\epsilon \varepsilon} H_{\delta \epsilon \
\varepsilon} H_{\mu}{}^{\theta \iota} H^{\mu \zeta \eta} \
R_{\zeta \theta \eta \iota} \nn\\&&-  \frac{293}{32} H_{\alpha \
\beta}{}^{\delta} H^{\alpha \beta \gamma} \
H_{\gamma}{}^{\epsilon \varepsilon} H_{\delta}{}^{\mu \zeta} \
H_{\epsilon}{}^{\eta \theta} H_{\varepsilon \eta}{}^{\iota} R_{\
\mu \theta \zeta \iota} + \frac{1}{12} H_{\alpha \
\beta}{}^{\delta} H^{\alpha \beta \gamma} \
H_{\gamma}{}^{\epsilon \varepsilon} H_{\delta}{}^{\mu \zeta} \
H_{\epsilon \varepsilon}{}^{\eta} H_{\eta}{}^{\theta \iota} R_{\
\mu \theta \zeta \iota} \nn\\&&-  \frac{1}{8} H_{\alpha}{}^{\delta \
\epsilon} H^{\alpha \beta \gamma} H_{\beta}{}^{\varepsilon \
\mu} H_{\gamma}{}^{\zeta \eta} H_{\delta \
\varepsilon}{}^{\theta} H_{\epsilon \zeta}{}^{\iota} R_{\mu \
\theta \eta \iota} + \frac{329}{48} H_{\alpha \
\beta}{}^{\delta} H^{\alpha \beta \gamma} \
H_{\gamma}{}^{\epsilon \varepsilon} H_{\delta \epsilon}{}^{\mu} \
H_{\varepsilon}{}^{\zeta \eta} H_{\zeta}{}^{\theta \iota} \
R_{\mu \theta \eta \iota}\nn\\&& + \frac{5}{8} H_{\alpha}{}^{\delta \
\epsilon} H^{\alpha \beta \gamma} H_{\beta}{}^{\varepsilon \
\mu} H_{\gamma}{}^{\zeta \eta} H_{\delta \
\varepsilon}{}^{\theta} H_{\epsilon \zeta}{}^{\iota} R_{\mu \
\iota \eta \theta}
 \eeqa
 They can be reproduced by studying the seven-point S-matrix element. We have found 91 couplings with structure $H^2(\nabla H)^2R$, \ie 
 \beqa
{\cal L}_3^{H^2(\prt H)^2R}&=& 14 H^{\alpha \beta \gamma} H^{\delta \epsilon \varepsilon} R_{\
\beta \varepsilon \gamma \eta} \
\nabla_{\delta}H_{\alpha}{}^{\mu \zeta} \
\nabla_{\zeta}H_{\epsilon \mu}{}^{\eta} + \frac{9}{2} \
H^{\alpha \beta \gamma} H^{\delta \epsilon \varepsilon} \
R_{\epsilon \mu \varepsilon \eta} \nabla_{\delta}H_{\gamma \
\zeta}{}^{\eta} \nabla^{\zeta}H_{\alpha \beta}{}^{\mu}\nn\\&& -  \
\frac{59}{24} H^{\alpha \beta \gamma} H^{\delta \epsilon \
\varepsilon} R_{\epsilon \mu \varepsilon \eta} \
\nabla_{\zeta}H_{\gamma \delta}{}^{\eta} \
\nabla^{\zeta}H_{\alpha \beta}{}^{\mu} -  \frac{29}{16} \
H^{\alpha \beta \gamma} H^{\delta \epsilon \varepsilon} \
R_{\gamma \eta \varepsilon \mu} \nabla_{\zeta}H_{\delta \
\epsilon}{}^{\eta} \nabla^{\zeta}H_{\alpha \beta}{}^{\mu}\nn\\&& + \
\frac{13}{16} H^{\alpha \beta \gamma} H^{\delta \epsilon \
\varepsilon} R_{\gamma \mu \varepsilon \eta} \nabla_{\zeta}H_{\
\delta \epsilon}{}^{\eta} \nabla^{\zeta}H_{\alpha \
\beta}{}^{\mu} + \frac{43}{8} H^{\alpha \beta \gamma} \
H^{\delta \epsilon \varepsilon} R_{\gamma \eta \epsilon \
\varepsilon} \nabla_{\zeta}H_{\delta \mu}{}^{\eta} \
\nabla^{\zeta}H_{\alpha \beta}{}^{\mu} \nn\\&&-  \frac{251}{12} \
H^{\alpha \beta \gamma} H^{\delta \epsilon \varepsilon} \
R_{\epsilon \mu \varepsilon \eta} \nabla_{\gamma}H_{\beta \
\zeta}{}^{\eta} \nabla^{\zeta}H_{\alpha \delta}{}^{\mu} -  H^{\
\alpha \beta \gamma} H^{\delta \epsilon \varepsilon} R_{\gamma \
\eta \varepsilon \mu} \nabla_{\epsilon}H_{\beta \
\zeta}{}^{\eta} \nabla^{\zeta}H_{\alpha \delta}{}^{\mu} \nn\\&&-  \
\frac{1}{4} H^{\alpha \beta \gamma} H^{\delta \epsilon \
\varepsilon} R_{\gamma \eta \varepsilon \zeta} \
\nabla_{\epsilon}H_{\beta \mu}{}^{\eta} \
\nabla^{\zeta}H_{\alpha \delta}{}^{\mu} -  \frac{29}{4} \
H^{\alpha \beta \gamma} H^{\delta \epsilon \varepsilon} \
R_{\gamma \mu \varepsilon \eta} \nabla_{\zeta}H_{\beta \
\epsilon}{}^{\eta} \nabla^{\zeta}H_{\alpha \delta}{}^{\mu}\nn\\&& -  \
\frac{17}{2} H^{\alpha \beta \gamma} H^{\delta \epsilon \
\varepsilon} R_{\gamma \eta \epsilon \varepsilon} \
\nabla_{\zeta}H_{\beta \mu}{}^{\eta} \nabla^{\zeta}H_{\alpha \
\delta}{}^{\mu} + \frac{1}{2} H_{\alpha}{}^{\delta \epsilon} \
H^{\alpha \beta \gamma} R_{\delta \mu \epsilon \eta} \nabla_{\
\varepsilon}H_{\gamma \zeta}{}^{\eta} \
\nabla^{\zeta}H_{\beta}{}^{\varepsilon \mu}\nn\\&& + \frac{5}{2} \
H_{\alpha}{}^{\delta \epsilon} H^{\alpha \beta \gamma} \
R_{\gamma \eta \epsilon \mu} \nabla_{\varepsilon}H_{\delta \
\zeta}{}^{\eta} \nabla^{\zeta}H_{\beta}{}^{\varepsilon \mu} + \
\frac{1}{2} H_{\alpha}{}^{\delta \epsilon} H^{\alpha \beta \
\gamma} R_{\gamma \mu \epsilon \eta} \
\nabla_{\varepsilon}H_{\delta \zeta}{}^{\eta} \
\nabla^{\zeta}H_{\beta}{}^{\varepsilon \mu} \nn\\&&-  \
H_{\alpha}{}^{\delta \epsilon} H^{\alpha \beta \gamma} \
R_{\delta \mu \epsilon \eta} \nabla_{\zeta}H_{\gamma \
\varepsilon}{}^{\eta} \nabla^{\zeta}H_{\beta}{}^{\varepsilon \
\mu} -  \frac{19}{2} H_{\alpha}{}^{\delta \epsilon} H^{\alpha \
\beta \gamma} R_{\gamma \eta \epsilon \mu} \
\nabla_{\zeta}H_{\delta \varepsilon}{}^{\eta} \
\nabla^{\zeta}H_{\beta}{}^{\varepsilon \mu} \nn\\&&+ \frac{31}{2} H_{\
\alpha}{}^{\delta \epsilon} H^{\alpha \beta \gamma} R_{\gamma \
\mu \epsilon \eta} \nabla_{\zeta}H_{\delta \
\varepsilon}{}^{\eta} \nabla^{\zeta}H_{\beta}{}^{\varepsilon \
\mu} + 3 H_{\alpha}{}^{\delta \epsilon} H^{\alpha \beta \
\gamma} R_{\gamma \eta \delta \epsilon} \
\nabla_{\zeta}H_{\varepsilon \mu}{}^{\eta} \
\nabla^{\zeta}H_{\beta}{}^{\varepsilon \mu}\nn\\&& + \frac{17}{24} \
H_{\alpha \beta}{}^{\delta} H^{\alpha \beta \gamma} R_{\gamma \
\mu \delta \eta} \nabla_{\zeta}H_{\epsilon \
\varepsilon}{}^{\eta} \nabla^{\zeta}H^{\epsilon \varepsilon \
\mu} - 7 H^{\alpha \beta \gamma} H^{\delta \epsilon \
\varepsilon} R_{\beta \epsilon \gamma \varepsilon} \
\nabla_{\zeta}H_{\delta \mu \eta} \nabla^{\eta}H_{\alpha}{}^{\
\mu \zeta}\nn\\&& + \frac{29}{8} H^{\alpha \beta \gamma} H^{\delta \
\epsilon \varepsilon} R_{\beta \epsilon \gamma \varepsilon} \
\nabla_{\eta}H_{\delta \mu \zeta} \nabla^{\eta}H_{\alpha}{}^{\
\mu \zeta} + \frac{27}{4} H^{\alpha \beta \gamma} H^{\delta \
\epsilon \varepsilon} R_{\gamma \mu \varepsilon \eta} \nabla^{\
\zeta}H_{\alpha \delta}{}^{\mu} \nabla^{\eta}H_{\beta \
\epsilon \zeta}\nn\\&& -  \frac{1}{4} H^{\alpha \beta \gamma} \
H^{\delta \epsilon \varepsilon} R_{\gamma \zeta \varepsilon \
\eta} \nabla^{\zeta}H_{\alpha \delta}{}^{\mu} \
\nabla^{\eta}H_{\beta \epsilon \mu} + \frac{265}{24} \
H^{\alpha \beta \gamma} H^{\delta \epsilon \varepsilon} \
R_{\gamma \eta \epsilon \varepsilon} \nabla^{\zeta}H_{\alpha \
\delta}{}^{\mu} \nabla^{\eta}H_{\beta \mu \zeta}\nn\\&& + \
\frac{59}{12} H^{\alpha \beta \gamma} H^{\delta \epsilon \
\varepsilon} R_{\epsilon \mu \varepsilon \eta} \
\nabla^{\zeta}H_{\alpha \beta}{}^{\mu} \nabla^{\eta}H_{\gamma \
\delta \zeta} + \frac{15}{2} H^{\alpha \beta \gamma} \
H^{\delta \epsilon \varepsilon} R_{\varepsilon \zeta \mu \eta} \
\nabla_{\delta}H_{\alpha \beta}{}^{\mu} \
\nabla^{\eta}H_{\gamma \epsilon}{}^{\zeta}\nn\\&& -  \frac{131}{12} \
H^{\alpha \beta \gamma} H^{\delta \epsilon \varepsilon} \
R_{\varepsilon \eta \mu \zeta} \nabla_{\delta}H_{\alpha \
\beta}{}^{\mu} \nabla^{\eta}H_{\gamma \epsilon}{}^{\zeta} -  \
\frac{53}{4} H_{\alpha}{}^{\delta \epsilon} H^{\alpha \beta \
\gamma} R_{\delta \mu \epsilon \eta} \
\nabla^{\zeta}H_{\beta}{}^{\varepsilon \mu} \
\nabla^{\eta}H_{\gamma \varepsilon \zeta} \nn\\&&+ \frac{25}{3} \
H_{\alpha}{}^{\delta \epsilon} H^{\alpha \beta \gamma} \
R_{\delta \zeta \epsilon \eta} \
\nabla^{\zeta}H_{\beta}{}^{\varepsilon \mu} \
\nabla^{\eta}H_{\gamma \varepsilon \mu} -  \frac{61}{24} \
H^{\alpha \beta \gamma} H^{\delta \epsilon \varepsilon} \
R_{\epsilon \zeta \varepsilon \eta} \nabla_{\delta}H_{\alpha \
\beta}{}^{\mu} \nabla^{\eta}H_{\gamma \mu}{}^{\zeta}\nn\\&& + \
\frac{23}{12} H^{\alpha \beta \gamma} H^{\delta \epsilon \
\varepsilon} R_{\gamma \eta \varepsilon \mu} \nabla^{\zeta}H_{\
\alpha \beta}{}^{\mu} \nabla^{\eta}H_{\delta \epsilon \zeta} \
-  \frac{7}{8} H^{\alpha \beta \gamma} H^{\delta \epsilon \
\varepsilon} R_{\gamma \mu \varepsilon \eta} \nabla^{\zeta}H_{\
\alpha \beta}{}^{\mu} \nabla^{\eta}H_{\delta \epsilon \zeta} \
\nn\\&&+ \frac{17}{48} H^{\alpha \beta \gamma} H^{\delta \epsilon \
\varepsilon} R_{\gamma \zeta \varepsilon \eta} \
\nabla^{\zeta}H_{\alpha \beta}{}^{\mu} \nabla^{\eta}H_{\delta \
\epsilon \mu} + \frac{3}{16} H^{\alpha \beta \gamma} \
H^{\delta \epsilon \varepsilon} R_{\gamma \eta \varepsilon \
\zeta} \nabla^{\zeta}H_{\alpha \beta}{}^{\mu} \
\nabla^{\eta}H_{\delta \epsilon \mu}\nn\\&& + H_{\alpha}{}^{\delta \
\epsilon} H^{\alpha \beta \gamma} R_{\gamma \eta \epsilon \
\mu} \nabla^{\zeta}H_{\beta}{}^{\varepsilon \mu} \
\nabla^{\eta}H_{\delta \varepsilon \zeta} -  \frac{63}{4} \
H_{\alpha}{}^{\delta \epsilon} H^{\alpha \beta \gamma} \
R_{\gamma \mu \epsilon \eta} \
\nabla^{\zeta}H_{\beta}{}^{\varepsilon \mu} \
\nabla^{\eta}H_{\delta \varepsilon \zeta}\nn\\&& + \frac{179}{16} H_{\
\alpha}{}^{\delta \epsilon} H^{\alpha \beta \gamma} R_{\gamma \
\zeta \epsilon \eta} \nabla^{\zeta}H_{\beta}{}^{\varepsilon \
\mu} \nabla^{\eta}H_{\delta \varepsilon \mu} -  \
\frac{101}{16} H_{\alpha}{}^{\delta \epsilon} H^{\alpha \beta \
\gamma} R_{\gamma \eta \epsilon \zeta} \
\nabla^{\zeta}H_{\beta}{}^{\varepsilon \mu} \
\nabla^{\eta}H_{\delta \varepsilon \mu}\nn\\&& + \frac{89}{48} \
H^{\alpha \beta \gamma} H^{\delta \epsilon \varepsilon} \
R_{\gamma \eta \epsilon \varepsilon} \nabla^{\zeta}H_{\alpha \
\beta}{}^{\mu} \nabla^{\eta}H_{\delta \mu \zeta} -  \
\frac{5}{6} H_{\alpha \beta}{}^{\delta} H^{\alpha \beta \
\gamma} R_{\gamma \mu \delta \eta} \nabla^{\zeta}H^{\epsilon \
\varepsilon \mu} \nabla^{\eta}H_{\epsilon \varepsilon \zeta} \nn\\&&- \
 \frac{37}{24} H^{\alpha \beta \gamma} H^{\delta \epsilon \
\varepsilon} R_{\gamma \zeta \mu \eta} \
\nabla_{\delta}H_{\alpha \beta}{}^{\mu} \
\nabla^{\eta}H_{\epsilon \varepsilon}{}^{\zeta} -  \
\frac{13}{48} H^{\alpha \beta \gamma} H^{\delta \epsilon \
\varepsilon} R_{\gamma \eta \mu \zeta} \
\nabla_{\delta}H_{\alpha \beta}{}^{\mu} \
\nabla^{\eta}H_{\epsilon \varepsilon}{}^{\zeta}\nn\\&& -  \frac{9}{2} \
H_{\alpha}{}^{\delta \epsilon} H^{\alpha \beta \gamma} \
R_{\gamma \zeta \mu \eta} \
\nabla_{\delta}H_{\beta}{}^{\varepsilon \mu} \
\nabla^{\eta}H_{\epsilon \varepsilon}{}^{\zeta} + \frac{25}{4} \
H_{\alpha}{}^{\delta \epsilon} H^{\alpha \beta \gamma} \
R_{\gamma \eta \mu \zeta} \
\nabla_{\delta}H_{\beta}{}^{\varepsilon \mu} \
\nabla^{\eta}H_{\epsilon \varepsilon}{}^{\zeta}\nn\\&& -  \
\frac{565}{24} H^{\alpha \beta \gamma} H^{\delta \epsilon \
\varepsilon} R_{\beta \varepsilon \gamma \eta} \
\nabla_{\delta}H_{\alpha}{}^{\mu \zeta} \
\nabla^{\eta}H_{\epsilon \mu \zeta} -  \frac{1}{12} H^{\alpha \
\beta \gamma} H^{\delta \epsilon \varepsilon} R_{\gamma \zeta \
\varepsilon \eta} \nabla_{\delta}H_{\alpha \beta}{}^{\mu} \
\nabla^{\eta}H_{\epsilon \mu}{}^{\zeta} \nn\\&&-  \frac{7}{8} \
H^{\alpha \beta \gamma} H^{\delta \epsilon \varepsilon} \
R_{\gamma \eta \varepsilon \zeta} \nabla_{\delta}H_{\alpha \
\beta}{}^{\mu} \nabla^{\eta}H_{\epsilon \mu}{}^{\zeta} + \
\frac{251}{288} H^{\alpha \beta \gamma} H^{\delta \epsilon \
\varepsilon} R_{\varepsilon \eta \mu \zeta} \nabla_{\delta}H_{\
\alpha \beta \gamma} \nabla^{\eta}H_{\epsilon}{}^{\mu \zeta} \nn\\&&
-  \frac{97}{96} H_{\alpha}{}^{\delta \epsilon} H^{\alpha \
\beta \gamma} R_{\varepsilon \eta \mu \zeta} \
\nabla_{\delta}H_{\beta \gamma}{}^{\varepsilon} \
\nabla^{\eta}H_{\epsilon}{}^{\mu \zeta} + \frac{331}{96} \
H_{\alpha \beta}{}^{\delta} H^{\alpha \beta \gamma} \
R_{\varepsilon \eta \mu \zeta} \
\nabla_{\delta}H_{\gamma}{}^{\epsilon \varepsilon} \
\nabla^{\eta}H_{\epsilon}{}^{\mu \zeta} \nn\\&&-  \frac{103}{16} \
H_{\alpha}{}^{\delta \epsilon} H^{\alpha \beta \gamma} \
R_{\gamma \eta \delta \epsilon} \
\nabla^{\zeta}H_{\beta}{}^{\varepsilon \mu} \
\nabla^{\eta}H_{\varepsilon \mu \zeta} -  \frac{51}{8} \
H_{\alpha}{}^{\delta \epsilon} H^{\alpha \beta \gamma} \
R_{\gamma \eta \epsilon \zeta} \
\nabla_{\delta}H_{\beta}{}^{\varepsilon \mu} \
\nabla^{\eta}H_{\varepsilon \mu}{}^{\zeta}\nn\\&& + \frac{89}{32} H_{\
\alpha}{}^{\delta \epsilon} H^{\alpha \beta \gamma} \
R_{\epsilon \eta \mu \zeta} \nabla_{\delta}H_{\beta \
\gamma}{}^{\varepsilon} \nabla^{\eta}H_{\varepsilon}{}^{\mu \
\zeta} + \frac{389}{48} H^{\alpha \beta \gamma} H^{\delta \
\epsilon \varepsilon} R_{\gamma \eta \mu \zeta} \
\nabla_{\epsilon}H_{\alpha \beta \delta} \
\nabla^{\eta}H_{\varepsilon}{}^{\mu \zeta}\nn\\&& -  \frac{23}{72} \
H_{\alpha}{}^{\delta \epsilon} H^{\alpha \beta \gamma} \
R_{\beta \delta \gamma \epsilon} \nabla_{\eta}H_{\varepsilon \
\mu \zeta} \nabla^{\eta}H^{\varepsilon \mu \zeta} + 8 \
H^{\alpha \beta \gamma} H^{\delta \epsilon \varepsilon} \
R_{\gamma \eta \epsilon \varepsilon} \nabla^{\zeta}H_{\alpha \
\delta}{}^{\mu} \nabla_{\mu}H_{\beta \zeta}{}^{\eta}\nn\\&& -  \
\frac{13}{2} H^{\alpha \beta \gamma} H^{\delta \epsilon \
\varepsilon} R_{\gamma \eta \epsilon \varepsilon} \
\nabla^{\zeta}H_{\alpha \beta}{}^{\mu} \nabla_{\mu}H_{\delta \
\zeta}{}^{\eta} + \frac{5}{9} H^{\alpha \beta \gamma} \
H^{\delta \epsilon \varepsilon} R_{\epsilon \zeta \varepsilon \
\eta} \nabla^{\eta}H_{\delta \mu}{}^{\zeta} \
\nabla^{\mu}H_{\alpha \beta \gamma}\nn\\&& + \frac{601}{48} \
H^{\alpha \beta \gamma} H^{\delta \epsilon \varepsilon} \
R_{\varepsilon \eta \mu \zeta} \nabla^{\eta}H_{\gamma \
\epsilon}{}^{\zeta} \nabla^{\mu}H_{\alpha \beta \delta} + \
\frac{469}{48} H^{\alpha \beta \gamma} H^{\delta \epsilon \
\varepsilon} R_{\epsilon \zeta \varepsilon \eta} \
\nabla^{\eta}H_{\gamma \mu}{}^{\zeta} \nabla^{\mu}H_{\alpha \
\beta \delta}\nn\\&& + \frac{15}{2} H^{\alpha \beta \gamma} \
H^{\delta \epsilon \varepsilon} R_{\gamma \zeta \mu \eta} \
\nabla^{\eta}H_{\epsilon \varepsilon}{}^{\zeta} \
\nabla^{\mu}H_{\alpha \beta \delta} + \frac{341}{96} \
H^{\alpha \beta \gamma} H^{\delta \epsilon \varepsilon} \
R_{\gamma \eta \mu \zeta} \nabla^{\eta}H_{\epsilon \
\varepsilon}{}^{\zeta} \nabla^{\mu}H_{\alpha \beta \delta}\nn\\&& -  \
\frac{379}{48} H^{\alpha \beta \gamma} H^{\delta \epsilon \
\varepsilon} R_{\gamma \zeta \varepsilon \eta} \
\nabla^{\eta}H_{\epsilon \mu}{}^{\zeta} \nabla^{\mu}H_{\alpha \
\beta \delta} + \frac{95}{16} H^{\alpha \beta \gamma} \
H^{\delta \epsilon \varepsilon} R_{\gamma \eta \varepsilon \
\zeta} \nabla^{\eta}H_{\epsilon \mu}{}^{\zeta} \
\nabla^{\mu}H_{\alpha \beta \delta}\nn\\&& + \frac{415}{48} \
H_{\alpha}{}^{\delta \epsilon} H^{\alpha \beta \gamma} \
R_{\varepsilon \zeta \mu \eta} \nabla^{\eta}H_{\delta \
\epsilon}{}^{\zeta} \nabla^{\mu}H_{\beta \
\gamma}{}^{\varepsilon} -  \frac{17}{32} H_{\alpha}{}^{\delta \
\epsilon} H^{\alpha \beta \gamma} R_{\varepsilon \eta \mu \
\zeta} \nabla^{\eta}H_{\delta \epsilon}{}^{\zeta} \
\nabla^{\mu}H_{\beta \gamma}{}^{\varepsilon}\nn\\&& + 3 H_{\alpha}{}^{\
\delta \epsilon} H^{\alpha \beta \gamma} R_{\epsilon \zeta \
\mu \eta} \nabla^{\eta}H_{\delta \varepsilon}{}^{\zeta} \
\nabla^{\mu}H_{\beta \gamma}{}^{\varepsilon} + \frac{209}{48} \
H_{\alpha}{}^{\delta \epsilon} H^{\alpha \beta \gamma} \
R_{\epsilon \eta \mu \zeta} \nabla^{\eta}H_{\delta \
\varepsilon}{}^{\zeta} \nabla^{\mu}H_{\beta \
\gamma}{}^{\varepsilon}\nn\\&& + \frac{155}{16} H_{\alpha}{}^{\delta \
\epsilon} H^{\alpha \beta \gamma} R_{\epsilon \zeta \
\varepsilon \eta} \nabla^{\eta}H_{\delta \mu}{}^{\zeta} \
\nabla^{\mu}H_{\beta \gamma}{}^{\varepsilon} -  \frac{499}{48} \
H_{\alpha}{}^{\delta \epsilon} H^{\alpha \beta \gamma} \
R_{\epsilon \eta \varepsilon \zeta} \nabla^{\eta}H_{\delta \
\mu}{}^{\zeta} \nabla^{\mu}H_{\beta \gamma}{}^{\varepsilon}\nn\\&& -  \
\frac{245}{48} H_{\alpha}{}^{\delta \epsilon} H^{\alpha \beta \
\gamma} R_{\delta \zeta \epsilon \eta} \
\nabla^{\eta}H_{\varepsilon \mu}{}^{\zeta} \
\nabla^{\mu}H_{\beta \gamma}{}^{\varepsilon} + \frac{7}{2} H_{\
\alpha}{}^{\delta \epsilon} H^{\alpha \beta \gamma} R_{\delta \
\zeta \epsilon \eta} \nabla_{\mu}H_{\varepsilon}{}^{\zeta \
\eta} \nabla^{\mu}H_{\beta \gamma}{}^{\varepsilon}\nn\\&& -  \
\frac{1}{4} H_{\alpha}{}^{\delta \epsilon} H^{\alpha \beta \
\gamma} R_{\epsilon \zeta \mu \eta} \
\nabla_{\varepsilon}H_{\gamma}{}^{\zeta \eta} \
\nabla^{\mu}H_{\beta \delta}{}^{\varepsilon} + \frac{1}{4} H_{\
\alpha}{}^{\delta \epsilon} H^{\alpha \beta \gamma} \
R_{\varepsilon \zeta \mu \eta} \nabla^{\eta}H_{\gamma \
\epsilon}{}^{\zeta} \nabla^{\mu}H_{\beta \
\delta}{}^{\varepsilon}\nn\\&& -  \frac{23}{24} H_{\alpha}{}^{\delta \
\epsilon} H^{\alpha \beta \gamma} R_{\varepsilon \eta \mu \
\zeta} \nabla^{\eta}H_{\gamma \epsilon}{}^{\zeta} \
\nabla^{\mu}H_{\beta \delta}{}^{\varepsilon} -  \frac{15}{4} \
H_{\alpha}{}^{\delta \epsilon} H^{\alpha \beta \gamma} \
R_{\epsilon \zeta \mu \eta} \nabla^{\eta}H_{\gamma \
\varepsilon}{}^{\zeta} \nabla^{\mu}H_{\beta \
\delta}{}^{\varepsilon}\nn\\&& + \frac{25}{6} H_{\alpha}{}^{\delta \
\epsilon} H^{\alpha \beta \gamma} R_{\epsilon \eta \mu \zeta} \
\nabla^{\eta}H_{\gamma \varepsilon}{}^{\zeta} \
\nabla^{\mu}H_{\beta \delta}{}^{\varepsilon} + \frac{13}{12} \
H_{\alpha}{}^{\delta \epsilon} H^{\alpha \beta \gamma} \
R_{\epsilon \zeta \varepsilon \eta} \nabla^{\eta}H_{\gamma \
\mu}{}^{\zeta} \nabla^{\mu}H_{\beta \delta}{}^{\varepsilon}\nn\\&& + \
\frac{11}{12} H_{\alpha}{}^{\delta \epsilon} H^{\alpha \beta \
\gamma} R_{\epsilon \eta \varepsilon \zeta} \
\nabla^{\eta}H_{\gamma \mu}{}^{\zeta} \nabla^{\mu}H_{\beta \
\delta}{}^{\varepsilon} + \frac{13}{12} H_{\alpha}{}^{\delta \
\epsilon} H^{\alpha \beta \gamma} R_{\gamma \zeta \epsilon \
\eta} \nabla^{\eta}H_{\varepsilon \mu}{}^{\zeta} \
\nabla^{\mu}H_{\beta \delta}{}^{\varepsilon}\nn\\&& -  \frac{241}{48} \
H_{\alpha \beta}{}^{\delta} H^{\alpha \beta \gamma} \
R_{\varepsilon \zeta \mu \eta} \nabla^{\eta}H_{\delta \
\epsilon}{}^{\zeta} \nabla^{\mu}H_{\gamma}{}^{\epsilon \
\varepsilon} + \frac{331}{48} H_{\alpha \beta}{}^{\delta} \
H^{\alpha \beta \gamma} R_{\varepsilon \eta \mu \zeta} \
\nabla^{\eta}H_{\delta \epsilon}{}^{\zeta} \
\nabla^{\mu}H_{\gamma}{}^{\epsilon \varepsilon}\nn\\&& -  \
\frac{11}{48} H_{\alpha \beta}{}^{\delta} H^{\alpha \beta \
\gamma} R_{\epsilon \zeta \varepsilon \eta} \
\nabla^{\eta}H_{\delta \mu}{}^{\zeta} \
\nabla^{\mu}H_{\gamma}{}^{\epsilon \varepsilon} + \
\frac{181}{24} H_{\alpha \beta}{}^{\delta} H^{\alpha \beta \
\gamma} R_{\delta \zeta \mu \eta} \nabla^{\eta}H_{\epsilon \
\varepsilon}{}^{\zeta} \nabla^{\mu}H_{\gamma}{}^{\epsilon \
\varepsilon}\nn\\&& + \frac{83}{96} H_{\alpha \beta}{}^{\delta} \
H^{\alpha \beta \gamma} R_{\delta \eta \mu \zeta} \
\nabla^{\eta}H_{\epsilon \varepsilon}{}^{\zeta} \
\nabla^{\mu}H_{\gamma}{}^{\epsilon \varepsilon} -  \
\frac{499}{48} H_{\alpha \beta}{}^{\delta} H^{\alpha \beta \
\gamma} R_{\delta \zeta \varepsilon \eta} \
\nabla^{\eta}H_{\epsilon \mu}{}^{\zeta} \
\nabla^{\mu}H_{\gamma}{}^{\epsilon \varepsilon}\nn\\&& + \
\frac{137}{16} H_{\alpha \beta}{}^{\delta} H^{\alpha \beta \
\gamma} R_{\delta \eta \varepsilon \zeta} \
\nabla^{\eta}H_{\epsilon \mu}{}^{\zeta} \
\nabla^{\mu}H_{\gamma}{}^{\epsilon \varepsilon} + \
\frac{29}{24} H_{\alpha \beta}{}^{\delta} H^{\alpha \beta \
\gamma} R_{\epsilon \zeta \varepsilon \eta} \
\nabla_{\mu}H_{\delta}{}^{\zeta \eta} \
\nabla^{\mu}H_{\gamma}{}^{\epsilon \varepsilon}\nn\\&& -  \frac{2}{3} \
H_{\alpha \beta}{}^{\delta} H^{\alpha \beta \gamma} R_{\delta \
\zeta \varepsilon \eta} \nabla_{\mu}H_{\epsilon}{}^{\zeta \
\eta} \nabla^{\mu}H_{\gamma}{}^{\epsilon \varepsilon} -  \
\frac{7}{144} H_{\alpha \beta \gamma} H^{\alpha \beta \gamma} \
R_{\varepsilon \eta \mu \zeta} \nabla^{\eta}H_{\delta \
\epsilon}{}^{\zeta} \nabla^{\mu}H^{\delta \epsilon \
\varepsilon} \nn\\&&-  \frac{3}{8} H_{\alpha \beta \gamma} H^{\alpha \
\beta \gamma} R_{\epsilon \zeta \varepsilon \eta} \
\nabla^{\eta}H_{\delta \mu}{}^{\zeta} \nabla^{\mu}H^{\delta \
\epsilon \varepsilon}
 \eeqa
 They can be reproduced by studying the five-point S-matrix element.  We have found 17 couplings with structure $H^4R^2$, \ie 
 \beqa
{\cal L}_3^{H^4R^2}&=& \frac{7}{2} H_{\alpha}{}^{\delta \epsilon} H^{\alpha \beta \
\gamma} H_{\beta}{}^{\varepsilon \mu} H_{\delta}{}^{\zeta \
\eta} R_{\gamma}{}^{\theta}{}_{\varepsilon \zeta} R_{\epsilon \
\theta \mu \eta} -  \frac{11}{4} H_{\alpha}{}^{\delta \
\epsilon} H^{\alpha \beta \gamma} H_{\beta}{}^{\varepsilon \
\mu} H^{\zeta \eta \theta} R_{\gamma \zeta \delta \epsilon} \
R_{\varepsilon \eta \mu \theta}\nn\\&& + \frac{20}{3} H_{\alpha \
\beta}{}^{\delta} H^{\alpha \beta \gamma} \
H_{\gamma}{}^{\epsilon \varepsilon} H_{\epsilon}{}^{\mu \zeta} \
R_{\delta}{}^{\eta}{}_{\mu}{}^{\theta} R_{\varepsilon \theta \
\zeta \eta} + \frac{95}{48} H_{\alpha \beta}{}^{\delta} \
H^{\alpha \beta \gamma} H_{\gamma}{}^{\epsilon \varepsilon} H^{\
\mu \zeta \eta} R_{\delta}{}^{\theta}{}_{\epsilon \mu} \
R_{\varepsilon \theta \zeta \eta} \nn\\&&+ \frac{29}{12} \
H_{\alpha}{}^{\delta \epsilon} H^{\alpha \beta \gamma} \
H_{\beta \delta}{}^{\varepsilon} H_{\gamma}{}^{\mu \zeta} \
R_{\epsilon}{}^{\eta}{}_{\mu}{}^{\theta} R_{\varepsilon \theta \
\zeta \eta} -  \frac{53}{24} H_{\alpha \beta}{}^{\delta} \
H^{\alpha \beta \gamma} H_{\gamma}{}^{\epsilon \varepsilon} H_{\
\delta}{}^{\mu \zeta} R_{\epsilon}{}^{\eta}{}_{\mu}{}^{\theta} \
R_{\varepsilon \theta \zeta \eta}\nn\\&& + \frac{5}{3} H_{\alpha}{}^{\
\delta \epsilon} H^{\alpha \beta \gamma} \
H_{\varepsilon}{}^{\eta \theta} H^{\varepsilon \mu \zeta} \
R_{\beta \delta \gamma \epsilon} R_{\mu \eta \zeta \theta} + \
\frac{171}{8} H_{\alpha}{}^{\delta \epsilon} H^{\alpha \beta \
\gamma} H_{\beta}{}^{\varepsilon \mu} H_{\delta \
\varepsilon}{}^{\zeta} \
R_{\gamma}{}^{\eta}{}_{\epsilon}{}^{\theta} R_{\mu \eta \zeta \
\theta} \nn\\&&-  \frac{859}{96} H_{\alpha \beta}{}^{\delta} \
H^{\alpha \beta \gamma} H_{\gamma}{}^{\epsilon \varepsilon} H_{\
\epsilon \varepsilon}{}^{\mu} R_{\delta}{}^{\zeta \eta \theta} \
R_{\mu \eta \zeta \theta} -  \frac{101}{24} H_{\alpha \
\beta}{}^{\delta} H^{\alpha \beta \gamma} \
H_{\gamma}{}^{\epsilon \varepsilon} H_{\epsilon}{}^{\mu \zeta} \
R_{\delta}{}^{\eta}{}_{\varepsilon}{}^{\theta} R_{\mu \eta \
\zeta \theta} \nn\\&&+ \frac{59}{24} H_{\alpha \beta}{}^{\delta} \
H^{\alpha \beta \gamma} H_{\gamma}{}^{\epsilon \varepsilon} H_{\
\delta}{}^{\mu \zeta} R_{\epsilon}{}^{\eta}{}_{\varepsilon}{}^{\
\theta} R_{\mu \eta \zeta \theta} + \frac{17}{24} \
H_{\alpha}{}^{\delta \epsilon} H^{\alpha \beta \gamma} \
H_{\beta \delta}{}^{\varepsilon} H_{\gamma \epsilon}{}^{\mu} \
R_{\varepsilon}{}^{\zeta \eta \theta} R_{\mu \eta \zeta \
\theta} \nn\\&&-  \frac{37}{24} H_{\alpha \beta}{}^{\delta} H^{\alpha \
\beta \gamma} H_{\gamma}{}^{\epsilon \varepsilon} H_{\delta \
\epsilon}{}^{\mu} R_{\varepsilon}{}^{\zeta \eta \theta} R_{\mu \
\eta \zeta \theta} -  \frac{55}{8} H_{\alpha}{}^{\delta \
\epsilon} H^{\alpha \beta \gamma} H_{\beta}{}^{\varepsilon \
\mu} H_{\delta}{}^{\zeta \eta} R_{\gamma \epsilon \
\varepsilon}{}^{\theta} R_{\mu \theta \zeta \eta}\nn\\&& - 2 \
H_{\alpha}{}^{\delta \epsilon} H^{\alpha \beta \gamma} \
H_{\beta}{}^{\varepsilon \mu} H_{\delta}{}^{\zeta \eta} \
R_{\gamma \varepsilon \epsilon}{}^{\theta} R_{\mu \theta \zeta \
\eta} -  \frac{161}{8} H_{\alpha}{}^{\delta \epsilon} \
H^{\alpha \beta \gamma} H_{\beta}{}^{\varepsilon \mu} \
H_{\delta \varepsilon}{}^{\zeta} \
R_{\gamma}{}^{\eta}{}_{\epsilon}{}^{\theta} R_{\mu \theta \
\zeta \eta} \nn\\&&+ \frac{5}{48} H_{\alpha \beta}{}^{\delta} \
H^{\alpha \beta \gamma} H_{\gamma}{}^{\epsilon \varepsilon} H_{\
\delta \epsilon \varepsilon} R_{\mu \eta \zeta \theta} R^{\mu \
\zeta \eta \theta}
 \eeqa
 They can be reproduced by studying the six-point S-matrix element.  We have finally found 13 couplings with structure $H^2R^3$, \ie
 \beqa
 {\cal L}_3^{H^2R^3}&=&\frac{20}{3} H_{\alpha}{}^{\delta \epsilon} H^{\alpha \beta \
\gamma} R_{\beta \delta}{}^{\varepsilon \mu} \
R_{\gamma}{}^{\zeta}{}_{\varepsilon}{}^{\eta} R_{\epsilon \zeta \
\mu \eta} -  \frac{34}{3} H_{\alpha}{}^{\delta \epsilon} \
H^{\alpha \beta \gamma} \
R_{\beta}{}^{\varepsilon}{}_{\delta}{}^{\mu} \
R_{\gamma}{}^{\zeta}{}_{\varepsilon}{}^{\eta} R_{\epsilon \zeta \
\mu \eta}\nn\\&& + 2 H_{\alpha}{}^{\delta \epsilon} H^{\alpha \beta \
\gamma} R_{\beta}{}^{\varepsilon}{}_{\gamma}{}^{\mu} \
R_{\delta}{}^{\zeta}{}_{\varepsilon}{}^{\eta} R_{\epsilon \zeta \
\mu \eta} + \frac{8}{3} H_{\alpha}{}^{\delta \epsilon} \
H^{\alpha \beta \gamma} \
R_{\beta}{}^{\varepsilon}{}_{\delta}{}^{\mu} \
R_{\gamma}{}^{\zeta}{}_{\varepsilon}{}^{\eta} R_{\epsilon \eta \
\mu \zeta}\nn\\&& -  \frac{20}{3} H_{\alpha}{}^{\delta \epsilon} \
H^{\alpha \beta \gamma} \
R_{\beta}{}^{\varepsilon}{}_{\gamma}{}^{\mu} \
R_{\delta}{}^{\zeta}{}_{\varepsilon}{}^{\eta} R_{\epsilon \eta \
\mu \zeta} -  \frac{5}{3} H_{\alpha \beta}{}^{\delta} \
H^{\alpha \beta \gamma} R_{\gamma}{}^{\epsilon \varepsilon \
\mu} R_{\delta}{}^{\zeta}{}_{\varepsilon}{}^{\eta} R_{\epsilon \
\eta \mu \zeta} \nn\\&&-  \frac{10}{3} H_{\alpha}{}^{\delta \
\epsilon} H^{\alpha \beta \gamma} R_{\beta}{}^{\varepsilon}{}_{\
\delta}{}^{\mu} R_{\gamma}{}^{\zeta}{}_{\epsilon}{}^{\eta} \
R_{\varepsilon \zeta \mu \eta} + \frac{5}{6} H_{\alpha \
\beta}{}^{\delta} H^{\alpha \beta \gamma} \
R_{\gamma}{}^{\epsilon \varepsilon \mu} R_{\delta \epsilon}{}^{\
\zeta \eta} R_{\varepsilon \zeta \mu \eta}\nn\\&& + \frac{4}{3} \
H_{\alpha}{}^{\delta \epsilon} H^{\alpha \beta \gamma} \
R_{\beta}{}^{\varepsilon}{}_{\gamma}{}^{\mu} \
R_{\delta}{}^{\zeta}{}_{\epsilon}{}^{\eta} R_{\varepsilon \zeta \
\mu \eta} + 4 H_{\alpha}{}^{\delta \epsilon} H^{\alpha \beta \
\gamma} R_{\beta \delta \gamma}{}^{\varepsilon} \
R_{\epsilon}{}^{\mu \zeta \eta} R_{\varepsilon \zeta \mu \
\eta}\nn\\&& -  \frac{5}{3} H_{\alpha \beta}{}^{\delta} H^{\alpha \
\beta \gamma} \
R_{\gamma}{}^{\epsilon}{}_{\delta}{}^{\varepsilon} \
R_{\epsilon}{}^{\mu \zeta \eta} R_{\varepsilon \zeta \mu \
\eta} + \frac{14}{3} H_{\alpha}{}^{\delta \epsilon} H^{\alpha \
\beta \gamma} R_{\beta}{}^{\varepsilon}{}_{\delta}{}^{\mu} \
R_{\gamma}{}^{\zeta}{}_{\epsilon}{}^{\eta} R_{\varepsilon \eta \
\mu \zeta}\nn\\&& -  \frac{5}{6} H_{\alpha}{}^{\delta \epsilon} \
H^{\alpha \beta \gamma} R_{\beta \delta \gamma \epsilon} \
R_{\varepsilon \zeta \mu \eta} R^{\varepsilon \mu \zeta \eta}
 \eeqa
They can be reproduced by studying the five-point S-matrix element. Such couplings in fact have been found in  \cite{Liu:2019ses} using a scheme for the couplings which is different from the scheme \reef{T54} that we use in this paper for finding the above couplings.

\section{Discussion}

In this paper, we speculate that the bosonic part of the classical effective action of type II superstring theories is invariant under the $Z_2$-subgroup of $O(1,1,R)$ after dimensionally reducing the theory on a circle. We have shown that  imposition of the gauge symmetries and this $Z_2$-symmetry on the effective action  for NS-NS fields  at order $\alpha'^3$, can fix the effective action, \ie \reef{S3}, up to an overall factor.  In fact, the gauge symmetries require to have 872 couplings at order $\alpha'^3$ with unfixed coefficients \cite{Garousi:2020mqn}, and the T-duality symmetry  fixes these 872 parameters in terms of only one parameter.

Most of the couplings in \reef{S3} are new couplings which have not been found in the literature by other methods in string theory. When $B$-field is zero, the   couplings \reef{S3}  reduce to two Riemann quartet terms that their coefficients, after using the cyclic symmetry of the Riemann curvature,  become exactly the same as the coefficients  that have been found in  \cite{Gross:1986mw,Grisaru:1986vi,Freeman:1986zh} by the S-matrix  and the sigma-model methods. These couplings are invariant under the field redefinitions. However, the couplings which have $B$-field are not invariant under the field redefinitions, except the parameters $c_1,c_2,c_5,c_7,c_8$ in the couplings with structure $H^8$ \cite{Garousi:2020mqn}. The couplings \reef{S3} that we have found are in one particular scheme. To compare these couplings with the couplings in the literature that are found by the S-matrix elements, one has to first reproduce the field theory S-matrix elements using the couplings in the scheme \reef{T54} and  compare them with the corresponding string theory S-matrix elements  to fix the parameters in \reef{T54}. Then the fixed couplings should be compared with the couplings \reef{S3} that the T-duality produces. Using the scheme \reef{T54}, some of the parameters in \reef{T54} have been found in \cite{Garousi:2020mqn} by S-matrix element of four NS-NS vertex operators. Those couplings are exactly the same as the couplings that the T-duality constraint produces. It would be interesting to fix some of other parameters in \reef{T54} by comparing them with the string theory S-matrix element of five NS-NS vertex operators calculated in \cite{Liu:2019ses} and compare the resulting  couplings with the couplings in \reef{S3}.

The number of gauge invariant couplings in the minimal scheme \reef{T54} is 872. However, it is not guaranteed that the number of  couplings in the  string theory, \ie the  couplings   \reef{S3}, is minimum. One can use field redefinitions, total derivative terms and Bianchi identities to rewrite the couplings \reef{S3} in other schemes that may have less couplings than in \reef{S3}. In fact the two gravity couplings in \reef{RRf} and the first two couplings in \reef{H8} are invariant under the field redefinitions and total derivative terms. All other terms are scheme dependent.  One way to find a scheme with less number of couplings than 445 couplings in the present scheme, is to use a minimal scheme in which there are maximum number of couplings involving gravity and dilaton. In the scheme \reef{T54} there are 36 such couplings. On the other hand, it has been argued in \cite{Razaghian:2018svg} that the T-duality constraint makes the coefficients of these terms in the minimal schemes to be zero. In fact the couplings \reef{S3} and the couplings in the bosonic string theory at order $\alpha'$, $\alpha'^2$  \cite{Garousi:2019wgz,Garousi:2019mca} have no such couplings. In that scheme the number of T-duality invariant couplings may be less than 445 that we have found in the scheme \reef{T54}. It would interesting to write the couplings \reef{S3} in a scheme which has minimum number of couplings.  

The gravity couplings \reef{RRf} can be written in another    scheme as 
\beqa
\bS_3(G,\Phi)&=&-\frac{2a}{\kappa^2}\int d^{10}x\,\sqrt{-G}e^{-2\Phi}(t_8t_8+\frac{1}{4}\epsilon_8\epsilon_8)R^4\labell{S31}
\eeqa
where now the constant factor is $a=-\z(3)/(3\times 2^{13})$ and the tensors $\epsilon_8\epsilon_8$ and $t_8$ are defined as
\beqa
\epsilon_8{}^{\mu_1\cdots\mu_8}\epsilon_8{}^{\nu_1\cdots\nu_8}&=&\frac{1}{2}\epsilon_{10}{}^{\mu_1\cdots\mu_8\alpha\beta}\epsilon_{10}{}^{\nu_1\cdots\nu_8}{}_{\alpha\beta}\\
t_8^{\mu_1\cdots\mu_8}M^1_{\mu1\mu2}M^2_{\mu3\mu4}M^3_{\mu5\mu6}M^4_{\mu7\mu8}&=&8\Tr(M^1M^2M^3M^4)+8\Tr(M^1M^3M^4M^2)\nn\\&&+8\Tr(M^1M^3M^2M^4)-2\Tr(M^1M^2)\Tr(M^3M^4)\nn\\&&-2\Tr(M^1M^3)\Tr(M^2M^4)-2\Tr(M^1M^4)\Tr(M^2M^3)\nn
\eeqa
where $M^1,\cdots, M^4$ are four arbitrary antisymmetric matrices. It has been speculated in \cite{Kehagias:1997cq,Liu:2013dna,Liu:2019ses} that the B-field couplings may be included in above action by extending the curvature tensor to include the torsion, \ie
\beqa
\bS_3(G,\Phi,B)&=&-\frac{2a}{\kappa^2}\int d^{10}x\,\sqrt{-G}e^{-2\Phi}(t_8t_8+\frac{1}{4}\epsilon_8\epsilon_8)\cR^4\labell{S31}
\eeqa
where the generalized curvature $\cR$ is 
\beqa
\cR_{\mu\nu}{}^{\alpha\beta}&= &R_{\mu\nu}{}^{\alpha\beta}+\nabla_{[\mu}H_{\nu]}{}^{\alpha\beta}+\frac{1}{2}H_{[\mu}{}^{\alpha\gamma}H_{\nu]\gamma}{}^{\beta}\labell{cR}
\eeqa
We have checked explicitly that the four-field couplings in \reef{RRf}, \reef{PH4} and \reef{R2PH2} are fully consistent with  \reef{S31} after using on-shell relations.   The presence of the last terms in \reef{cR} has been supported by studying couplings with structure $H^2R^3$ in \cite{Liu:2019ses}. Having all couplings in scheme \reef{T54}, \ie \reef{S3},  one may use appropriate field redefinitions, total derivative terms and Bianchi identities to check if  the couplings \reef{S3} can be written in another scheme as  \reef{S31}. It has been observed in \cite{Garousi:2020mqn} that the couplings with coefficients  $c_1,c_2,c_5,c_7,c_8$ in \reef{T54} do not change under the  field redefinitions. The T-duality constraint produces $c_2=c_5=c_7=0$ and $c_1=-9/128,\, c_8=1/48$. In fact the first and the second  terms  in \reef{H8} do not change under field redefinitions. Hence one can check the proposal \reef{S31}  with these coefficients. If one   replaces $\cR_{\mu\nu}{}^{\alpha\beta}$ in \reef{S31} with $\frac{1}{2}H_{[\mu}{}^{\alpha\gamma}H_{\nu]\gamma}{}^{\beta}$, one would find $c_2\neq 0$ and $c_8=0$  which are different from what the T-duality constraint produces. Hence, our calculations indicate that the effective action \reef{S3} can not  be written in any other scheme as \reef{S31}.

Our calculations fix all parameters in the gauge invariant Lagrangian \reef{T54} up to an overall factor and fix many parameters in the T-duality transformation \reef{T22}. The parameters in the T-duality transformation dependent on the scheme that we use in \reef{T54}. One may consider the corrections to the Buscher rules as field redefinitions in the base space. Then one may ask if there is a scheme in which the T-duality transformation is only the standard  Buscher rules \reef{T2}? In other words, is there a scheme in which the T-duality requires no field redefinition in the base space? The answer is no because  the field redefinitions in the base space involve  much more parameters than the  parameters in the field redefinition in the original space, and this in turn is a result of the fact that  the number of fields in the base space is more than the number fields in the original space. The explicit  calculation to answer this question has been also done in  \cite{Garousi:2019wgz} for the couplings at order $\alpha'$ in the bosonic string theory. If one uses the field redefinition, then there would be  8 independent couplings in the original space. However, if one does not use the field redefinition, then there would be  20 independent couplings. The corresponding T-duality transformations have been found in \cite{Garousi:2019wgz} which has 11 arbitrary parameters. For no  choice of these parameters, the T-duality transformation is the standard Buscher rules.  

Assuming the reduction of the type II effective action at order $\alpha'^3$ on a circle has the $Z_2$ symmetry, we have fixed all 872 parameters in \reef{T54} up to one overall factor. One may ask if there is such symmetry in the effective action in the first place? One can verify the presence of this symmetry for the case that there is no B-field. In this case the effective action is given by \reef{RRf}. To simplify the calculation we assume the only non-zero field in the base space is $\vp$. Then one finds the reduction of \reef{RRf} is not invariant under  the Buscher rules. Its transformation produces the following anomalous terms: 
\beqa
S_3(\vp)-S_3(-\vp)=\alpha\int d^9x\Big[\frac{1}{2}\prt_a\prt_b\vp\prt^a\vp\prt^b\vp\prt_c\prt_d\vp\prt^c\prt^d\vp-\frac{1}{2}\prt_a\prt^c\vp\prt^a\vp\prt_b\prt^d\vp\prt^b\vp\prt_c\prt_d\vp\Big]\labell{ap1}
\eeqa 
where $\alpha$ is an overall factor. If one considers the following corrections to the Buscher rules:
\beqa
\vp&\rightarrow& -\vp\nn\\
\bphi&\rightarrow&\alpha'^3\Big(\frac{1}{2}\prt_a\prt_b\prt^c\prt_c\vp\prt^a\vp\prt^b\vp+\frac{1}{4}\prt_a\prt^c\prt_c\vp\prt^a\vp\prt^b\prt_b\vp-\frac{1}{16}\prt_a\vp\prt^a\vp\prt^b\prt_b\prt^c\prt_c\vp\nn\\
&&\qquad+2\prt_a\prt_b\prt_c\vp\prt^a\vp\prt^b\prt^c\vp+\frac{3}{4}\prt^a\prt_a\vp\prt_b\prt_c\vp\prt^b\prt^c\vp\Big)
\eeqa
Then one finds the T-duality of the leading order effective action produces also the following anomalous couplings at order $\alpha'^3$:
\beqa
\delta^3S_0(-\vp)&=&\alpha\int d^9x\Big[-\frac{1}{2}\prt^d\vp\prt_d\vp\Big(\frac{1}{2}\prt_a\prt_b\prt^c\prt_c\vp\prt^a\vp\prt^b\vp+\frac{1}{4}\prt_a\prt^c\prt_c\vp\prt^a\vp\prt^b\prt_b\vp\labell{ap2}\\
&&\qquad\qquad-\frac{1}{16}\prt_a\vp\prt^a\vp\prt^b\prt_b\prt^c\prt_c\vp+2\prt_a\prt_b\prt_c\vp\prt^a\vp\prt^b\prt^c\vp+\frac{3}{4}\prt^a\prt_a\vp\prt_b\prt_c\vp\prt^b\prt^c\vp\Big)\Big]\nn
\eeqa
These two anomalous couplings \reef{ap1} and \reef{ap2}  cancel each other up to some total derivative terms, \ie
\beqa
S_3(\vp)-S_3(-\vp)-\delta^3S_0(-\vp)&=&\alpha\int d^9x \prt_a J_3^a
\eeqa
where the vector $J_3^a$ is the following:
\beqa
J_3^a&=&-\frac{1}{32} \prt^a\prt^d\prt_d\vp\prt_b\vp\prt^b\vp\prt_c\vp\prt^c\vp-\frac{1}{4}\prt^a\vp\prt_b\prt^d\vp\prt^b\vp\prt_c\prt_d\vp\prt^c\vp\nn\\
&&+\frac{1}{4}\prt^a\vp\prt_b\vp\prt^b\vp\prt_c\prt_d\vp\prt^c\prt^d\vp+\frac{1}{4}\prt^a\prt_c\prt_d\vp\prt_b\vp\prt^b\vp\prt^c\vp\prt^d\vp\nn\\
&&+\frac{1}{8}\prt^a\prt_c\vp\prt_b\vp\prt^b\vp\prt^c\vp\prt^d\prt_d\vp
\eeqa
Using the Stokes's theorem, the total derivative terms become zero for the closed spacetime manifold. Hence, assuming there are higher derivative corrections to the Buscher rules, one finds the $Z_2$ symmetry for the effective action of type II superstring theories at order $\alpha'^3$. If the spacetime manifold has boundary, the total derivative terms can not be ignored. In that case the $Z_2$ symmetry restores if one includes some couplings in the boundary \cite{Garousi:2019xlf}. It would be interesting to find the boundary terms in the effective action \reef{S3}.



\end{document}